\documentclass[preprint]{revtex4-1}

\usepackage{graphicx}
\usepackage{pstricks}
\usepackage{amsmath}
\usepackage{bm}
\usepackage{tabularray}

\usepackage[position=top]{subfig}
\usepackage{setspace}
\usepackage{floatrow}
\usepackage{epstopdf}

\begin{document}

\title{Electronic, optical and topological properties of defects in bismuthene}

\author{Gabriel Elyas Gama Araújo}
\author{Andreia Luisa da Rosa}
\affiliation{Universidade Federal de Goi\'as, Institute of Physics, Campus Samambaia, 74960600 Goi\^ania, Brazil}
\author{Mauricio Chagas da Silva}
\affiliation{Université de Lorraine, CNRS, LPCT, 54000 Nancy, France}
\author{Alexandre Cavalheiro Dias}
\affiliation{University of Bras{\'{i}}lia, Institute of Physics and International Center of Physics, Bras{\'{i}}lia $70919$-$970$, Distrito Federal, Brazil}
\author{Thomas Frauenheim}
\affiliation{Bremen Center for Computational Materials Science, University of Bremen, Am Fallturm 1, 28359 Bremen, Germany}
\affiliation{Shenzhen Computational Science and Applied Research Institute, Shenzhen, China}
\affiliation{Beijing Computational Science Research Center, Beijing, China}

\begin{abstract}

In this work we use first principles density-functional theory and Bethe-Salpeter equation together with tight-binding based maximally localized wannier functions (MLWF-TB) to investigate the electronic, optical and topological properties of two-dimensional bismuth (bismuthene) containing vacancy defects. We demonstrate that these properties depend on the shape and size of the nanopores. Furthermore, ab-initio molecular dynamics  (AIMD) simulations show that all pores are thermally stable at room temperature.  Finally, adsorption of gas phase small molecules indicate that these pores behave as sensors, opening the path for further applications in gas separation and sensing.

\end{abstract}


\maketitle

\section{Introduction}

Solid state nanopore sensors have arisen  as a promising tool for detection of organic and biomolecules, energy conversion and storage and gas separation\,\cite{HE2024215816,NPGAsia2021,NatureReview2020,D4NR03966A}, paving the path for applications in advanced eletronic devices\,\cite{acs.langmuir.4c04961,HE2024215816}. 
Advances in solid-state nanopores fabrication have been reported in several monolayered materials, such as 
transition metal dichalcogenides\,\cite{NatCommMoS22023,NatureMoS2}, graphene\cite{acsnano.1c07960,D4NR03966A,acsnano.9b09353,pnas.1119827109} and SiN\cite{YAN2024156560,acsnano.2c07240}. Current atomic-scale techniques allow the pore sizes to be adjusted by removing only a few atoms.  
Solid-state nanopores are robust, flexible, and can relatively easily be integrated into other solid-state electronic devices, compared to their biological counterparts.

Layered bismuthene has been successfully synthesized on SiC
substrates\,\cite{Reis2017,Langmuir_Yaegashi_2022}, Ag (111) \cite{PhysRevB.108.L161407} or in its free-standing form\,\cite{APLMAT2022,NatComm2020}. More recently hollow bismuthene sheets with layers thicknesses of few layers
layers were directly fabricated on stainless-steel  substrates\,\cite{hollow_bi}. 

Bismuth monolayers have earlier predicted to be a topological insulator (TI),\cite{Rivelino2015,AsiaNat,Reis2017} with 2D
robust edge states against applied strain and under the
application of an external electric field\,\cite{C7RA05787C,NL_heine,Wittemeier_2022,Sarker_2020}. In addition, Quantum Spin Hall (QSH) states have been predicted to emerge as the bismuth layers films are adsorbed with hydrogen and other small functional
groups\,\cite{ROSA2021121849,Rivelino2015,NL_heine,JPCC2020}.

Earlier theoretical molecular dynamics calculations show that hydrogenated 
bismuthene undergo a phase transition to more open, amorphous structures ate higher temperatures stable. Surprisingly, hydrogenated bismuthene retains its topological
insulating behavior,\cite{Costa2019}. 
However, it remains unclear whether other type of defective layers retain their topological properties or are stable under realistic environments.

In addition, excitons (electron-hole pairs) have quantum degrees of freedom, such as spins, that can be optically manipulated for control and readout. Bismuthene, with large SOC, provides an interesting platform to explore the internal spin structure of excitons\,\cite{pnas.2307611120}.  Hydrogenated bismuthene demonstrate that time reversal symmetry breaking associated with  spin degree of freedom  results in total dichroism.\cite{sciadv.aay2730} Measuring circular dichroism from a 2D system allows
one to directly determine its topological property.
\cite{sciadv.aay2730} far-infrared reflectivity and Kerr angle spectra on a high-quality crystal of pure semi-metallic bismuth  as a function of magnetic field, and the conductivity for left- and right-handed circular polarizations, which allows further understanding towards valley polarized
effects\,\cite{PhysRevLett.117.017402}. 

Indeed, theoretical calculations suggested valley-pollarization features giving origin to quantum anomalous Hall states
emerge as bismuthene are adsorbed with hydrogen\,\cite{Niu2015,ROSA2021121849,PhysRevB.100.165417,npj2018} with an exciton with binding energy of 0.18\,eV \cite{Ciraci2019}. This
indicates that bismuthene is promising for applications in solar
cells, light-emitting devices and quantum information.

In this paper we perform first principles density-functional theory calculations to investigate the electronic, optical and topological properties of bismuthene containing vacancy defects forming subnanonoter pores. We demonstrate that these properties depend on the shape and size of the nanopores allowing to distinguish among different pore terminations. AIMD simulations show that all pores are thermally stable at room temperature.  Adsorption of gas phase small molecules indicate that these pores can be perform as sensors, opening the path for further investigations on gas separation and gas sensing.

\section{Methodology}

In this work we use density-functional
theory\,\cite{Hohenberg:64,Kohn:65} within the generalized gradient
approximation (GGA) according to the  parameterization of PBE (Perdew-Burke-Ernzerhof)\,\cite{Perdew:96} for the exchange an correlation potential to investigate the electronic, optical and thermal stability of small size pores in bismuthene. The projected augmented wave
method (PAW)\,\cite{Bloechel:94,Kresse:99} has been used to described the electronic potential. A
($6\times6\times1$) supercell with a (4$\times$4$\times$1) {\bf k}-point
sampling and an energy cutoff of 400\,eV is used to calculate the
atomic relaxation, electronic structure and molecular dynamics simulations of the two-dimensional
nanopores.  Spin-orbit coupling (SOC) is included in the solution of the
Kohn-Sham (KS) equations. The calculations of Z$_2$
topological invariants were performed with the Z2Pack
package\,\cite{PhysRevB.83.235401,PhysRevB.95.075146}.

AIMD simulations have been pursued using a NVT ensemble with the structures in contact with an Andersen thermostat at 300\,K. Simulation time of 10\,ps has been performed.  Forces on atoms have been converged up to 10$^{-6}$ eV/{\AA}. All these methods have been implemented in the VASP code\,\cite{Kresse:99}. Structure optimizations assured the minimization of stress tensor and atomic forces until the atomic forces fell below 0.01 eV/{\AA}. To eliminate any spurious interactions between the monolayer and its periodic images in the $z$-direction, we incorporated a vacuum layer with a thickness of 10\,{\AA}. For all calculations, \textbf{k}-meshes were automatically generated utilizing the Monkhorst-Pack method \cite{Monkhorst_5188_1976}, ensuring a \textbf{k}-points density of 40\,{\AA} in the directions of the in-plane lattice vectors. 
The optical response of the materials was computed to include excitonic effects by solving the Bethe–Salpeter equation (BSE)\cite{Salpeter_1232_1951} and also in the independent particle approximation  (IPA) by employing the WanTiBEXOS package\cite{Dias_108636_2022}. We first created a maximally localized Wannier function tight-binding (MLWF-TB) Hamiltonian to conduct these evaluations derived from DFT@GGA calculations including SOC through the Wannier90 package\,\cite{Mostofi_685_2008}. For the solution of the Bethe-Salpeter equation (BSE), we employed the Rytova–Keldysh 2D Coulomb Potential (V2DRK) \cite{Rozzi_205119_2006}, considering the monolayers in a vacuum setting. Each monolayer effective dielectric screening parameter was determined by averaging the in-plane diagonal components of the macroscopic dielectric constant, with a unit cell vacuum spacing of 15\,{\AA} in the non-periodic direction. Our calculations were performed with a 120{\AA} density of \textbf{k}-points to  determine the imaginary parts of the dielectric function. The optical properties were calculated at the Independent Particle Approximation (IPA) and BSE levels, considering the lowest 8-14 conduction bands and the highest 6-12 valence bands. 
Production of images have been provided by the VESTA\,\cite{VESTA}, Phonopy\,\cite{phonopy1,phonopy2} and xmgrace softwares\cite{xmgrace}. 

\section{Results and Discussions}

\begin{figure}[ht!]
  \centering
   \subfloat[]{\includegraphics[width = 5cm,scale=1, clip = true]{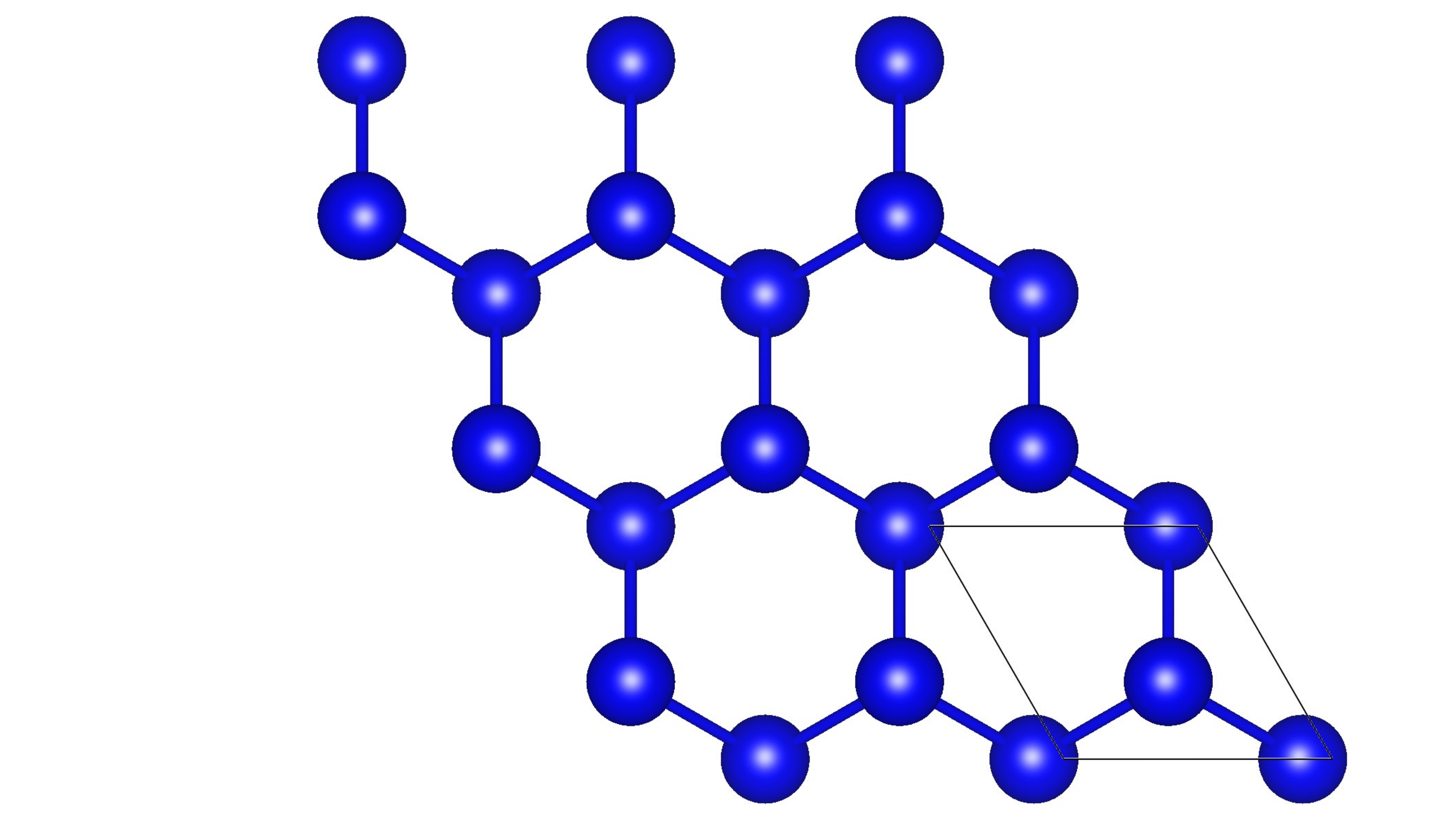}}
    \subfloat[]{\includegraphics[width =3cm,scale=1, clip = true]{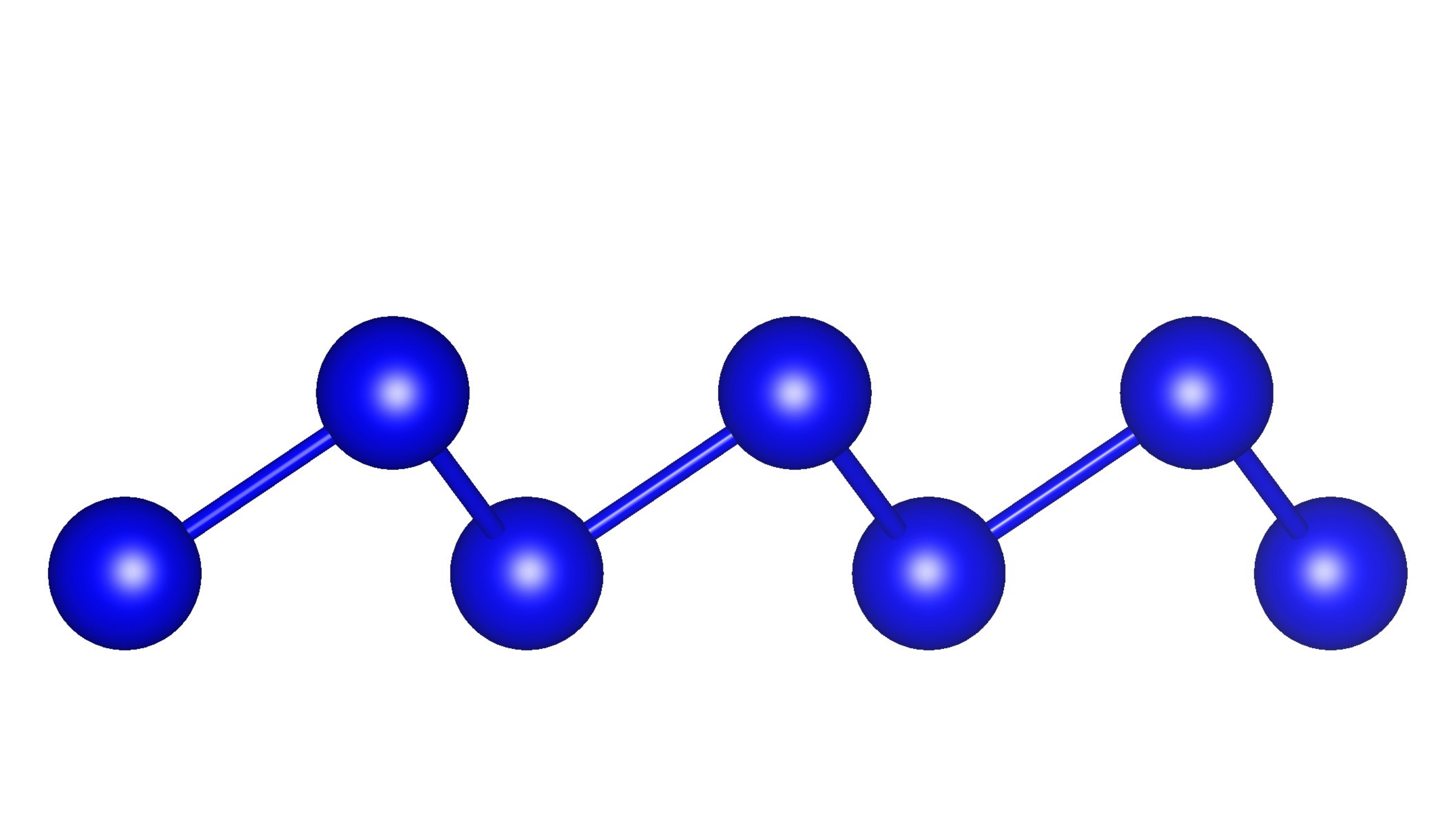}}
    \hspace{0.3cm}
     \subfloat[]{\includegraphics[width = 5cm,scale=1, clip = true]{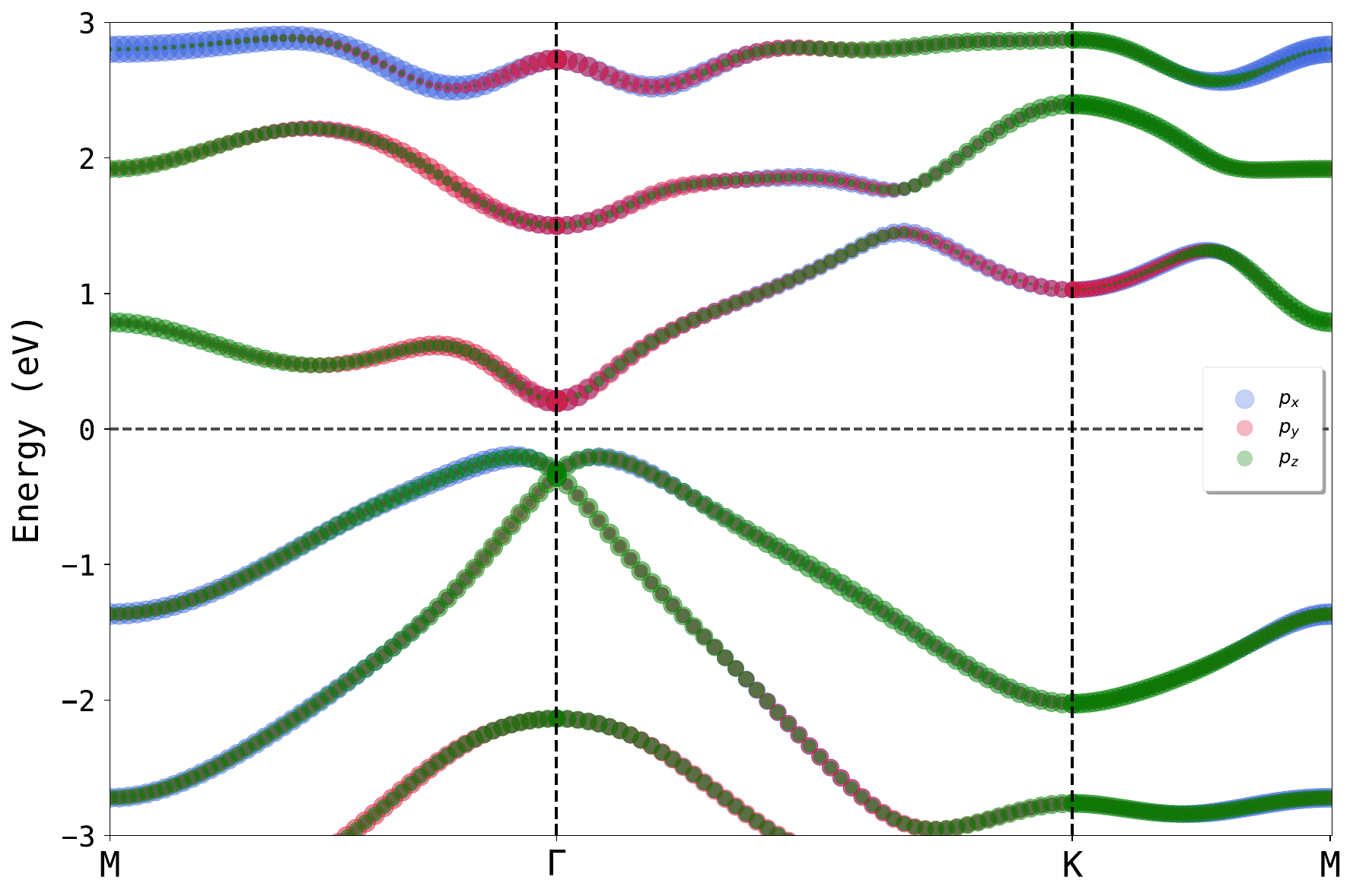}} 
\caption{\label{fig:bare_band} a) Top view, b) side view and c) band structure of bare buckled bismuthene calculated within DFT-GGA. The unit cell is shown in solid lines.}
\end{figure}

The stability of the buckled and planar structure has reported in previous publications.\cite{Rivelino2015,ROSA2021121849,Ciraci2019}. The free-stading buckled bismuthene was found to be more stable than the flat one with in-plane lattice constant of 5.30 {\AA}. The inclusion of SOC does not change significantly the lattice parameters, but it lowers the total energy. The optmized structure is shown in Fig. \ref{fig:bare_band} (a) together with its orbital decomposed band structure (b).

\begin{figure}[ht!]
  \centering
  \subfloat[]{\includegraphics[width = 3.5cm,scale=1, clip = true]{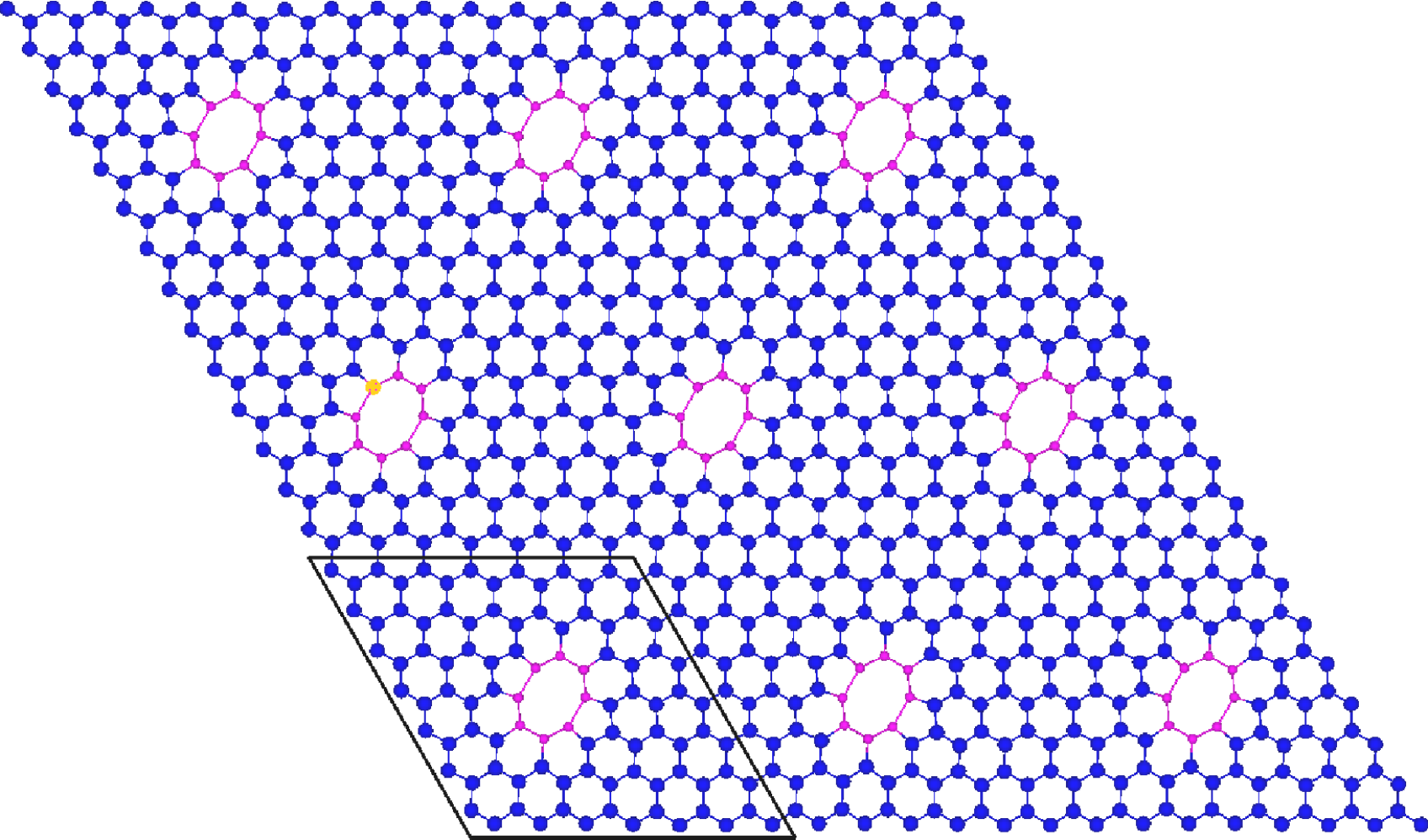}} 
\subfloat[]{\includegraphics[width = 3.5cm,scale=1, clip = true]{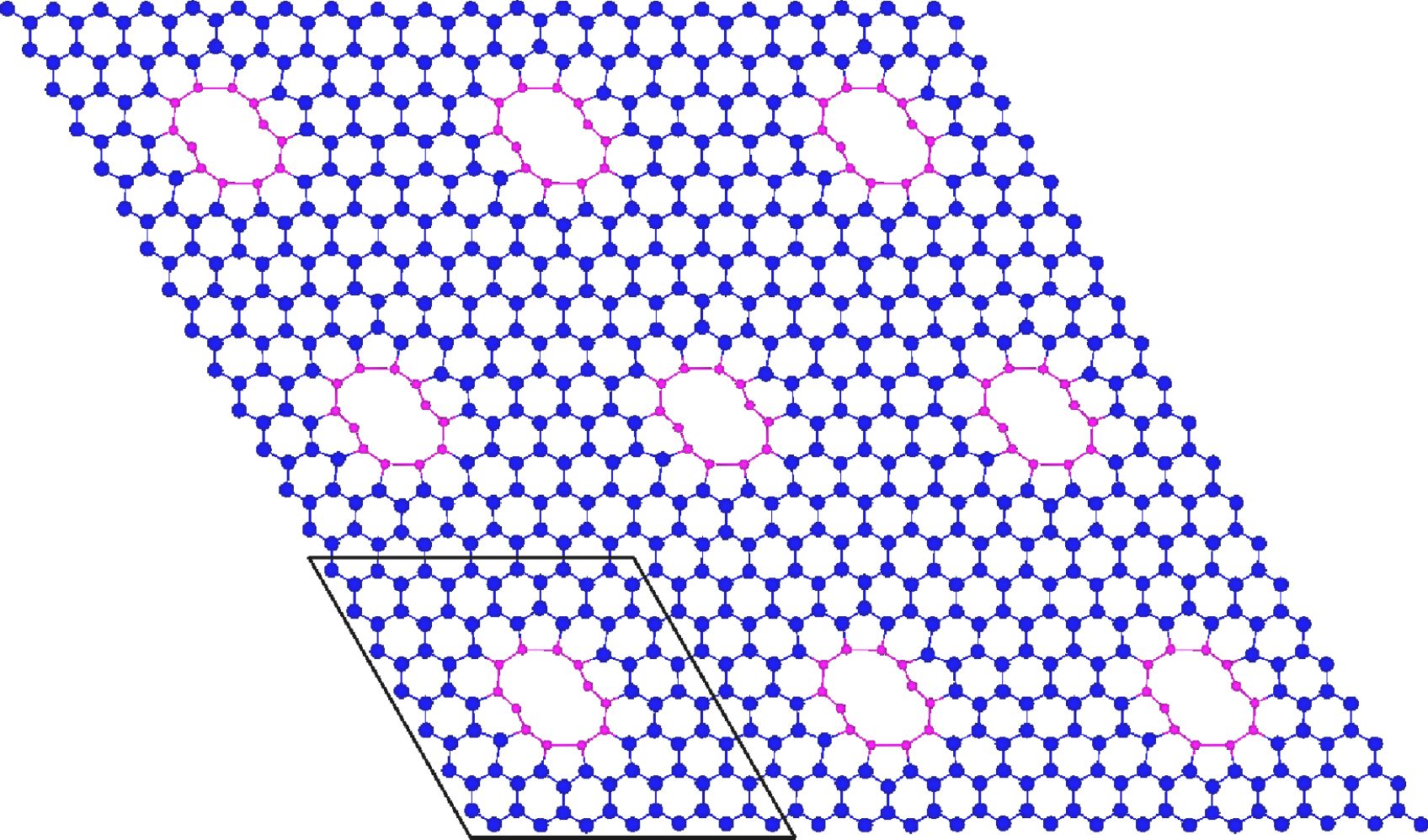}}
\subfloat[]{\includegraphics[width = 3.5cm,scale=1, clip = true]{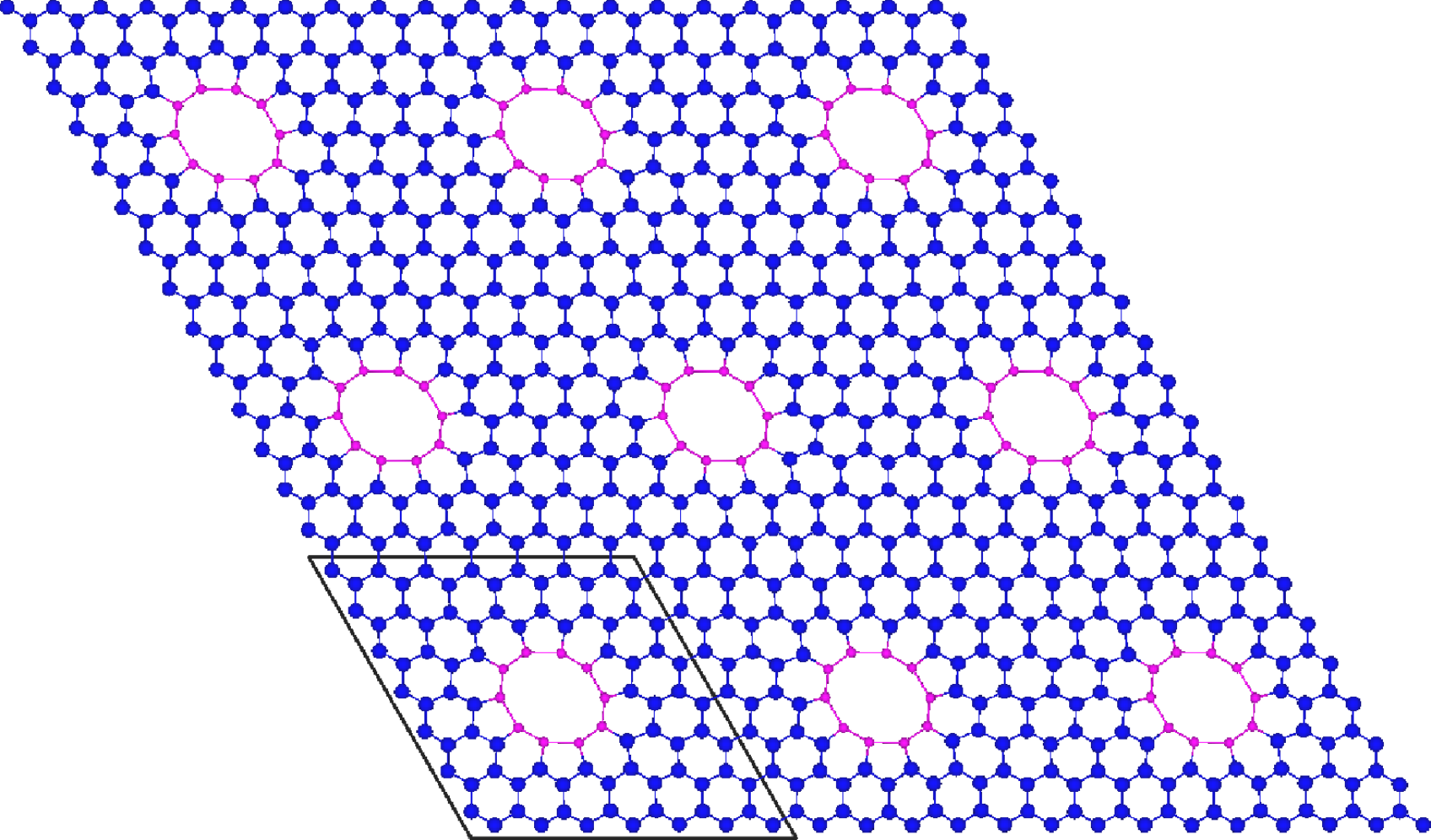}} 
\subfloat[]{\includegraphics[width = 3.5cm,scale=1, clip = true]{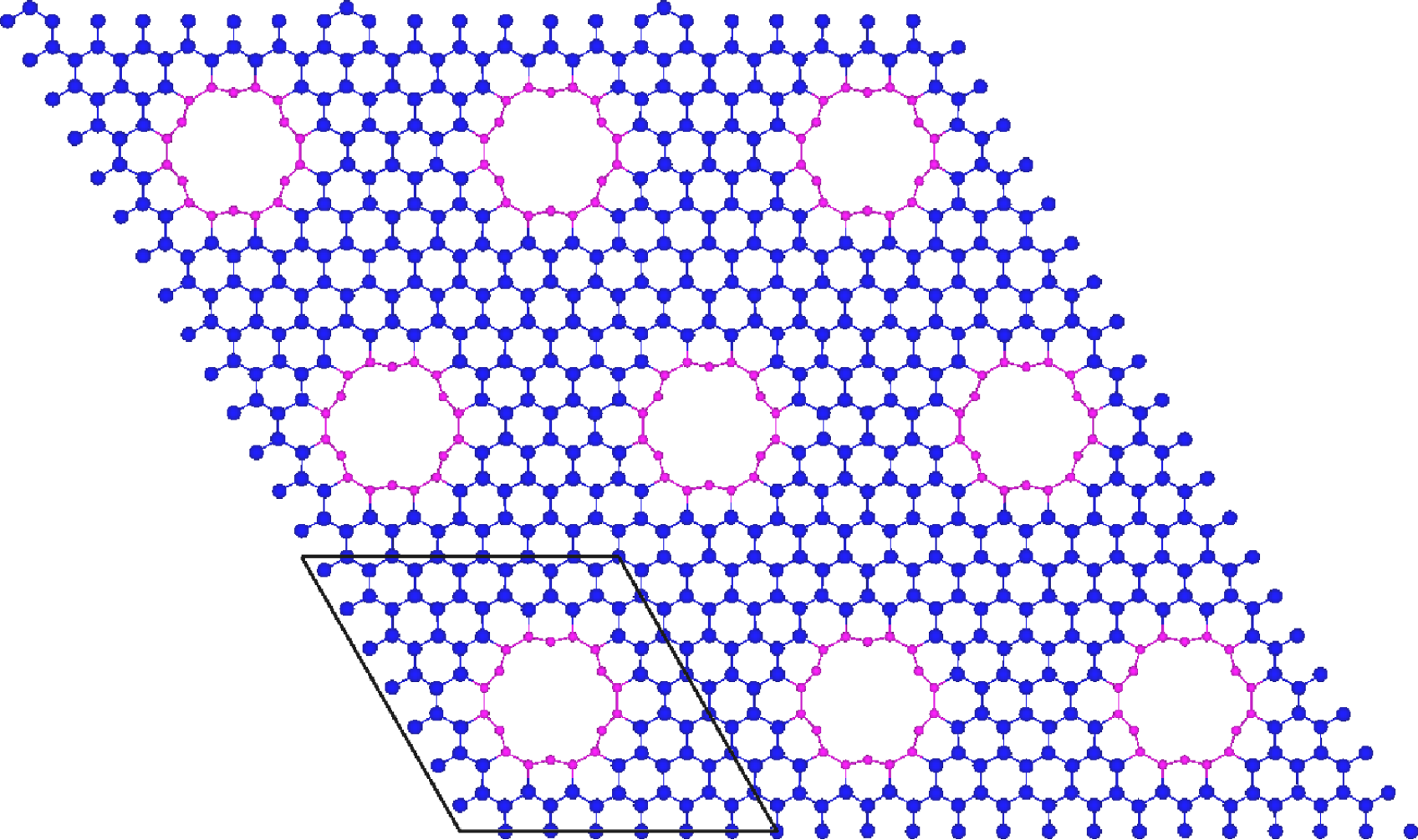}}\\ 
\subfloat[]{\includegraphics[width = 3.5cm,scale=1, clip = true]{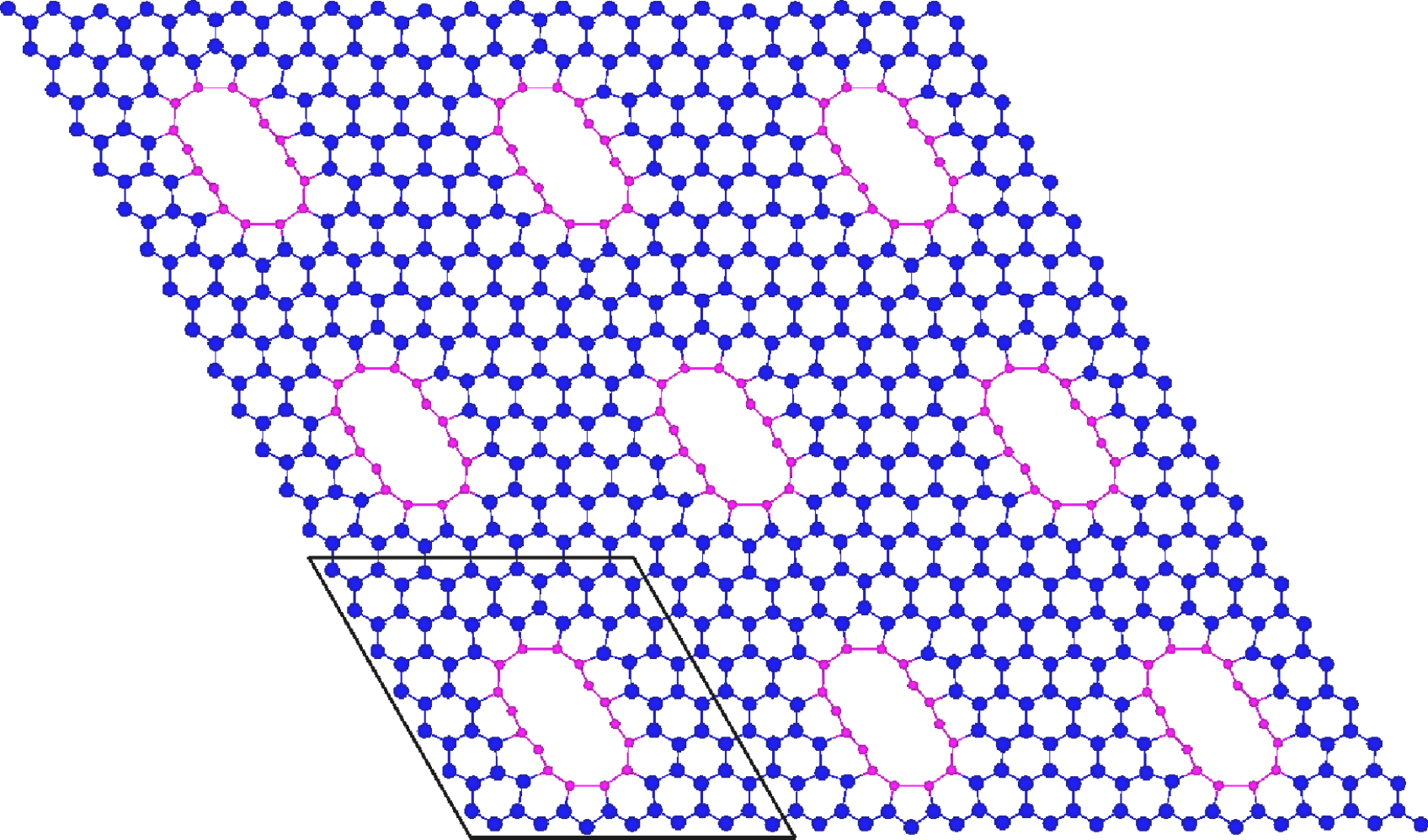}}
\subfloat[]{\includegraphics[width = 3.5cm,scale=1, clip = true]{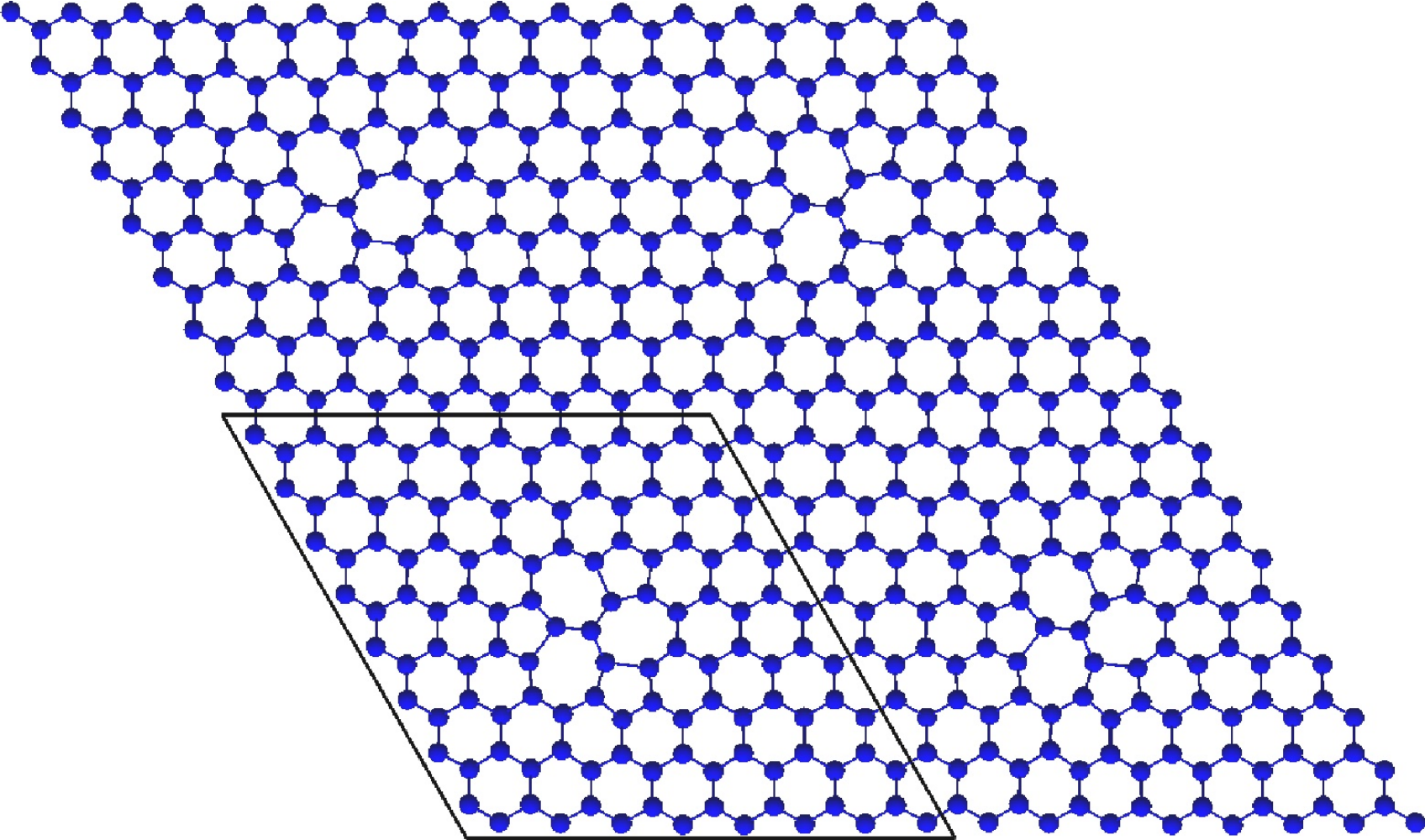}}
\subfloat[]{\includegraphics[width = 4.5cm,scale=1, clip = true]{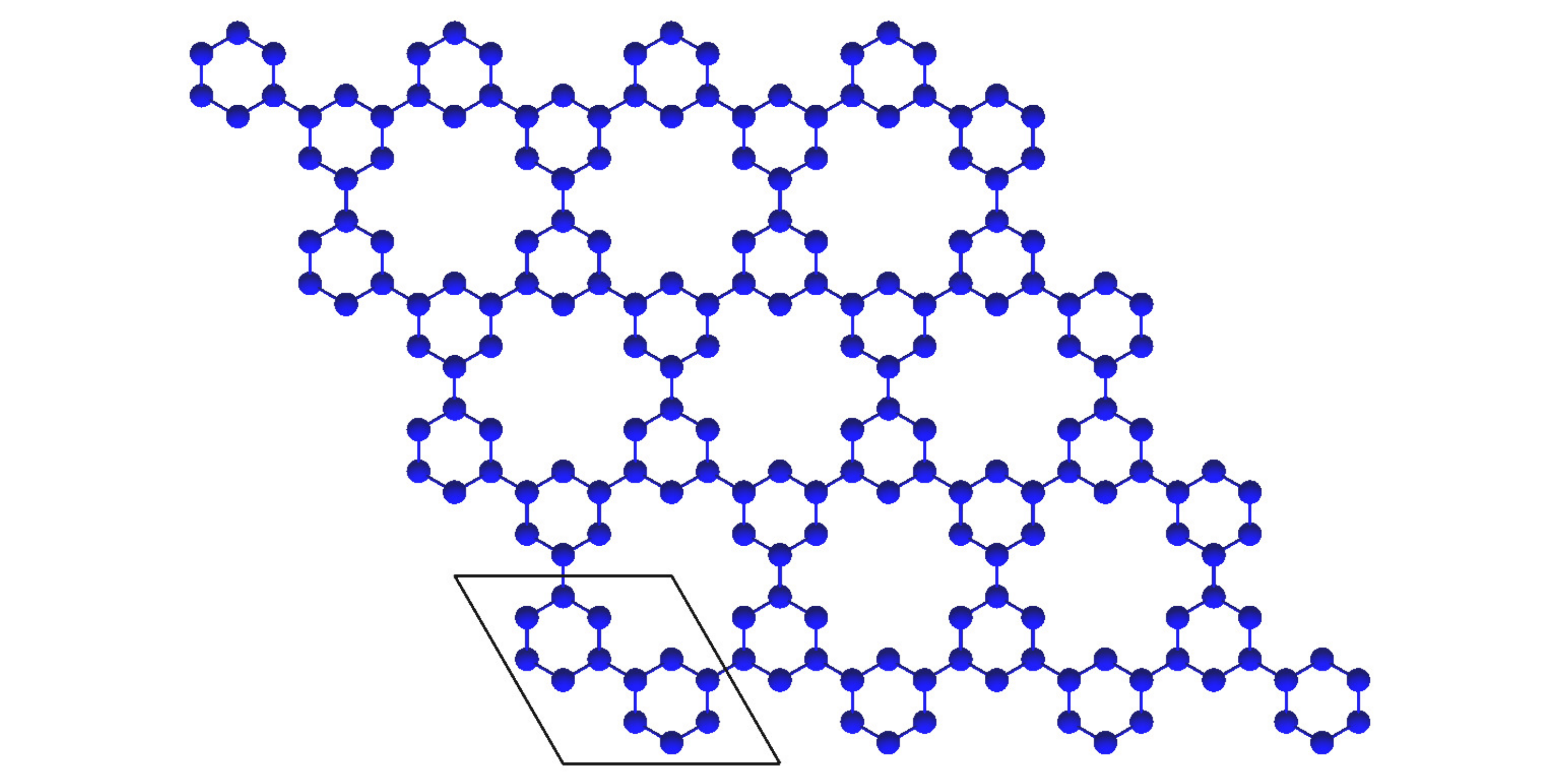}} 
\subfloat[]{\includegraphics[width = 4.5cm, scale=1, clip = true]{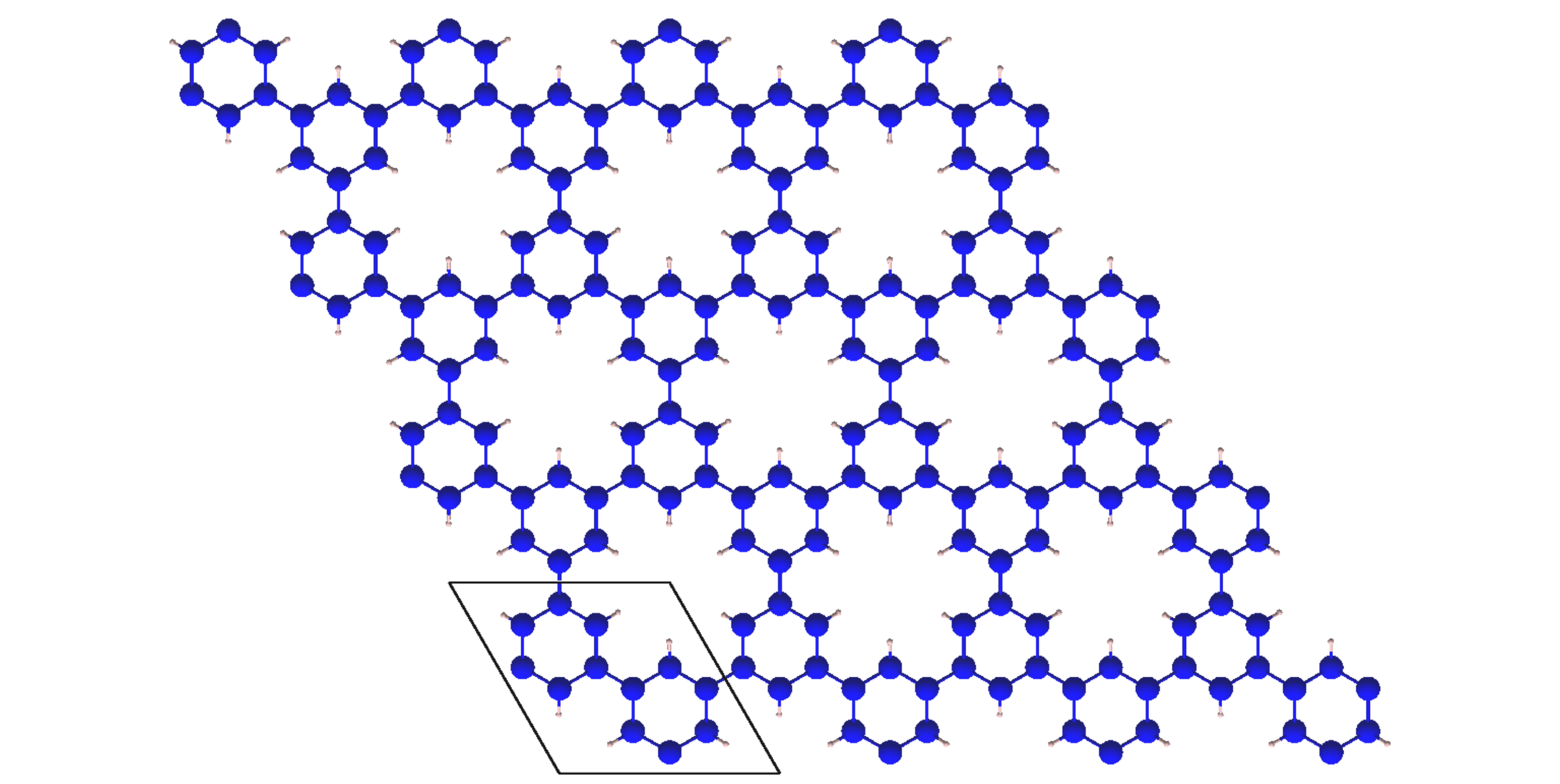}}  
\caption{\label{fig:structures} Top views of bismuthene nanopores. a) divacancy (P2 or 585), b) tetravacancy (P4), c) hexavacancy P6a , d) P6b hexavacancy P6b, e) hexavacancy P6c, f) pentagon-heptagon (P555-777), g) hexagonal bare (type-2) and h) hexagonal hydrogenated (type-2). The unit cell used in the calculation is shown in solid lines.}
\end{figure}

The structural, electronic and topological properties of the buckled and planar bismuthene structure have been reported in a previous publication\,\cite{JPCC2020,ROSA2021121849,Rivelino2015}. Free-stading buckled bismuthene
was found to be more stable than the flat one by more than 1.5eV \cite{ROSA2021121849}. Buckled
bismuthene is a topological insulator, whereas planar bismuthene is a trivial
semiconductor with a gap of 0.5\,eV.

In this work we have investigated pores of different sizes by removing 2-6 atoms from a perfect monolayer. Additional structure weren also considered. The relaxed structures are shown in Fig.\,\ref{fig:structures}.  The pore size has been considered as the largest atom-atom separation in the pore. The pores have sizes of: 585/P2 (8.15 {\AA}), P4 (6.30
{\AA}), P6a (10.9 {\AA}), P6b (13.3\,{\AA}), P6c (15.2 {\AA}),
555-777 (5.5-6.3 {\AA}). Additionally we  have investigated  additional defective structures  which we call bare hexagonal pore type-2 and
hydrogenated hexagonal pore type-2. In these layers
the bismuthene nanopores are connected by two benzene-like rings forming a network. The bare (hydrogenated) hexagonal pore type-2  has width of 10.3 (7.1) {\AA}.

Fig.\,\ref{fig:structures}(a) shows the
divacancy (P2/585 defect). The Bi-Bi bond length range is
3.07-3.12\,{\AA}. P4 is shown in Fig.\,\ref{fig:structures}(b) with
Bi-Bi distances of 3.04, 3.12, 3.29, 3.06 and 3.07\,{\AA}. The smallest distance of 3.04 {\AA} belongs to the inner side of the pore.

The P6a hexavacancy
shown in Fig.\,\ref{fig:structures}(c) leads to Bi-Bi bond lengths of
of 3.23, 3.07 and 3.31\,{\AA}. The 3.23 \,{\AA} belongs to the
distorted hexagon in the pore. P6b shown in has Bi-Bi distances of
3.02, 3.07, 3.06 as shown in Fig.\,\ref{fig:structures}(d). P6c in
Fig.\,\ref{fig:structures}(e) has bond lengths of 3.04, 3.09, 3.06 and
3.12. We have also investigate periodic porous structures as shown in
Figs.\,\ref{fig:structures}(f) and (h). In the later cases the lattice
parameter is 3.07 {\AA}. Fig.\,\ref{fig:structures}(e) shows the 585 which is a bismuth double
vacancy as initial structure and relaxes to a almost spherical
pore. Fig.\,\ref{fig:structures}(f) shows the bismuth pore 555-777,
which consists of pentagons-heptagons. 

\begin{figure*}[ht!]
\centering
\subfloat[]{\includegraphics[width = 8cm,scale=1, clip = true]{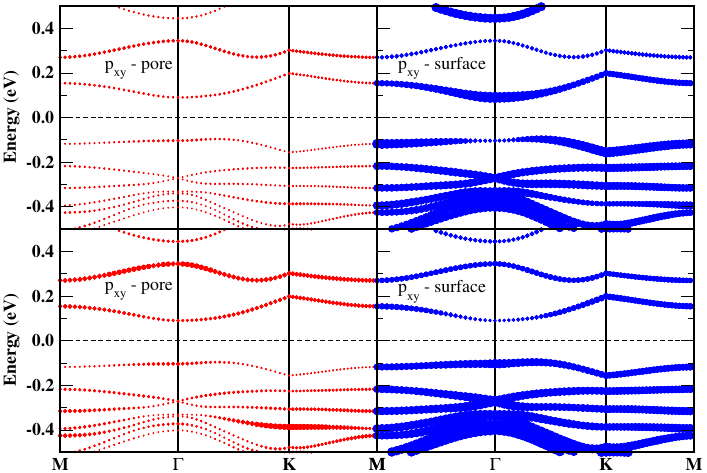}}
\subfloat[]{\includegraphics[width = 8cm,scale=1, clip = true]{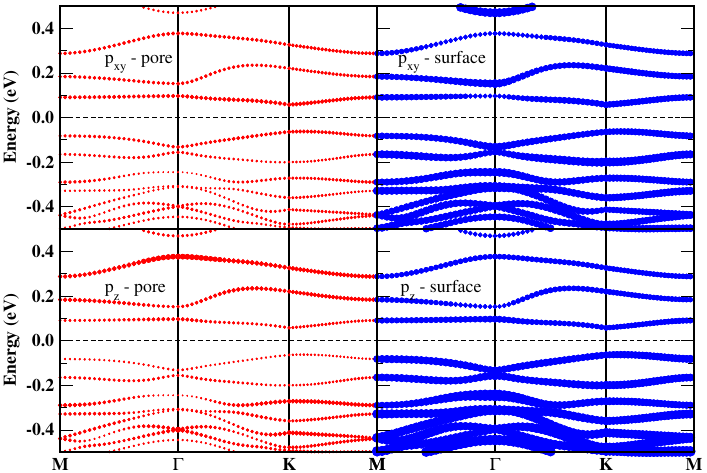}}\\
\subfloat[]{\includegraphics[width = 8cm,scale=1, clip = true]{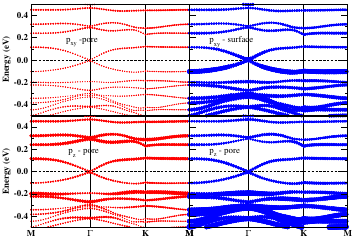}} 
\subfloat[]{\includegraphics[width = 8cm,scale=1, clip = true]{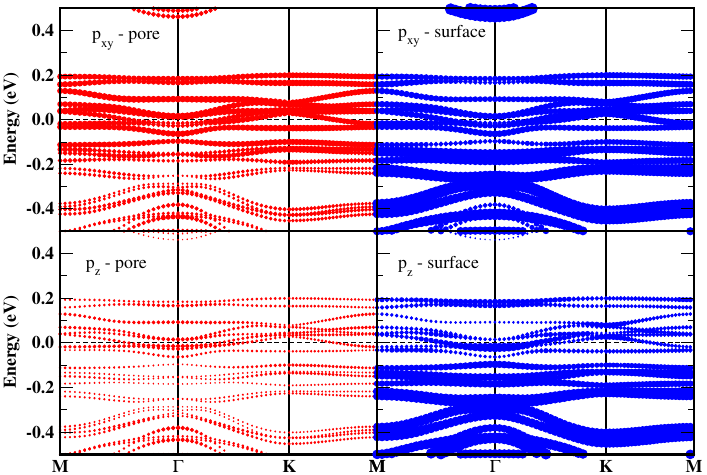}} \\
\subfloat[]{\includegraphics[width = 8cm,scale=1, clip = true]{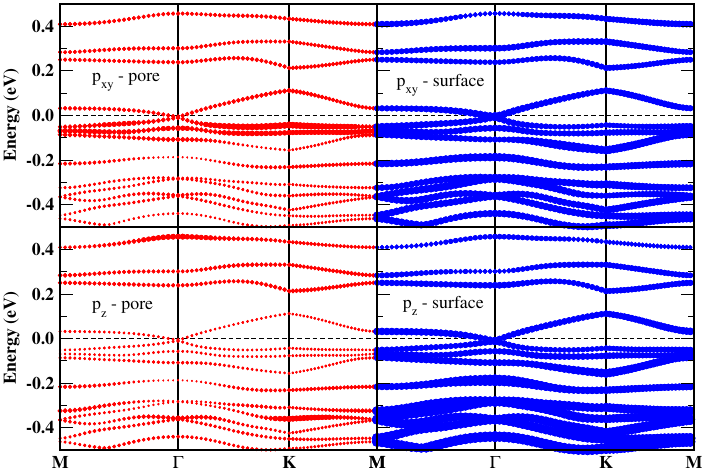}} 
\caption{Orbital decomposed band structure of bismuthene nanopores. a) 585, b) tetravacancy P4, c) hexagonal vacancy P6a, d) hexagonal vacancy P6b, e) hexagonal vacancy P6c.}
\label{fig:projected_bands1}
\end{figure*}

AIMD simulations have been pursued using a NVT ensemble with the structures in contact with an Andersen thermostat at 300\,K. Simulation times between 5 and 10\,ps have been performed. Results are shown in Fig. S1.

\begin{figure*}[ht!]
\centering
\subfloat[]{\includegraphics[width = 6cm,scale=1, clip = true]{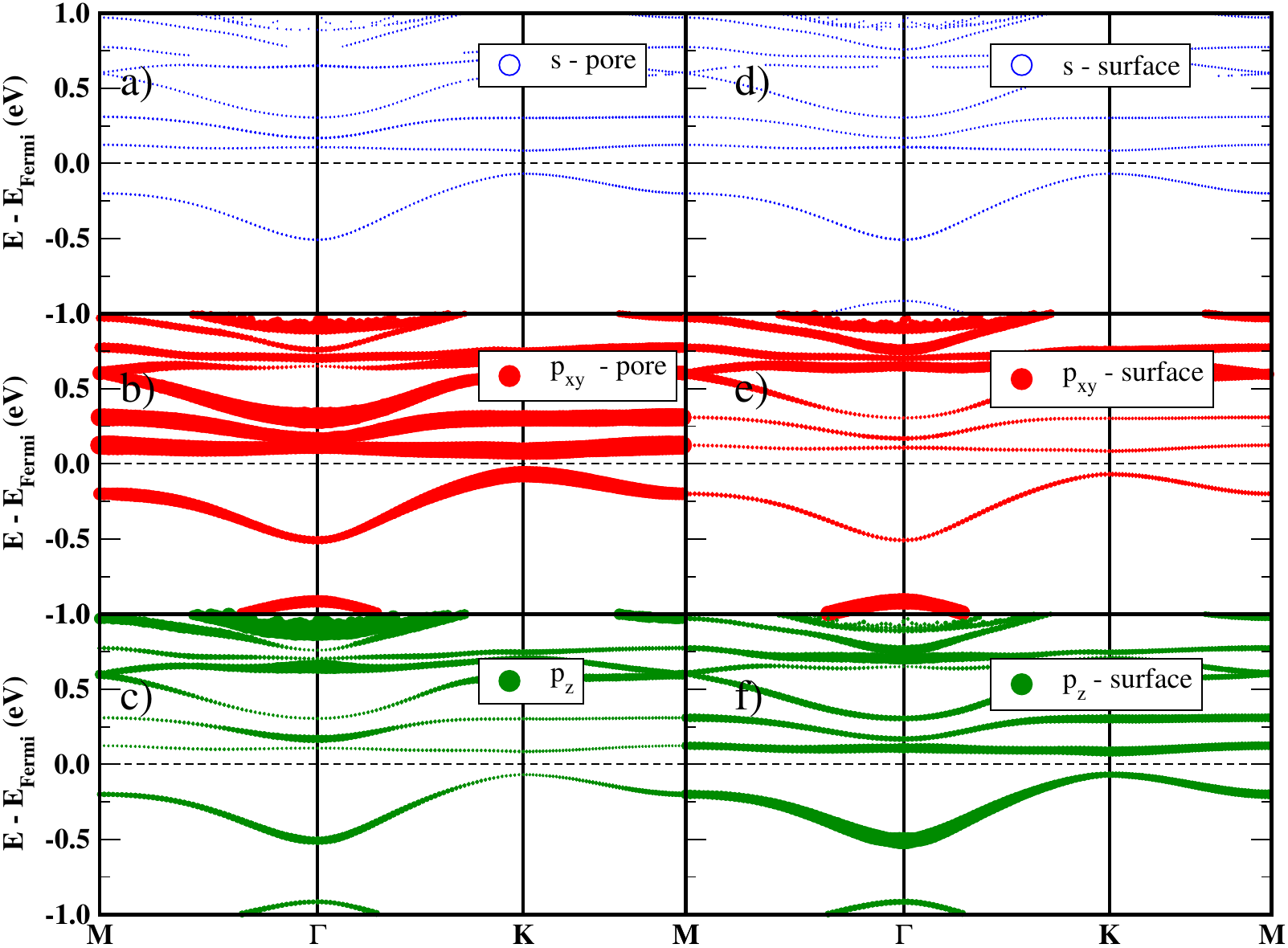}} 
\subfloat[]{\includegraphics[width = 6cm, scale=1, clip = true]{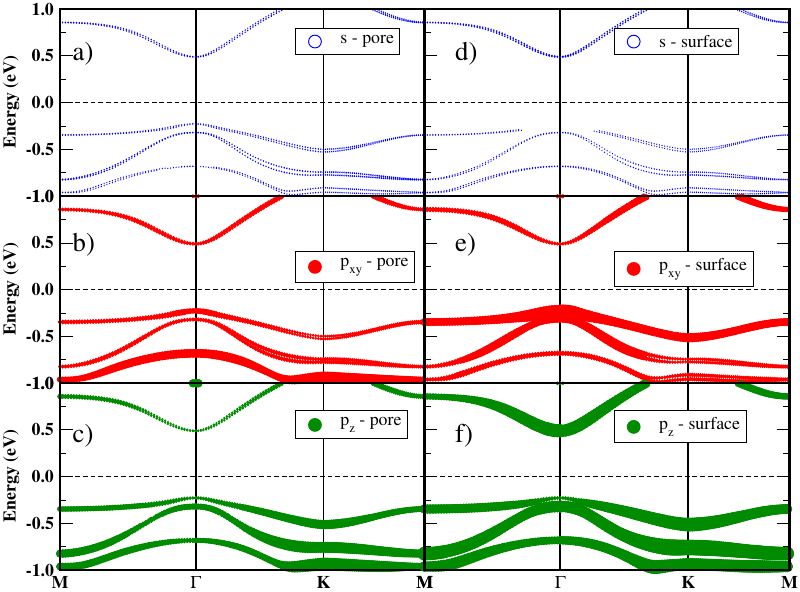}}  \\
\caption{Band structure of bismuthene nanopores. a) bare hexagonal pore type-2 and b) hydrogenated hexagonal pore type-2.}
\label{fig:projected_bands2}
\end{figure*}

In order to further understand the individual orbital pore contributions to
the band edges, we have calculated the orbital projected band
structures. The orbital decomposed band structure is shown in Figs. \ref{fig:bare_band}\,\ref{fig:projected_bands1},\ref{fig:projected_bands2} and .
The top of the valence band (VBM) lies at the
M point. The orbital decomposed
band structure shows that the bottom of the conduction band possesses mainly
$p_x$ and $p_y$ character. The VBM has mixed $p_y$ and $p_z$
characters along the $\Gamma$-K direction and $p_z$ and $p_x$ characters along the
$\Gamma$-M direction. Our results agree very well with the previous ones \cite{NL_heine,Rivelino2015}. The topology of the investigated pores was determined by calculating the Z$_2$ topological number according to Ref.\,\cite{Z2}. The results are shown in Table\,\ref{table:topology}.

\begin{table*}
\begin{center}
\caption{\label{table:topology} Nanopore size, bismuth bond lengths d$_{\rm Bi-Bi}$, topological invariant Z$_2$ and electronic band gap E$_g$, calculated within GGA-PBE.}
\begin{tabular*}{14cm}{@{\extracolsep{\fill}}lccccc}
\hline
defect        & size (\AA) & d$_{\rm Bi-Bi}$ & Z$_2$  & top. &   E$_g$ (eV) \\
\hline
divacancy P2 (585)                      &  8.2   & 3.07,3.10,3.12,&   0      & trivial        &  0.2 \\
tetravacancy        P4              &  6.3   & 3.04,3.12,3.29,3.06,3.07&   0      & trivial        & 0.1  \\
hexavacancy P6a &  10.9  & 3.23,3.07,3.31&   1      & metal             &    \\
hexavacancy P6b &  13.3  & 3.02,3.07,3.06&   1      & metal             &     \\
hexavacancy P6c &  15.2  & 3.04,3.09,3.06,3.12 &   1      & metal             &     \\
bare hexagonal pore type-2               &  10.3  & 3.07 &   1      & TI             &    0.2    \\
hydrogenated hexagonal pore type-2             &   7.1  & 3.07 &   1      & TI             &    0.7   \\
pentagon-heptagon (555-777) & & 3.17,3.12,3.08,3.05 &  1 &  TI & \\
\hline
\end{tabular*}
\end{center}
\end{table*}

Although it is more challenging to distinguish between surface and pore atoms, in this case we can easily see from Fig.\ref{fig:projected_bands2}(a) that there are indeed two kind of contributions, with p$_{xy}$ coming from pore and p$_z$ coming from
surface states. In Fig.\,\ref{fig:projected_bands2}(b) we have the opposite behavior, p$_{xy}$ from surface and p$_z$ from surface states.

Finally, we have calculated the Z$_2$ invariants for the nanopores. The results are shown in Table\,\ref{table:topology}. Layers with bare and hydrogenated hexagonal pore type-2 retain the TI character, while P2 and P4 are trivial small gap insulators. Hexagonal vacancies are metallic.
This implies that topological properties depende not only on the pore size but also on its termination.

\section{Optical properties and Berry curvature}

The orthogonal tight-binding Hamiltonian, $H(\mathbf{k})$, obtained within the Wannier90 framework \cite{wannier90,PhysRevB.74.195118} can be written as follows:

\begin{equation}
     H(\mathbf{k}) = H_{0}+ \sum_{i=1}^{N} e^{i \mathbf{k} \cdot \mathbf{R_{i}}} H_{\mathbf{R_{i}}}~,
\end{equation}

\noindent where $H_{0}$ corresponds to the Hamiltonian in unit cell, which contains the on-site energies and hopping parameters inside the cell. $H_{\mathbf{R_{i}}}$ corresponds to the hopping matrices, representing the interaction between unit cell and neighbors cells, while the matrix elements of $H_{0}$ and $H_{\mathbf{R_{i}}}$ are the output from the Wannier90 package \cite{wannier90}. The electronic energy levels are obtained as: 

\begin{equation}
    H(\mathbf{k}) |n,\mathbf{k} \rangle = E_{n,\mathbf{k}} |n,\mathbf{k} \rangle,
\end{equation}

\noindent where $E_{n,\mathbf{k}}$ and $|n,\mathbf{k} \rangle$ are the eigenvalues and eigenvectors respectively, $n$ corresponds to the band index, $v$ for the valence (occupied) states and $c$ for the conduction (unoccupied) states at each $\textbf{k}$-point in the BZ.

To obtain the optical properties, we calculated the real and imaginary parts of the frequency dependent dielectric tensor, $\epsilon_{1,\alpha,\beta}(\omega)$ and $\epsilon_{2,\alpha,\beta}(\omega)$, respectively, through the following expressions:

\begin{align}
     \epsilon_{1,\alpha,\beta}(\omega) =  \delta_{\alpha,\beta} +  \frac{e^{2} S_{p}}{\epsilon_{0}\Omega N_{\mathbf{k}}} \sum_{\mathbf{k},c,v} F_{\alpha,\beta}^{c,v,\mathbf{k}} \frac{(E_{c,\mathbf{k}}-E_{v,\mathbf{k}})-\hbar \omega}{\left(\hbar \omega - (E_{c,\mathbf{k}}-E_{v,\mathbf{k}})\right)^{2}+\eta^{2}}~,
\\
   \epsilon_{2,\alpha,\beta}(\omega)=  \frac{ e^{2} S_{p}}{\epsilon_{0}\Omega N_{\mathbf{k}}} \sum_{\mathbf{k},c,v}  F_{\alpha,\beta}^{c,v,\mathbf{k}}   \frac{\eta}{\left(\hbar \omega - (E_{c,\mathbf{k}}-E_{v,\mathbf{k}})\right)^{2}+\eta^{2}}~,
\end{align}

\noindent where $\delta_{\alpha,\beta}$ is a  kroenecker delta, $S_{p}$ is the spin factor (for spin and non-spin polarized cases). In the independent particle approximation (IPA), corresponds to the oscillator force:

\begin{equation}
    F_{\alpha,\beta}^{c,v,\mathbf{k}} =  \frac{\langle c,\mathbf{k} |P_{\alpha}|v,\mathbf{k} \rangle \langle v,\mathbf{k} |P_{\beta}|c,\mathbf{k} \rangle}{\left(E_{c,\mathbf{k}}-E_{v,\mathbf{k}}-i\eta_{1}\right) \left(E_{c,\mathbf{k}}-E_{v,\mathbf{k}}+i\eta_{1}\right)}~.
\end{equation}

\noindent $\Omega$ is the volume of unit cell, $\epsilon_{0}$ is the vacuum permittivity constant, $N_{\mathbf{k}}$ is the number of \textbf{k}-points employed for the BZ integration, $\omega$ is the incident photon frequency, $c(v)$ corresponds to the conduction (valence) states. $\eta$ is a parameter to smooth the dielectric function. $\alpha$ and $\beta$ correspond to the $x, y and z$ components in the dielectric tensor. $P_{\alpha}$ corresponds to the light-matter interaction operator:

\begin{equation}
    P_{\alpha} = \frac{\partial H(\mathbf{k})}{\partial k_{\alpha}}~,
\end{equation}

\noindent where $H(\mathbf{k})$ corresponds to the electronic Hamiltonian. For circularly polarized light, the light-matter interaction operator for $\sigma_{\pm}$ are written by the following expression:

\begin{equation}
    P_{\sigma_{\pm}} = \frac{1}{\sqrt{2}} \left( \frac{\partial H(\mathbf{k})}{\partial k_{x}}\pm i\frac{\partial H(\mathbf{k})}{\partial k_{y}}\right)~.
\end{equation}

The absorption coefficient $A_{\alpha,\beta}(\omega)$ is then obtained as:

\begin{equation}
    A_{\alpha,\beta}(\omega)= \frac{\sqrt2\omega}{c}\left[\frac{\sqrt{\epsilon_{1,\alpha,\beta}^{2}(\omega)+\epsilon_{2,\alpha,\beta}^{2}(\omega)}-\epsilon_{1,\alpha,\beta}(\omega)}{2}\right]^{\frac{1}{2}},
\end{equation}

\noindent being $c$ the light speed.

As a showcase we have  calculated bare and  hydrogenated bismuthene, previously reported in our Ref.\cite{ROSA2021121849}. Bismuthene and bismuthane has both time reversal and spatial inversion symmetries. The inclusion of SOC opens a gap of 0.5 eV at K and K' points.  When the SOC is turned off, the band structure shows the typical Dirac cone structure of a honeycomb lattice. Bismuthene and bismuthane are non-magnetic, so the Berry curvature has the same sign at both K and K'.   Our results are in agreement with previous results using a different approach using TD-DFT\cite{sciadv.adk3897}

The quantum geometrical properties of the Bloch wave function such as Berry curvature\cite{Berry1984} and band topology is of great importance, as it provides further information on the interplay
between quantum mechanics and materials electronic properties. Recently, the Berry curvature has been used to understand the relation between the quantum geometry and light-matter\cite{PhysRevB.104.064306,sciadv.adk3897,sciadv.1501524}. For materials with (without) inversion symmetry, the Berry curvatures of the Bloch states are zero (nonzero)\,\cite{Xiao2010,Zhang2011,Kormanyos2018}. Moreover, time reversal symmetry requires all Berry-phase related physical quantities to be valley-contrasting. To gain further insight into the electronic band structure and optical properties, we explore the Berry curvature $\Omega_{n}(\textbf{k})$, which is defined as:  $\Omega_{\rm n}(\bf k)= i \langle \nabla_{k} u_{n}\vert \times \vert \nabla_{k} u_{n}\rangle~$,
where $\vert u_{n} \rangle$ is the Bloch state.
One can then obtain the total Berry curvature in a semiconductor written as: 

\begin{equation}
    \Omega_{\alpha,\beta} (\mathbf{k}) = -2 \text{Im} \sum_{v,c} \frac{\langle c,\mathbf{k} |P_{\alpha}|v,\mathbf{k} \rangle \langle v,\mathbf{k} |P_{\beta}|c,\mathbf{k} \rangle}{\left(E_{c,\mathbf{k}}-E_{v,\mathbf{k}}\right)^{2}}~.
\end{equation}

Finally we have calculated the absorption spectra within the Bethe--Salpeter equation (BSE) is a many-body equation that describes the interaction between an electron and a hole. The excitonic states are obtained through the solution of the two-particle problem via BSE \cite{Salpeter_1232_1951,Dias_108636_2022}. The exciton Hamiltonian, $H_{exc}$ is composed by the electron, $H_{e}$, and hole, $H_{h}$, single particle Hamiltonians plus the  Coulomb potential, $V_{eh}$, for the electron-hole pairs interaction $H_{exc} = H_{e} + H_{h}+V_{eh}~$.
The excitonic states with momentum center of mass $\mathbf{Q}$ can be expanded in terms of the product of electron and hole pairs wave functions as follows:

\begin{equation}
\Psi_{ex}^{n}(\mathbf{Q}) = \sum_{c,v,\mathbf{k}} A_{c,v,\mathbf{k},\mathbf{Q}}^{n} \ \left( |c,\mathbf{k}+\mathbf{Q} \rangle \otimes |v,\mathbf{k} \rangle \right)~, \label{eq:Exc_Basis}
\end{equation}

\noindent where the index $c$ and $v$ corresponds to the conduction and valence bands states, with momentum $\mathbf{k}+\mathbf{Q}$ and $\mathbf{k}$, respectively. The problem of excitons eigenvalues, can be transformed into BSE, \cite{Salpeter_1232_1951} according to:

\begin{eqnarray}
    \left( E_{c,\mathbf{k}+\mathbf{Q}}-E_{v,\mathbf{k}}\right) A_{c,v,\mathbf{k},\mathbf{Q}}^{n}  +  \frac{1}{N_{k}}  \sum_{\mathbf{k'},v',c'} W_{(\mathbf{k},v,c),(\mathbf{k'},v',c'),\mathbf{Q}} \ A_{c',v',\mathbf{k'},\mathbf{Q}}^{n}  = E^{n}_{\mathbf{Q}} A_{c,v,\mathbf{k},\mathbf{Q}}^{n} \label{bse}~,
\end{eqnarray}

where $E^{n}_{\mathbf{Q}}$ are the energy of the excitonic state with momentum $\mathbf{Q}, A_{c,v,\mathbf{k},\mathbf{Q}}^{n}$ are the exciton wave function. $E_{c,\mathbf{k} + \mathbf{Q}} - E_{v,\mathbf{k}}$ are the single-particle energy difference between a conduction band state c with momentum $\mathbf{k}+\mathbf{Q}$ and a valence band state v with momentum $\mathbf{k}$, and $W_{(\mathbf{k},v,c),(\mathbf{k'},v',c'),\mathbf{Q}}$ are the many-body Coulomb interaction matrix element, which can be divided into two parts, direct interaction, $W^d$, and exchange interaction, $W^x$, respectively.

\begin{equation}
    W_{(\mathbf{k},v,c),(\mathbf{k'},v',c'),\mathbf{Q}}=W^d_{(\mathbf{k},v,c),(\mathbf{k'},v',c'),\mathbf{Q}}+W^x_{(\mathbf{k},v,c),(\mathbf{k'},v',c'),\mathbf{Q}}~.
\end{equation}

Since the Coulomb potential varies slightly inside unit cell in comparison with Bloch functions, we can approximate the orbital character of Coulomb term in the following way:

\begin{equation}
    W^{d}_{(\mathbf{k},v,c),(\mathbf{k'},v',c'),\mathbf{Q}}=V(\mathbf{k}-\mathbf{k'}) \ \langle c,\mathbf{k}+\mathbf{Q}|c',\mathbf{k'}+\mathbf{Q} \rangle \ \langle v',\mathbf{k'}|v,\mathbf{k} \rangle
\end{equation}

and  $N_{k}$ is the number of $\textbf{k}$-points in the BZ.

In the BSE formalism, the real and imaginary parts of the frequency dependent dielectric tensor $\epsilon^{BSE}_{1,\alpha,\beta}(\omega)$ and $\epsilon^{BSE}_{2,\alpha,\beta}(\omega)$ are obtained as follows:

\begin{align}
\epsilon^{BSE}_{1,\alpha,\beta}(\omega) =  \delta_{\alpha,\beta} +  \frac{e^{2} S_{p}}{\epsilon_{0}\Omega N_{\mathbf{k}}} \sum_{n} F_{\alpha,\beta}^{n,BSE}
&\frac{E_{0}^{n}-\hbar \omega}{\left(\hbar \omega - E_{0}^{n}\right)^{2}+\eta^{2}}~,
\end{align}

\begin{align}
\epsilon^{BSE}_{2,\alpha,\beta}(\omega)=  \frac{e^{2} S_{p}}{\epsilon_{0}\Omega N_{\mathbf{k}}}  \sum_{n}  F_{\alpha,\beta}^{n,BSE}   \frac{\eta}{\left(\hbar \omega - E_{0}^{n}\right)^{2}+\eta^{2}}~,
\end{align}

where $E_{0}^{n}$ is the direct $(\mathbf{Q}=0)$ excitonic state energy, $F_{\alpha,\beta}^{n,BSE}$ is the excitonic modulated oscillator 
force, calculated as:

\begin{align}
F_{\alpha,\beta}^{n,BSE} = \left( \sum_{c,v,\mathbf{k}} \frac{A^{n}_{c,v,\mathbf{k},0}\langle c,\mathbf{k} |P_{\alpha}|v,\mathbf{k} \rangle}{\left(E_{c,\mathbf{k}}-E_{v,\mathbf{k}}+i\eta_{1}\right)} \right) \left(\sum_{c',v',\mathbf{k'}} \frac{A^{n*}_{c',v',\mathbf{k'},0}\langle v',\mathbf{k'} |P_{\beta}|c',\mathbf{k'} \rangle}{\left(E_{c',\mathbf{k'}}-E_{v',\mathbf{k'}}-i\eta_{1}\right)} \right)\label{eq:bse-fosc} ~.
\end{align}

\begin{figure*}[ht!]
\centering
\subfloat[]{\includegraphics[width = 7.5cm, scale=1, clip = true]{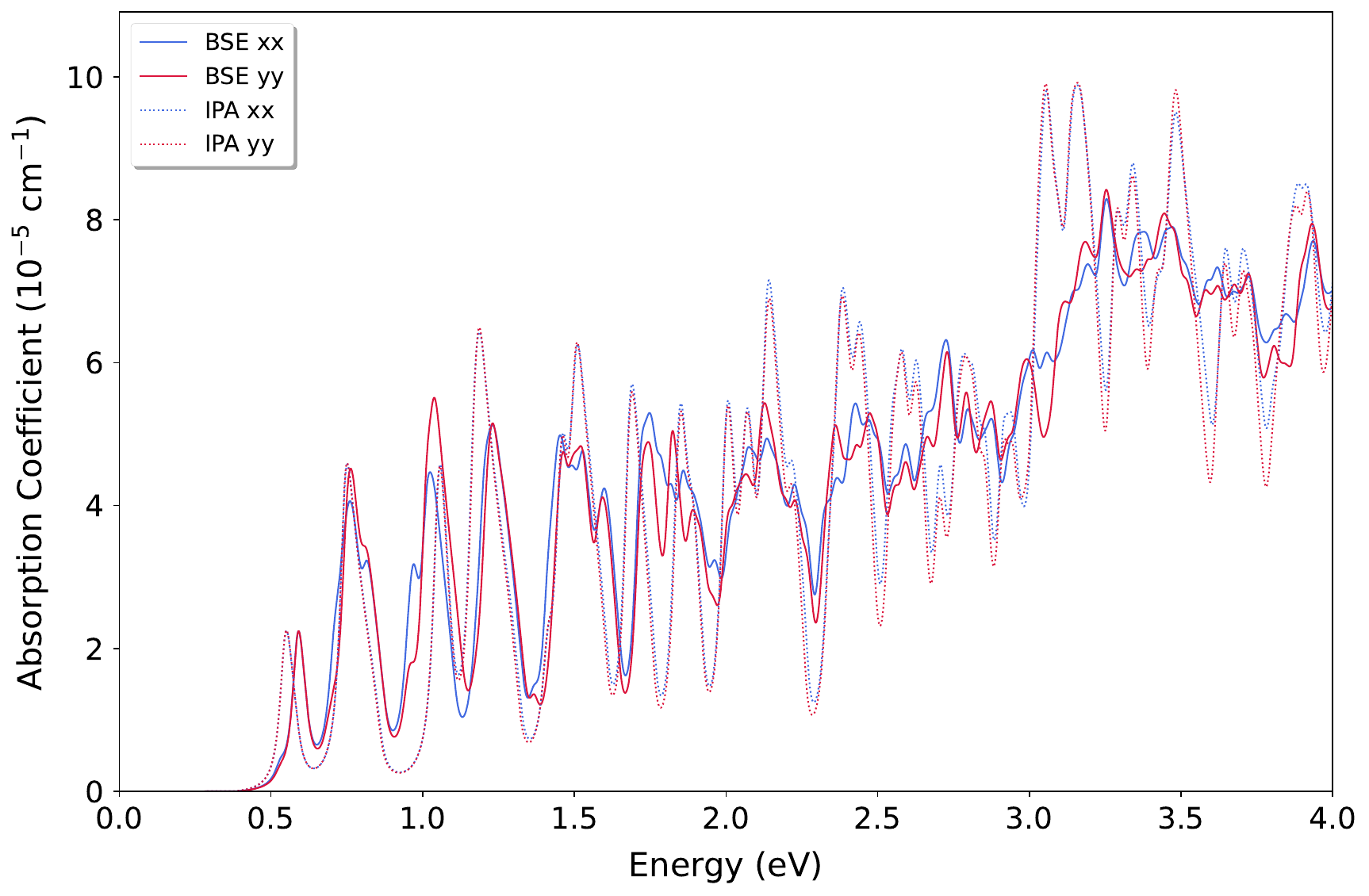}} \hspace{1cm}
\subfloat[]{\includegraphics[width = 7.5cm, scale=1, clip = true]{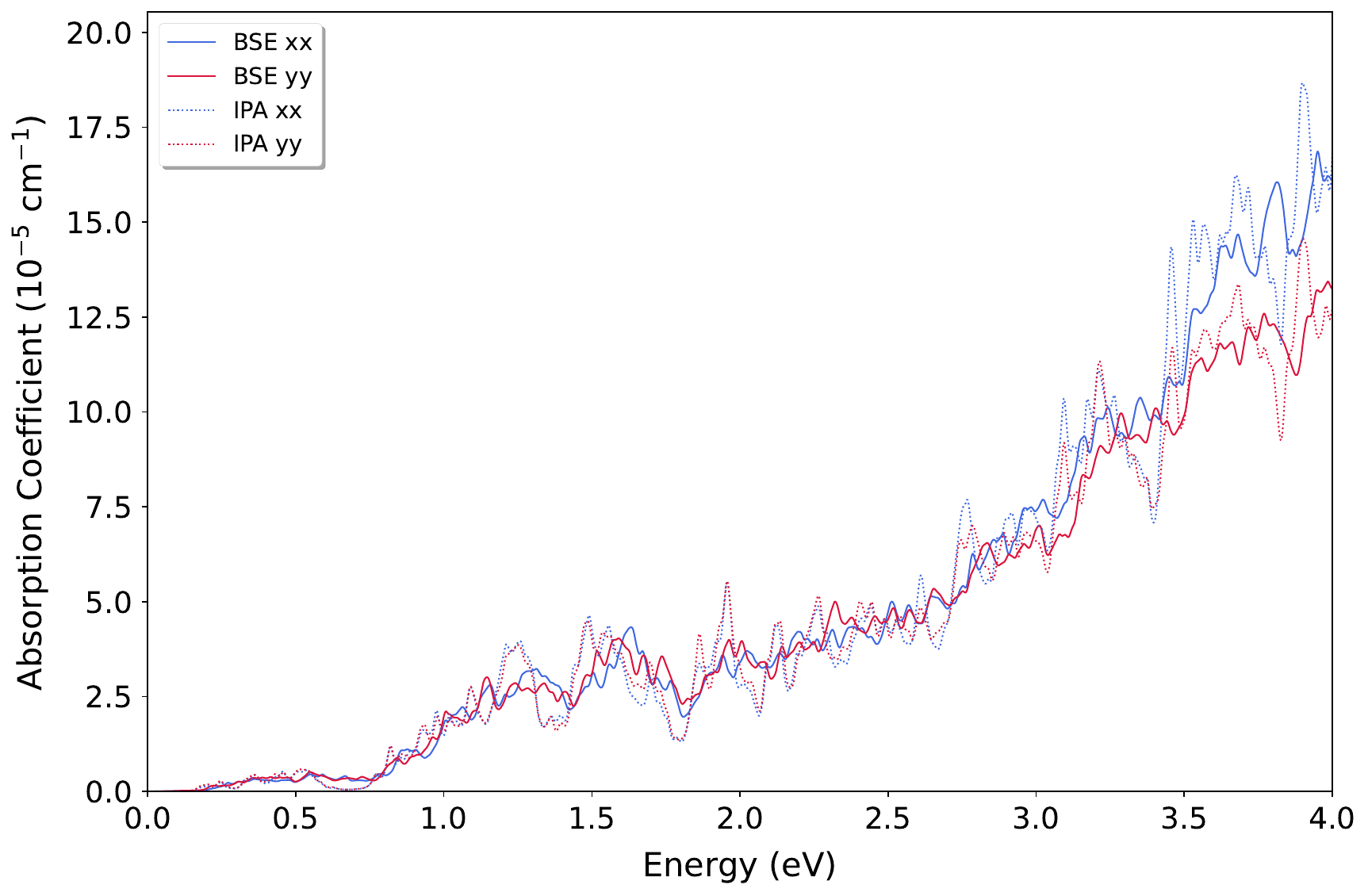}}\\
\caption{Absorption coeficient of bismuthene nanopores  calculated in the IPA (dashed lines) and BSE (solid lines) together with the MLWF-TB@DFT-GGA formalism considering light polarization along $\hat{x}$ (blue curves) and $\hat{y}$ (red curves) directions. a) pristine buckled bismuthene and b) bare hexagonal pore type-2. }
\label{fig:optics}
\end{figure*}

In Fig.\,\ref{fig:optics} we show the absorption coefficients for pure and bare hexagonal pore type-2  bismuthene monolayers within BSE (with excitonic effects) and IPA (without excitonic effects) approximations. Comparing the dashed (IPA) and solid (BSE) curves, we find that excitons binding energy of 0.10 and of xx for pristine buckled bismuthene . No significant optical anisotropy within both the IPA and BSE levels is found. 

\begin{figure*}[ht!]
\centering
\subfloat[]{\includegraphics[width = 8cm, scale=1, clip = true]{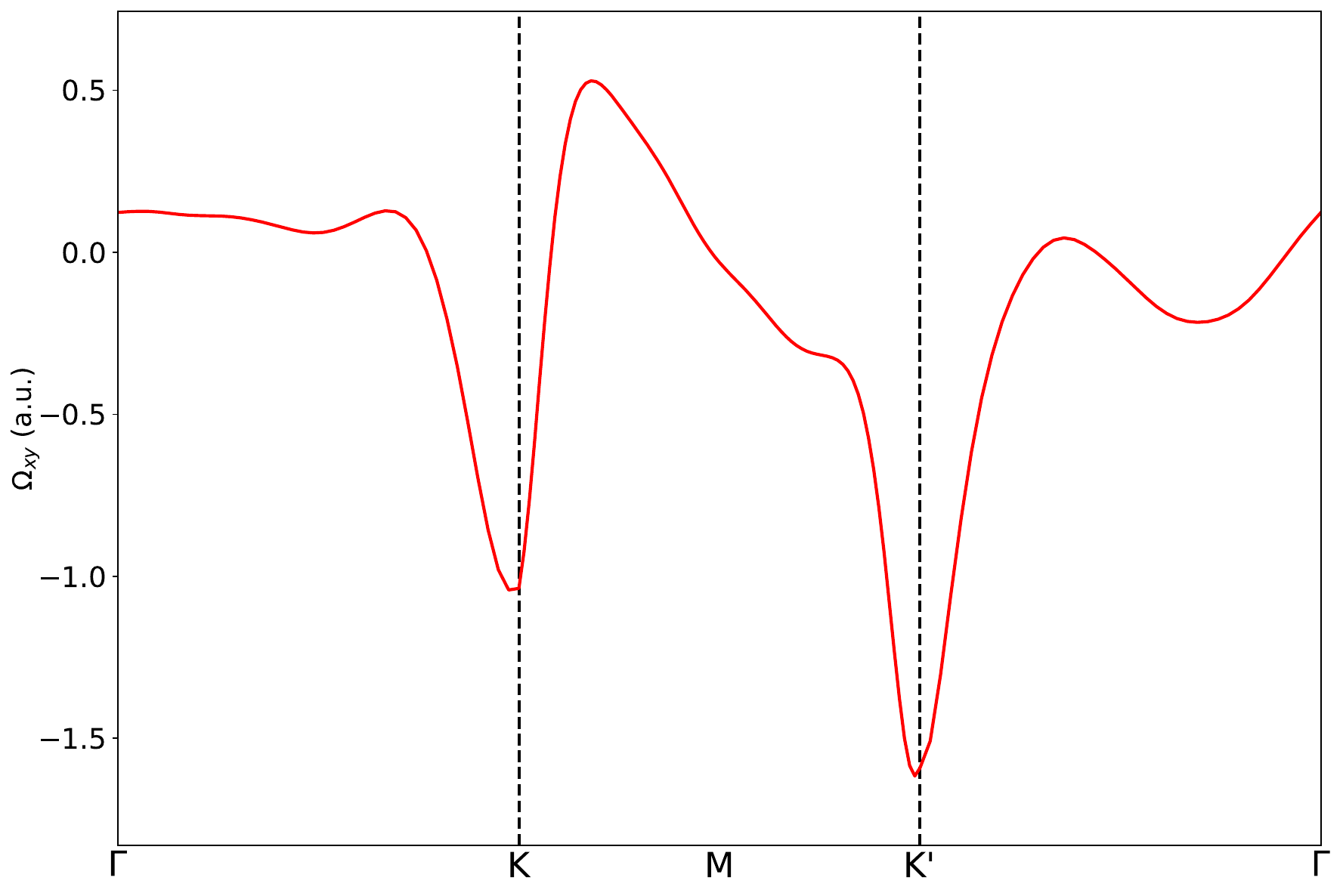}} 
\subfloat[]{\includegraphics[width = 8cm, scale=1, clip = true]{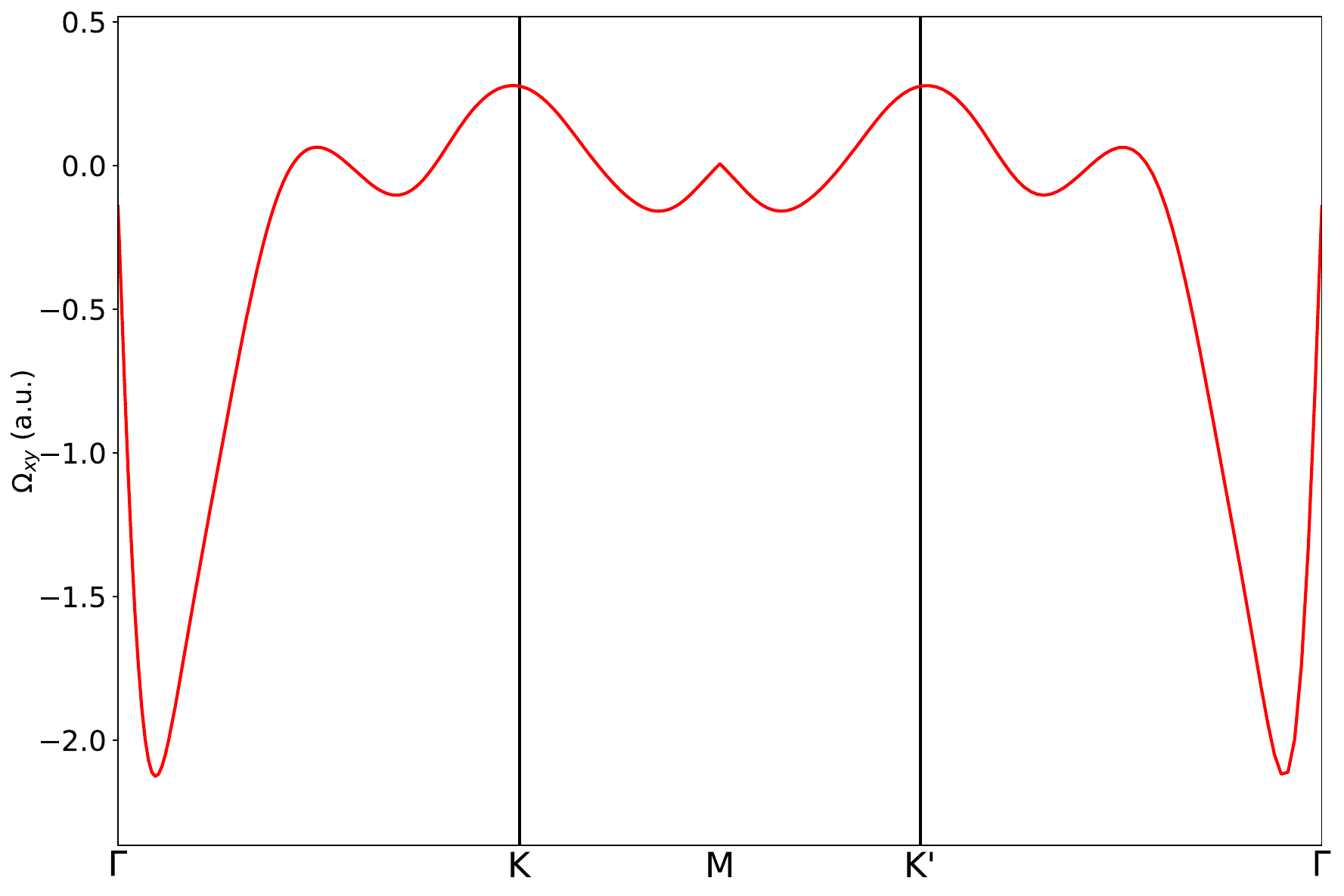}} 
\caption{Berry curvature of bismuthene using MLWF-TB@DFT-GGA. a) pristine buckled bismuthene and b) hydrogenated  hexagonal pore type-2.}
\label{fig:berry}
\end{figure*}

Fig.\,\ref{fig:berry}(a) shows emission at K point, indicating the valley asymetry of topological insulators. Additionally, symmetry breaking changes drastically the Berry curvature, as one can see from Fig.\,\ref{fig:optics}(b). The emission is highly localized around the M points.

\subsection{Adsorption of small molecules}

\begin{figure*}[ht!]
\centering
\subfloat[]{\includegraphics[width = 3cm, scale=1, clip = true]{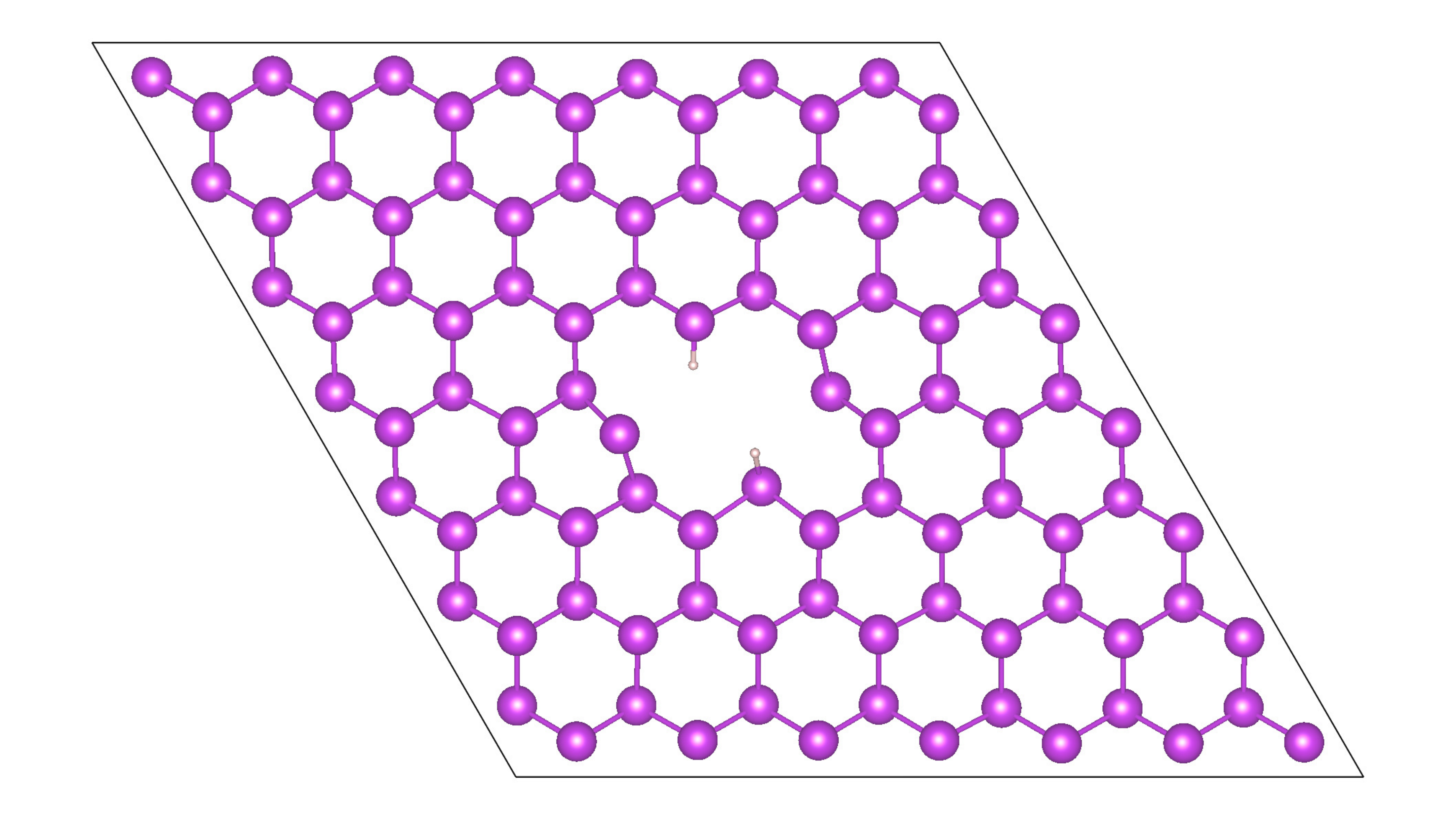}} 
\subfloat[]{\includegraphics[width = 3cm, scale=1, clip = true]{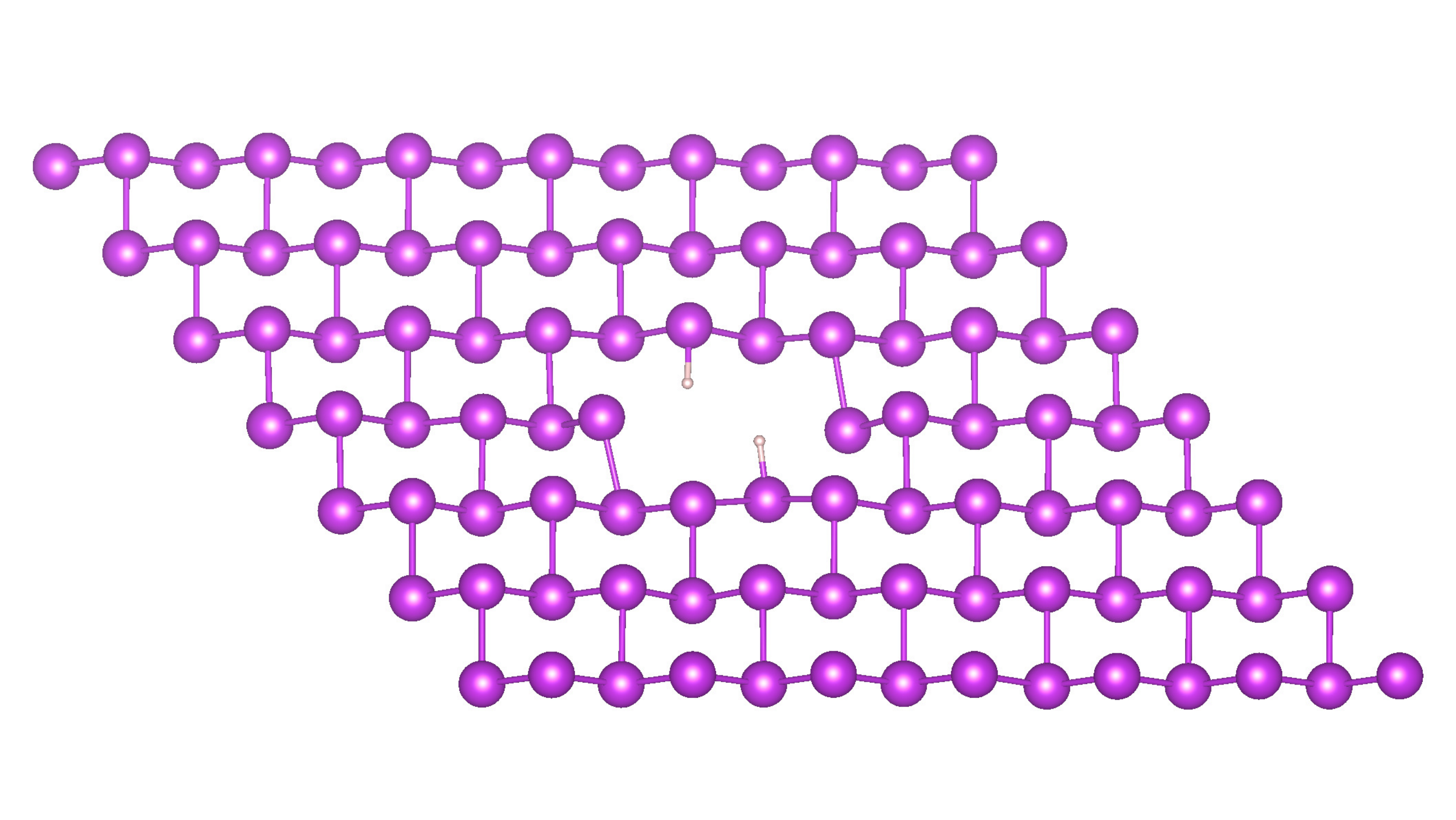}} 
\subfloat[]{\includegraphics[width = 3cm, scale=1, clip = true]{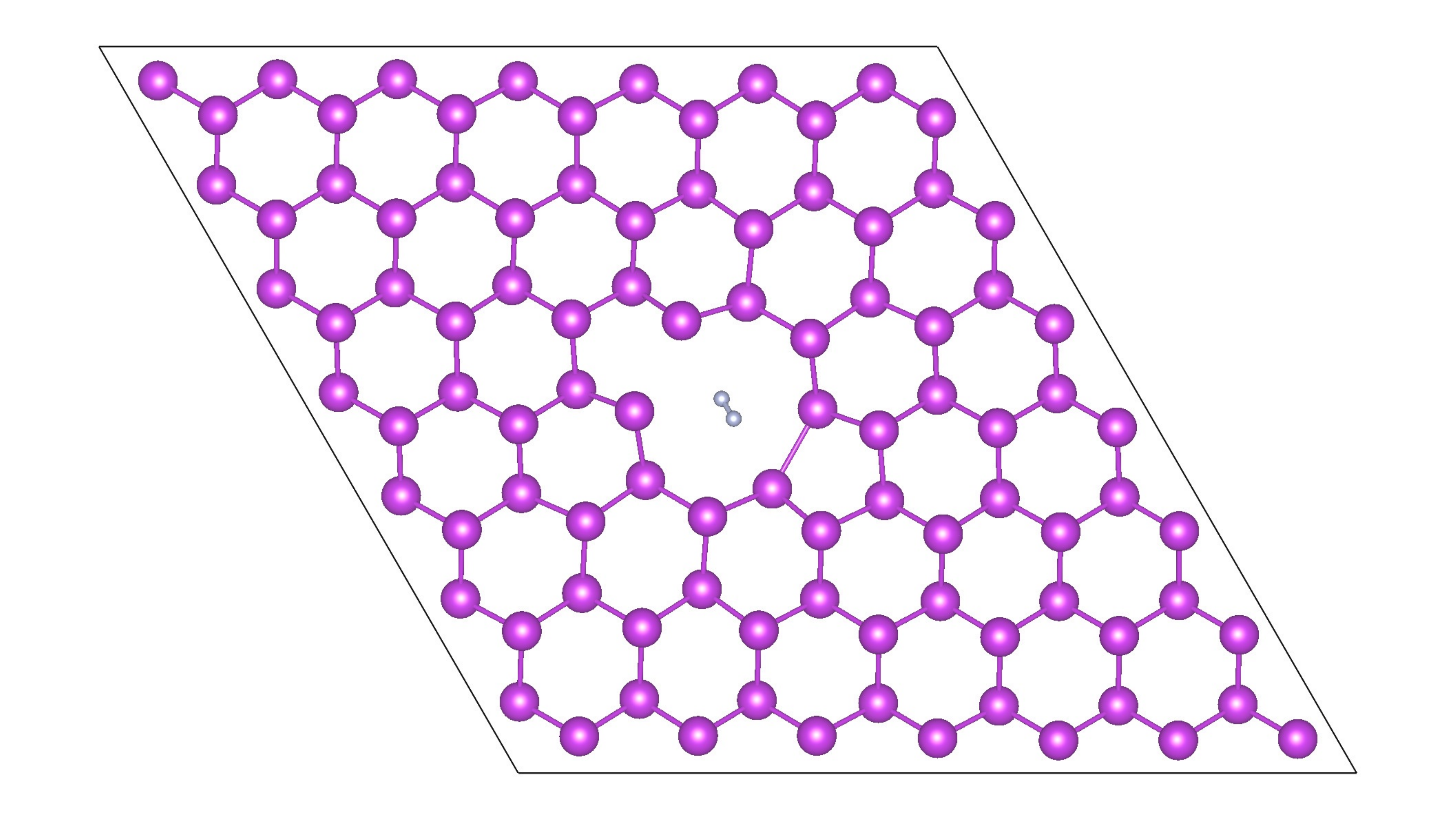}} 
\subfloat[]{\includegraphics[width = 3cm, scale=1, clip = true]{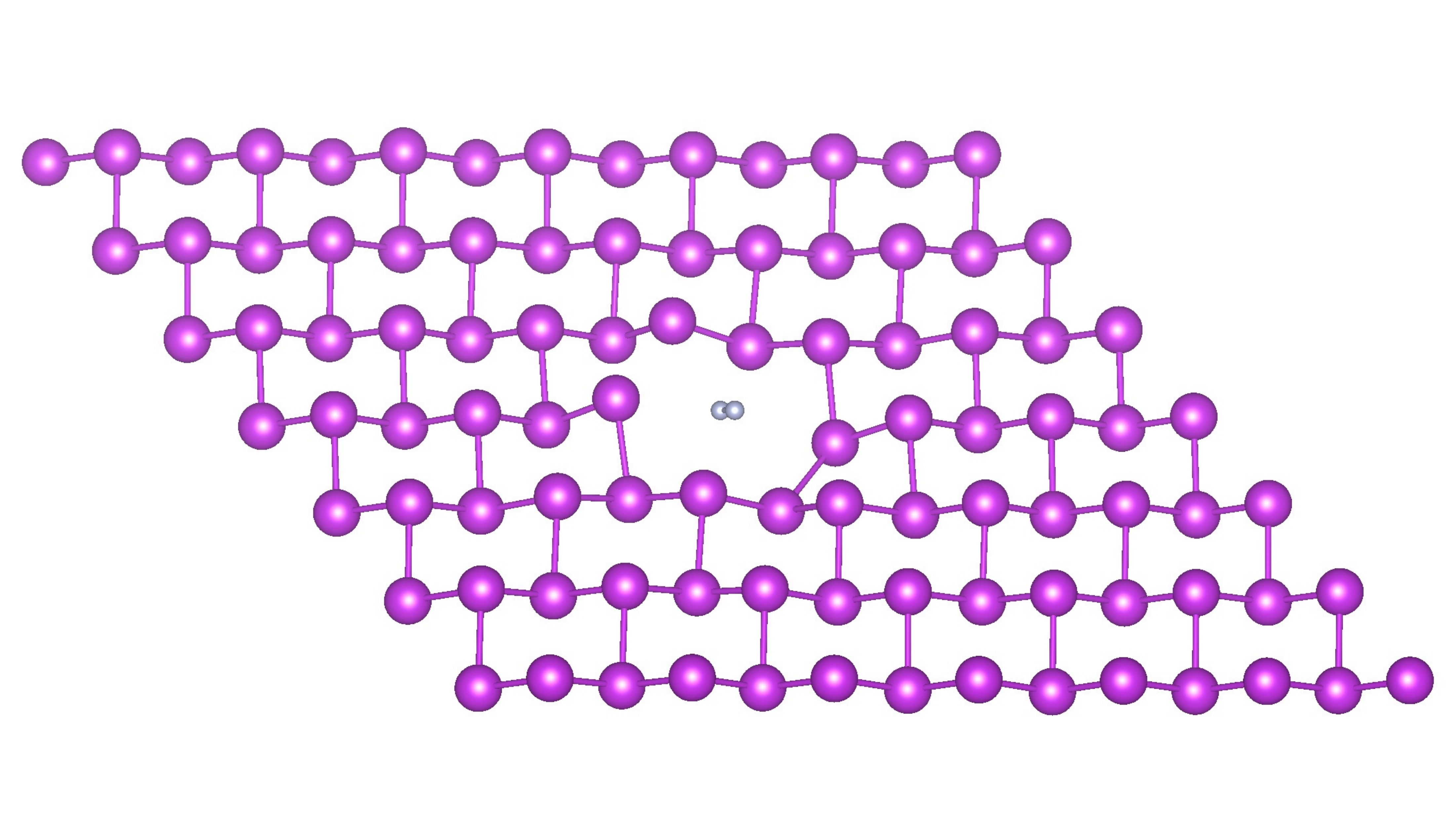}} \\
\subfloat[]{\includegraphics[width = 3cm, scale=1, clip = true]{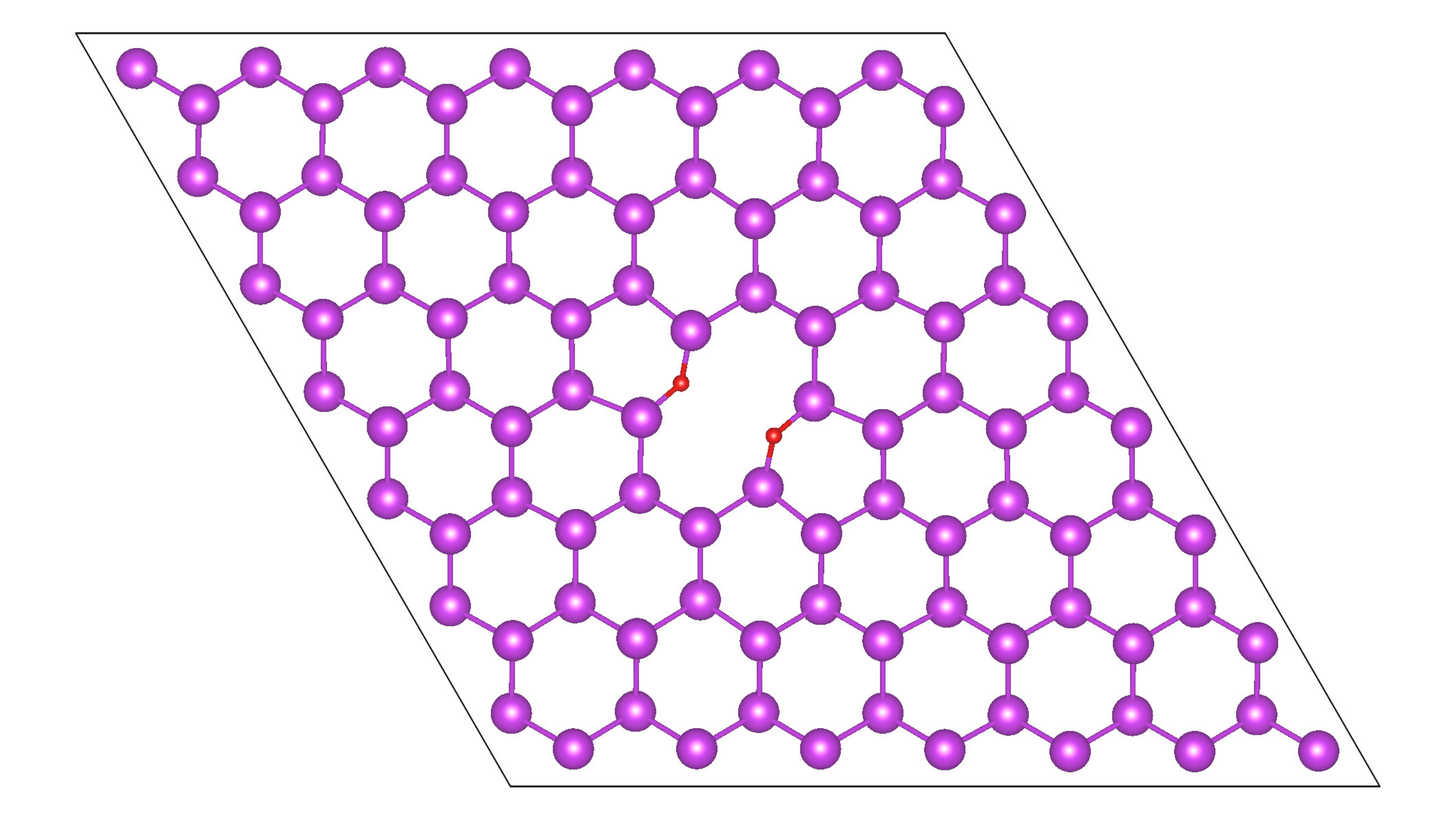}} 
\subfloat[]{\includegraphics[width = 3cm, scale=1, clip = true]{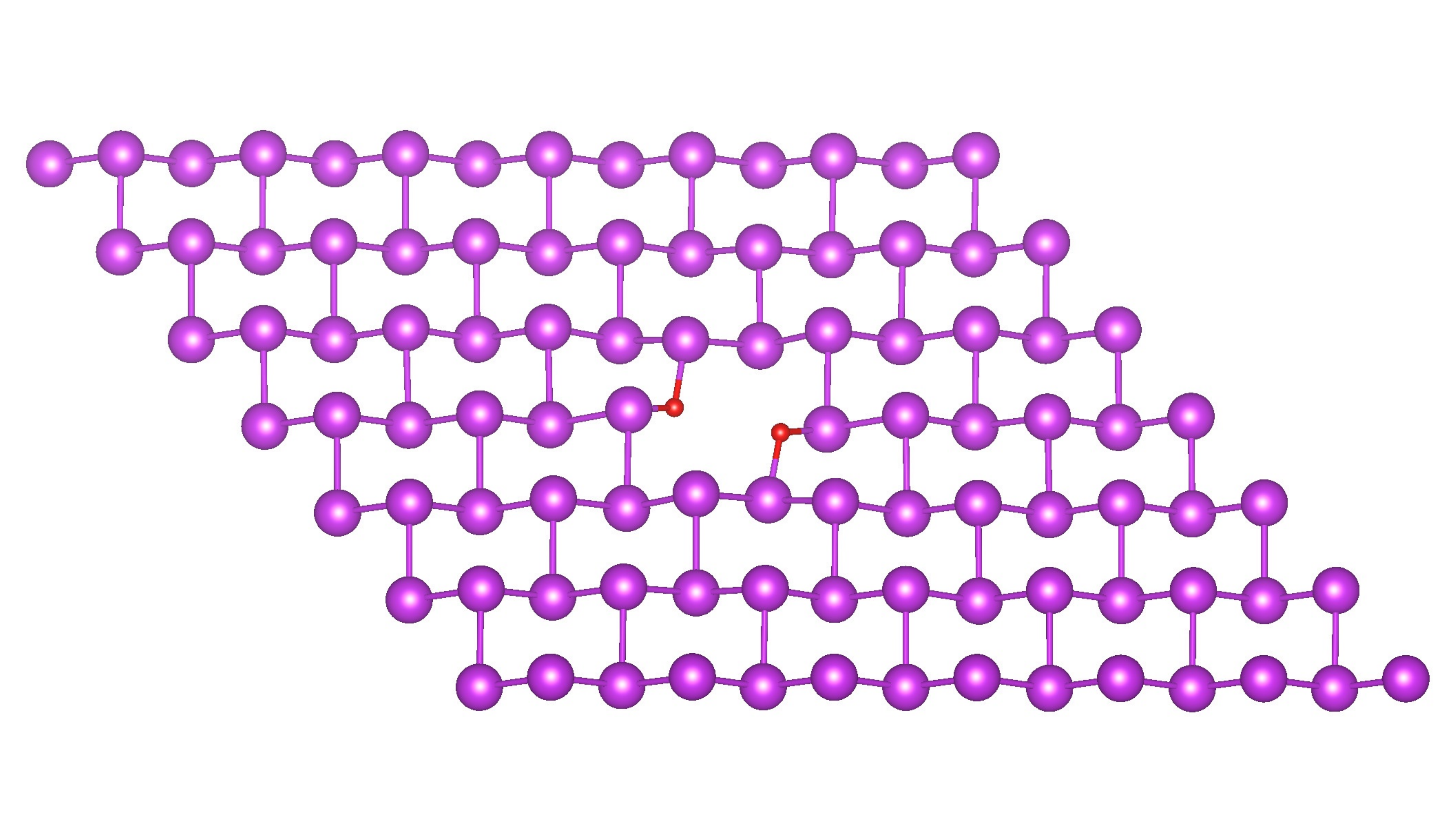}} 
\subfloat[]{\includegraphics[width = 3cm, scale=1, clip = true]{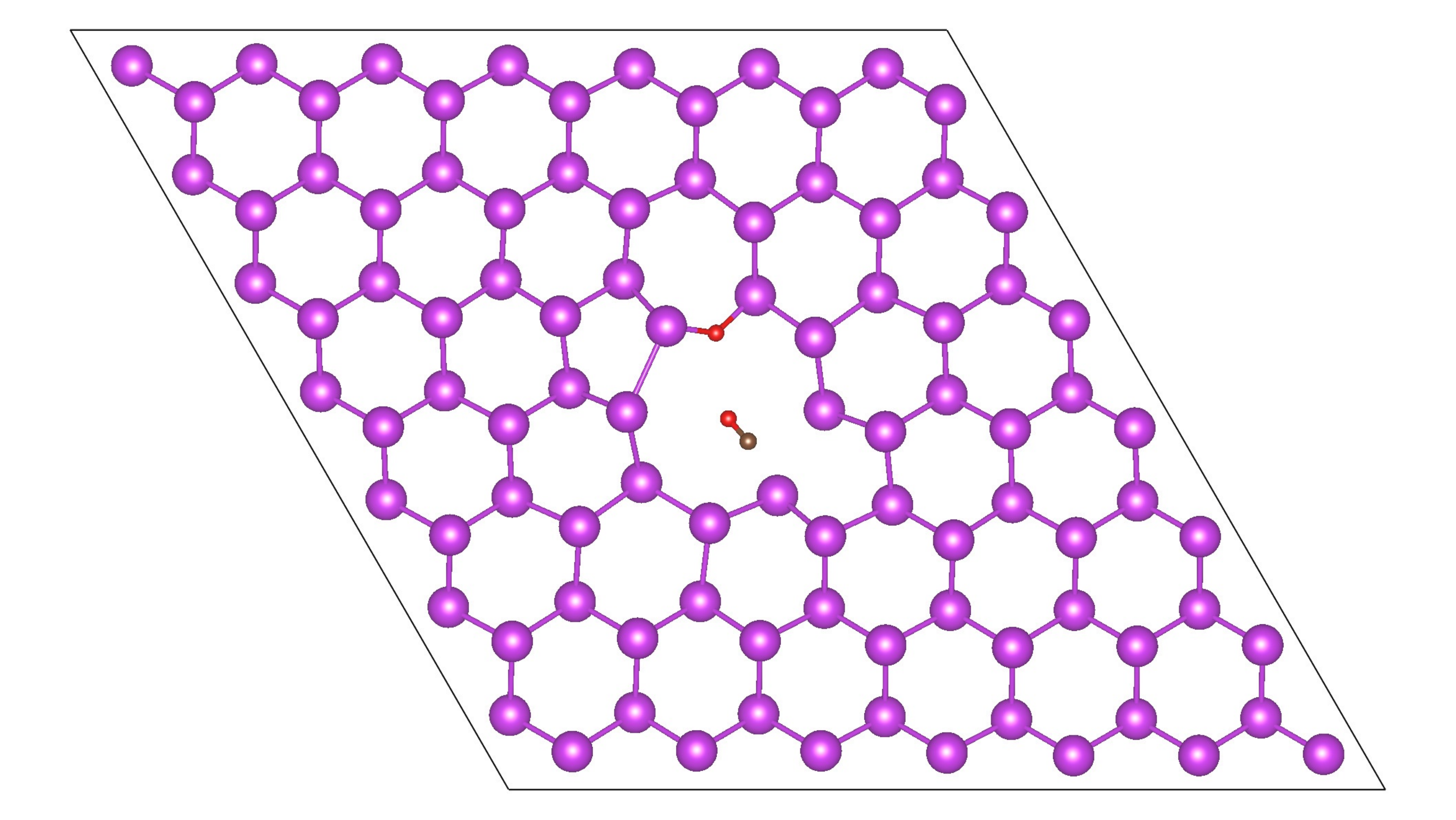}} 
\subfloat[]{\includegraphics[width = 3cm, scale=1, clip = true]{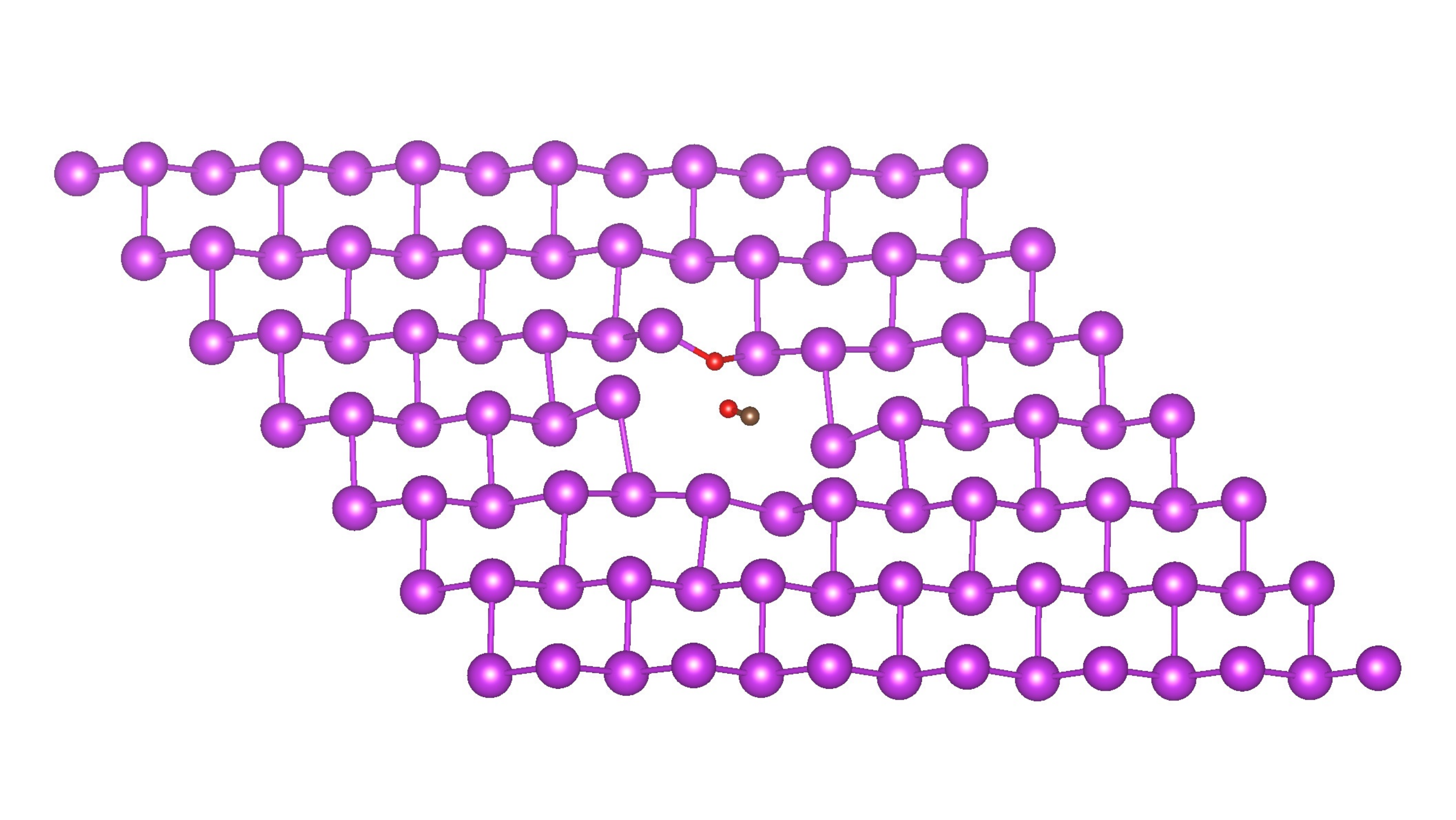}}\\
\subfloat[]{\includegraphics[width = 3cm, scale=1, clip = true]{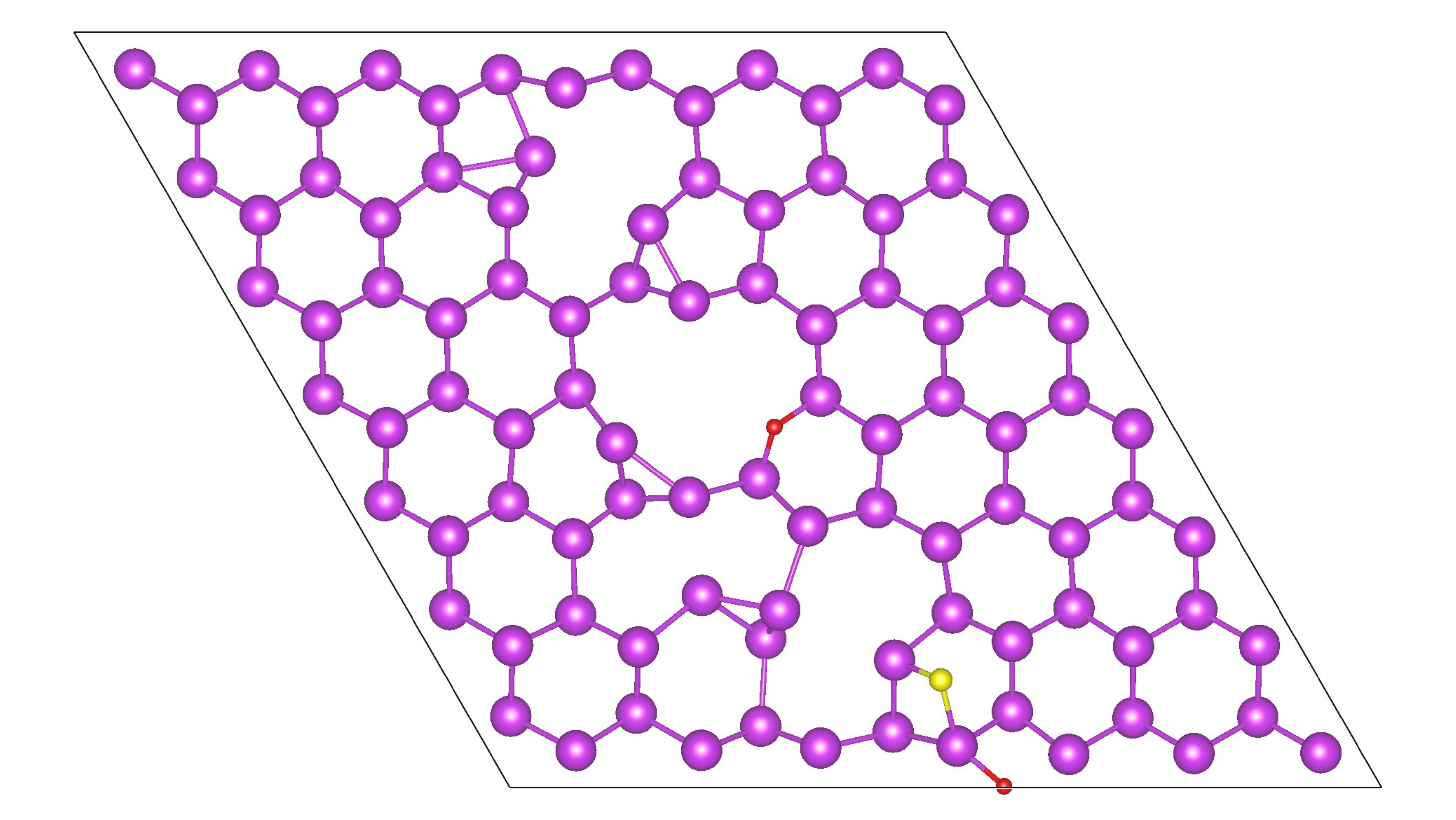}} 
\subfloat[]{\includegraphics[width = 3cm, scale=1, clip = true]{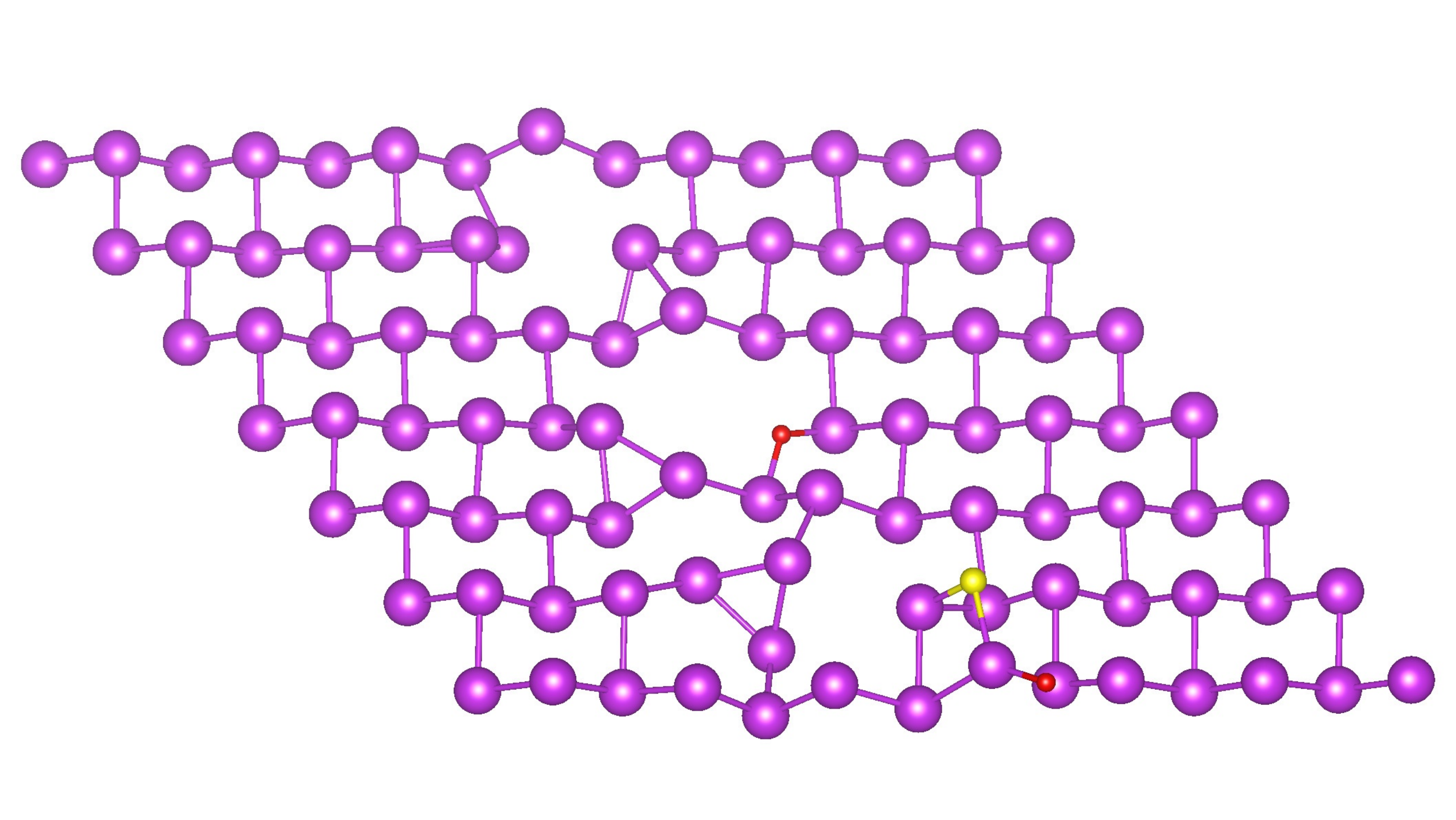}} 
\subfloat[]{\includegraphics[width = 3cm, scale=1, clip = true]{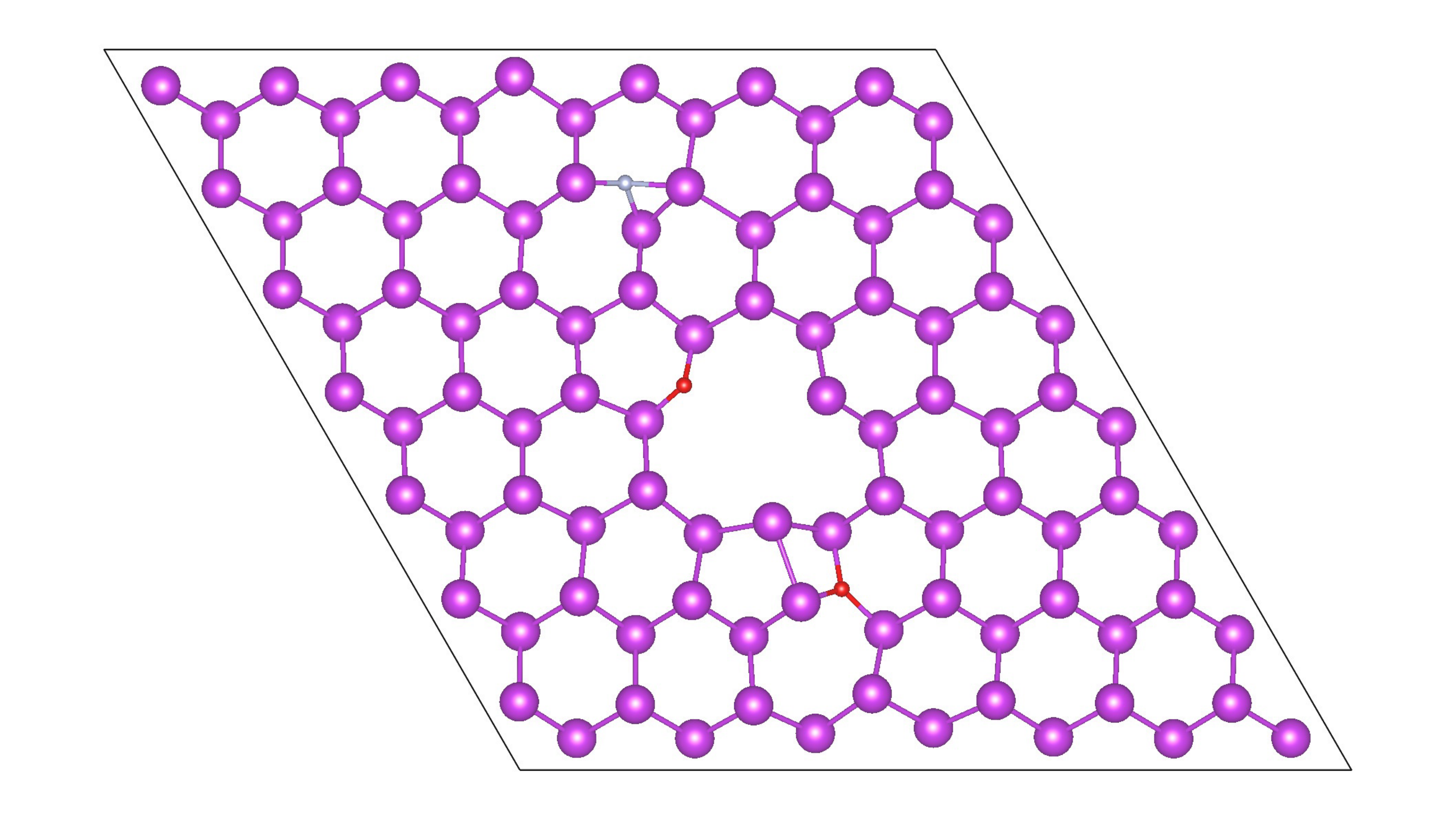}} 
\subfloat[]{\includegraphics[width = 3cm, scale=1, clip = true]{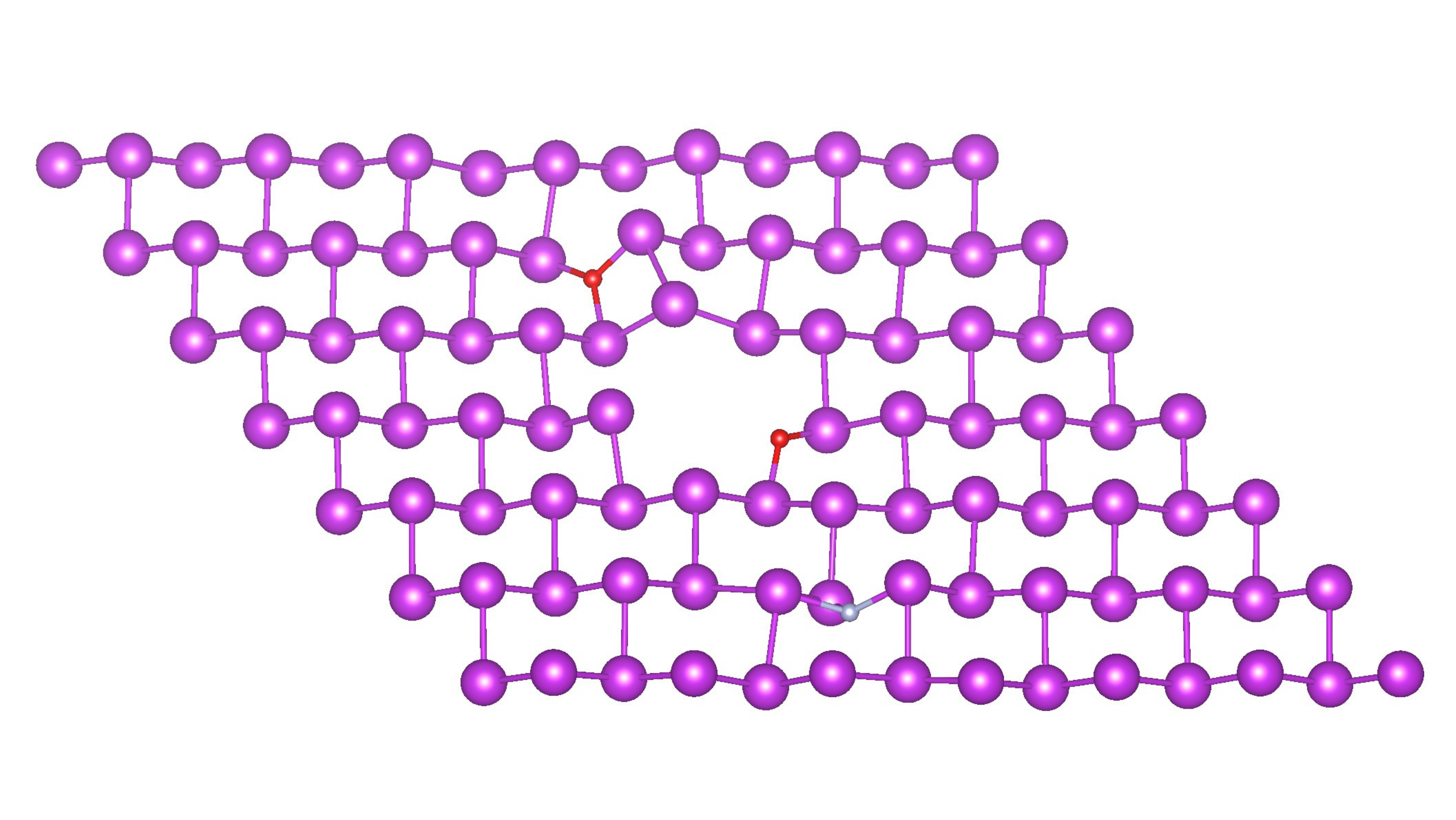}} \\
\subfloat[]{\includegraphics[width = 3cm, scale=1, clip = true]{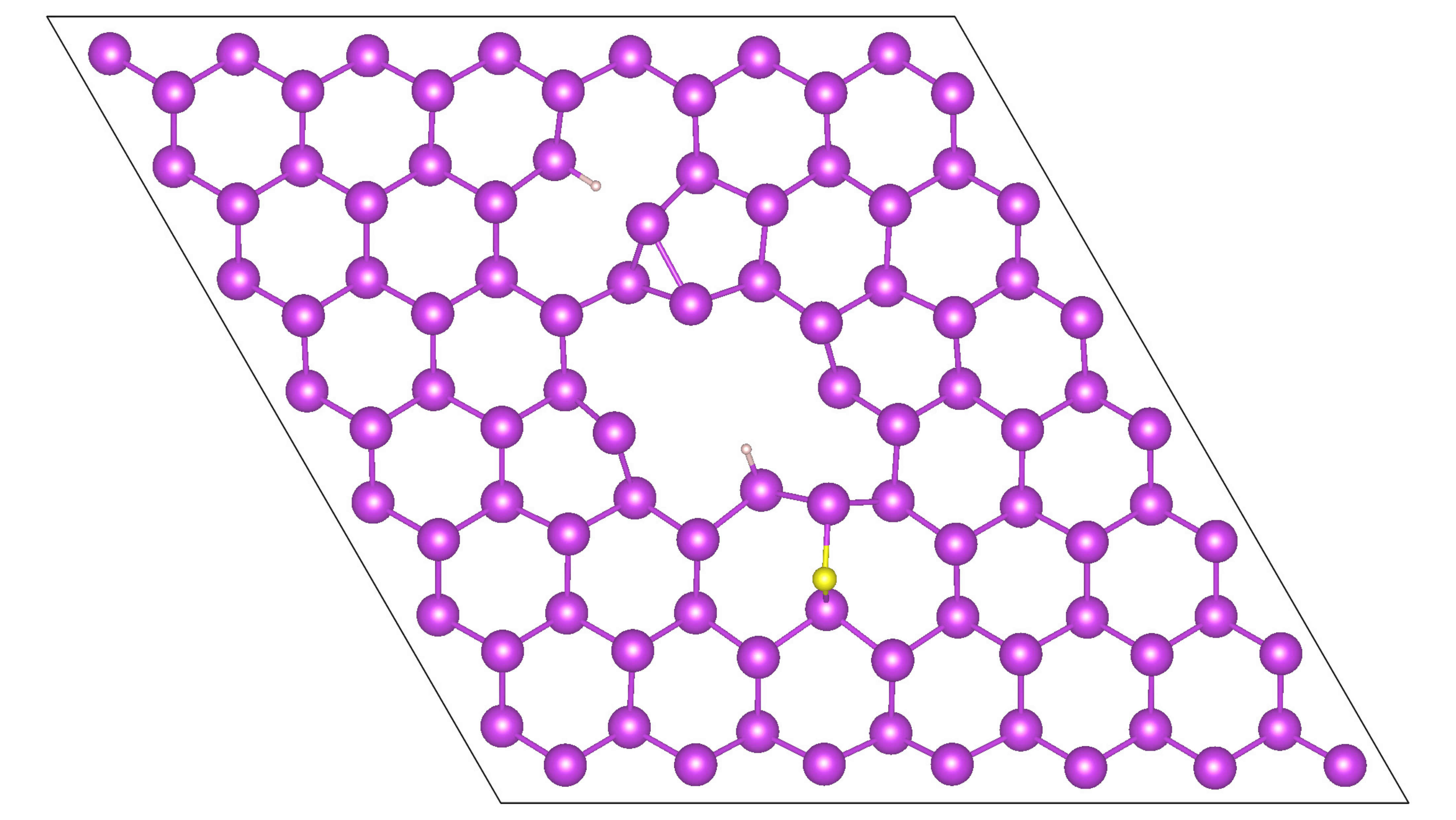}} 
\subfloat[]{\includegraphics[width = 3cm, scale=1, clip = true]{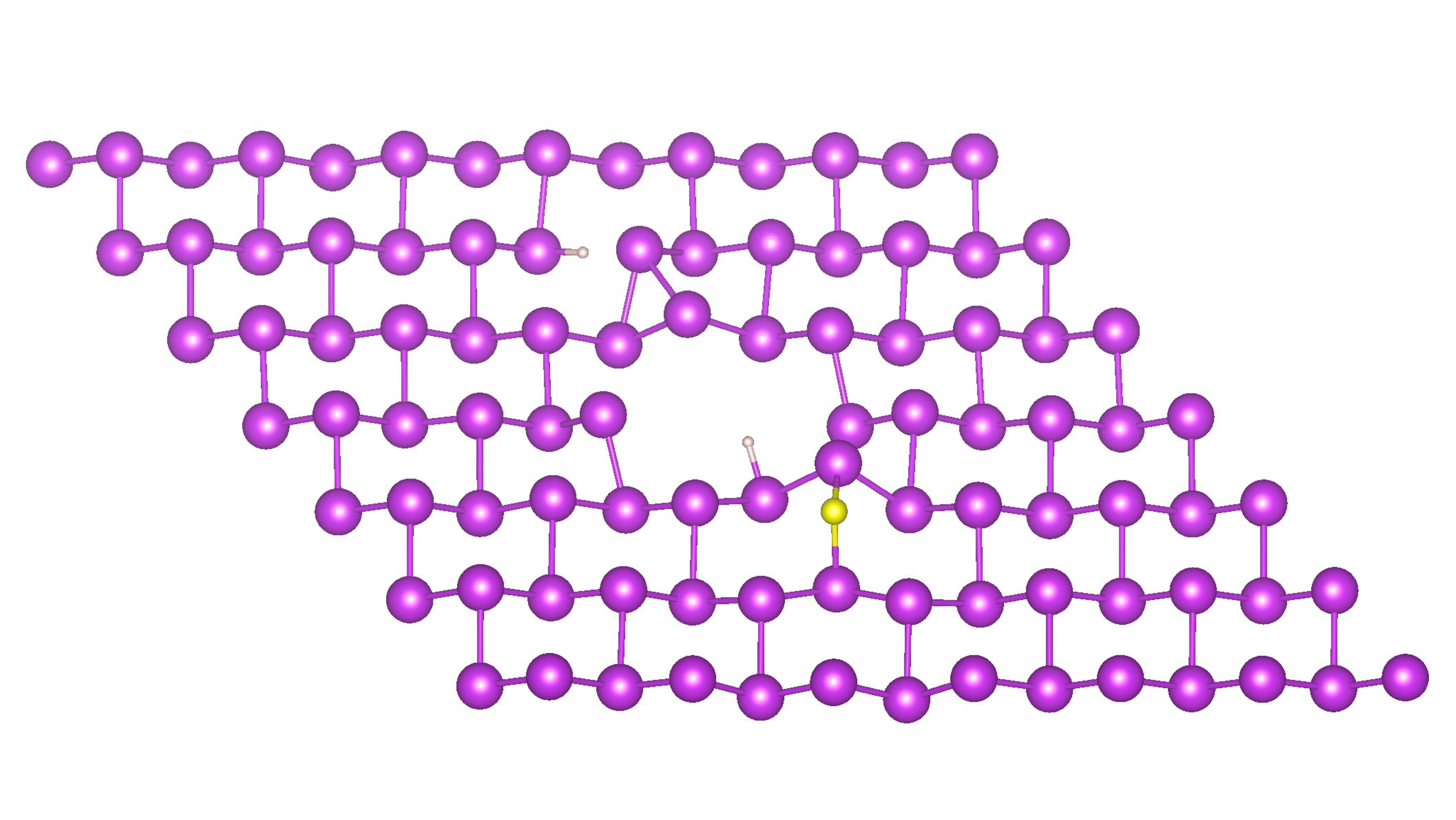}} 
\subfloat[]{\includegraphics[width = 3cm, scale=1, clip = true]{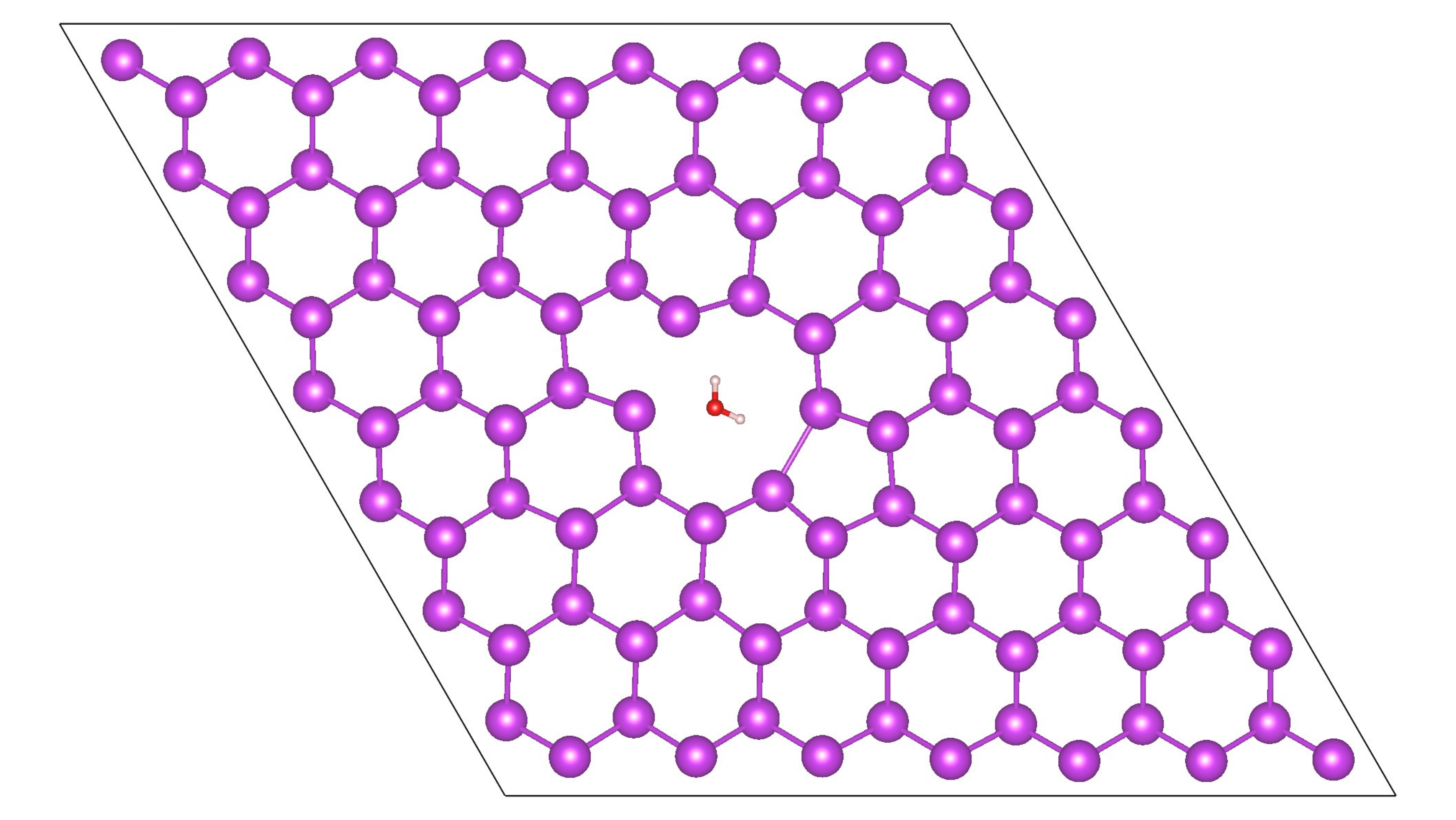}} 
\subfloat[]{\includegraphics[width = 3cm, scale=1, clip = true]{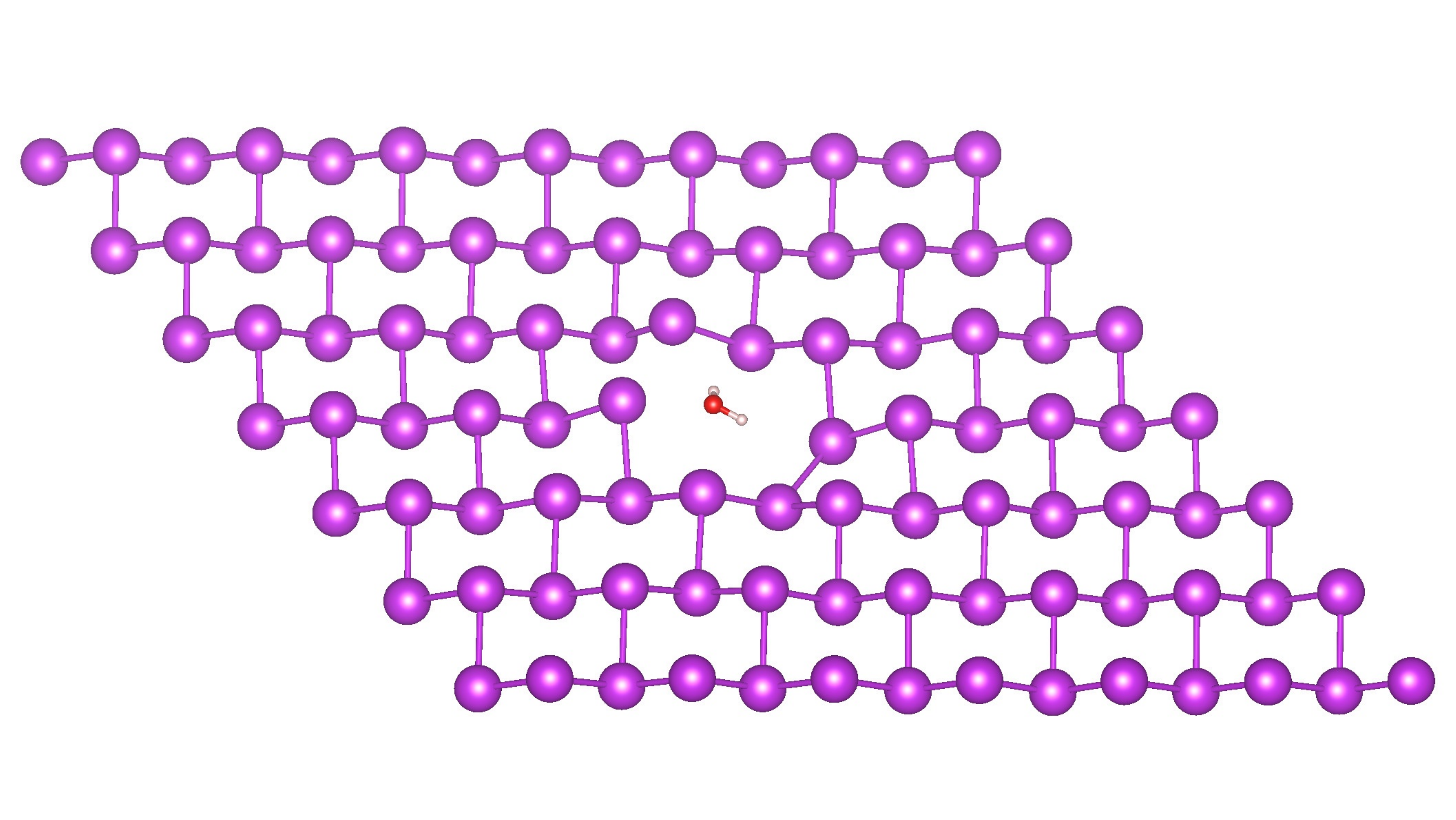}} \\
\subfloat[]{\includegraphics[width = 3cm, scale=1, clip = true]{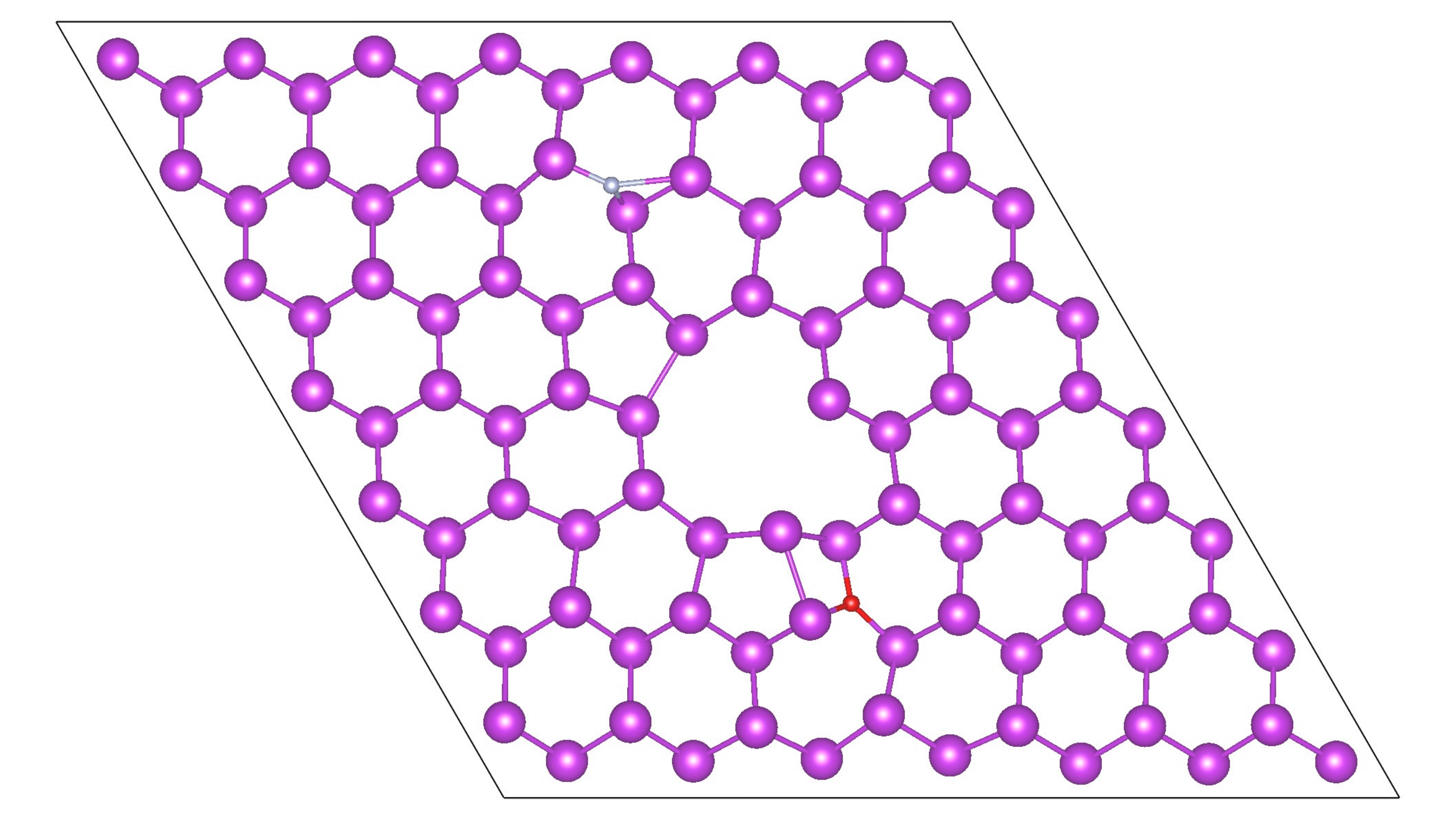}} 
\subfloat[]{\includegraphics[width = 3cm, scale=1, clip = true]{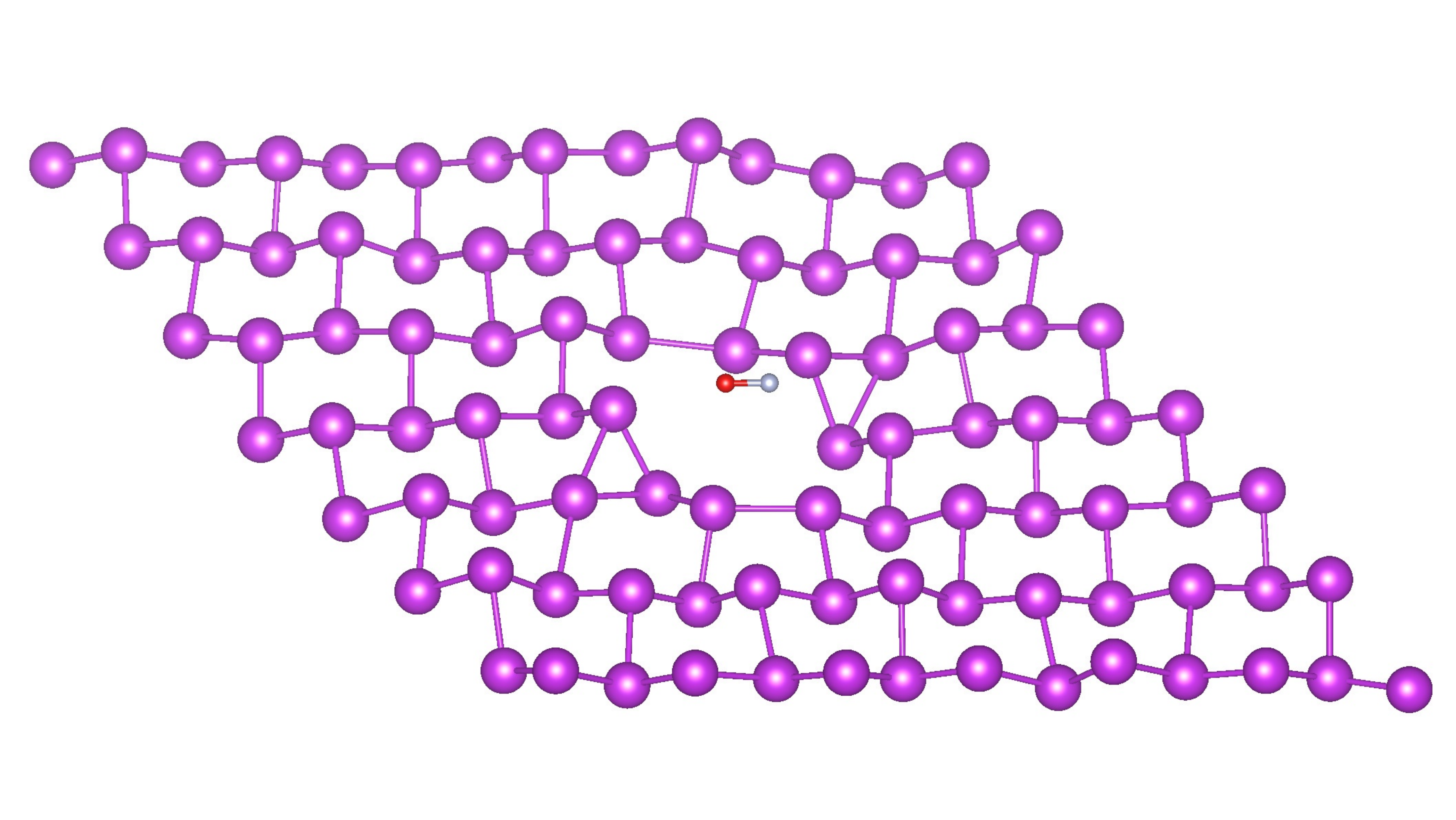}} 
\caption{Adsorption of small molecules in bismuthene nanopores on a divacancy (P2/585) calculated within GGA-PBE. a) H$_2$, b) N$_2$, c) O$_2$, d) NO, e) CO$_2$, f) SO$_2$, g) NO$_2$, h) H$_2$S and i) H$_2$O.}
\label{fig:P2_ads_small_molecules}
\end{figure*}

In order to understand the role of interaction of small molecules with bismuthene nanopores we have calculated the adsorption within the pore of H$2$, N$2$, O$2$, NO, CO$2$, SO$2$, NO$2$, H$2$S and H$2$O.
On bare surfaces, all  molecules spontaneoulsy dissociate (not shown). As show cases we chose P2 and P4. 
The optimized geometry of the adsorbed molecules on defective bismuthene is shown for P2 and P4 defects in Fig.\,\ref{fig:P2_ads_small_molecules} and Fig. S2, respectively. 
Calculations with vdW interactions show no significant changes in the dissociation process.
This also implies that the molecule atoms have a strong interaction with bismuth atoms. We have started from several initial configurations, including but not restricted to single molecules at distances of a) 1.0\,{\AA} above the surface, b) 2.0\,{\AA} above the surface, c) in the middle of the pore and d) close to the border of the pore. All lead to dissociation. Although the phase space is large, these configurations should be representative of reactions between the gases and the pores. On the other hand, P4 defects show little selectivity towards the investigated small molecules, with exception for CO$_2$ as shown in Fig. S2.

\section{Conclusions}

We perform first-principles calculations of electronic, optical and 
topological properties of bismuthene subnanometer pores.  We show that these properties depend on the shape and size of the nanopores. Adsorption of gas phase molecules indicate that these pores can be serve as sensors, but the sensitivity depends on the pore termination. Our results open the path for further investigations on bismuthene as membranes for gas separation and sensing.

\section{Acknowledgements}

We acknowledge the financial support from the Brazilian funding agency CNPq under grant numbers 305174/2023-1, 313081/2017-4, 305335/2020-0, 309599/2021-0, 408144/2022-0, 305952/2023-4, 444069/2024-0 and 444431/2024-1. A.C.D also acknowledge FAPDF grants numbers 00193-00001817/2023-43 and 00193-00002073/2023-84. A.C.D and A.L.R also acknowledges DPG-FAPDF-CAPES Centro-Oeste grant 00193-00000867/2024-94. We thank computational resources from Computer Cluster Bremen, LaMCAD/UFG, Santos Dumont/LNCC, CENAPAD-SP/Unicamp (project number 897 and 761) and Lobo Carneiro HPC (project number 133).



\newpage
\centering{\bf Supplementary Information}

{\bf Molecular Dynamics simulations}

In Fig.S1 we show the total energy profile of the AIMD calculations for 10\,ps simulation time.

\begin{figure*}[ht!]
\centering
\subfloat[]{\includegraphics[width = 6cm, scale=1, clip = true]{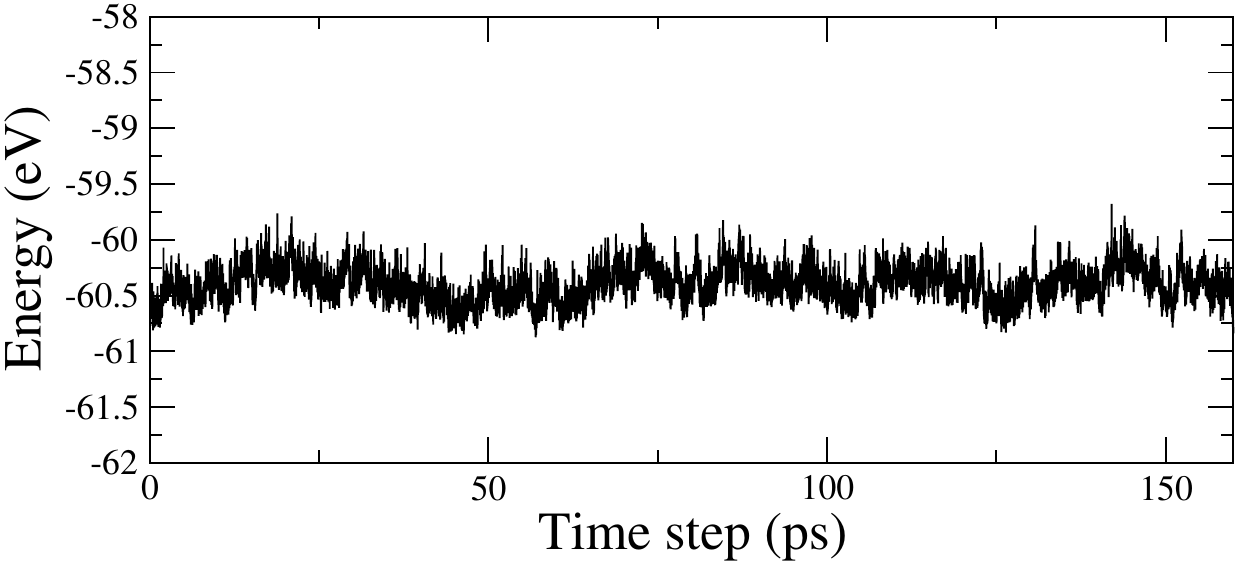}} 
\subfloat[]{\includegraphics[width = 6cm, scale=1, clip = true]{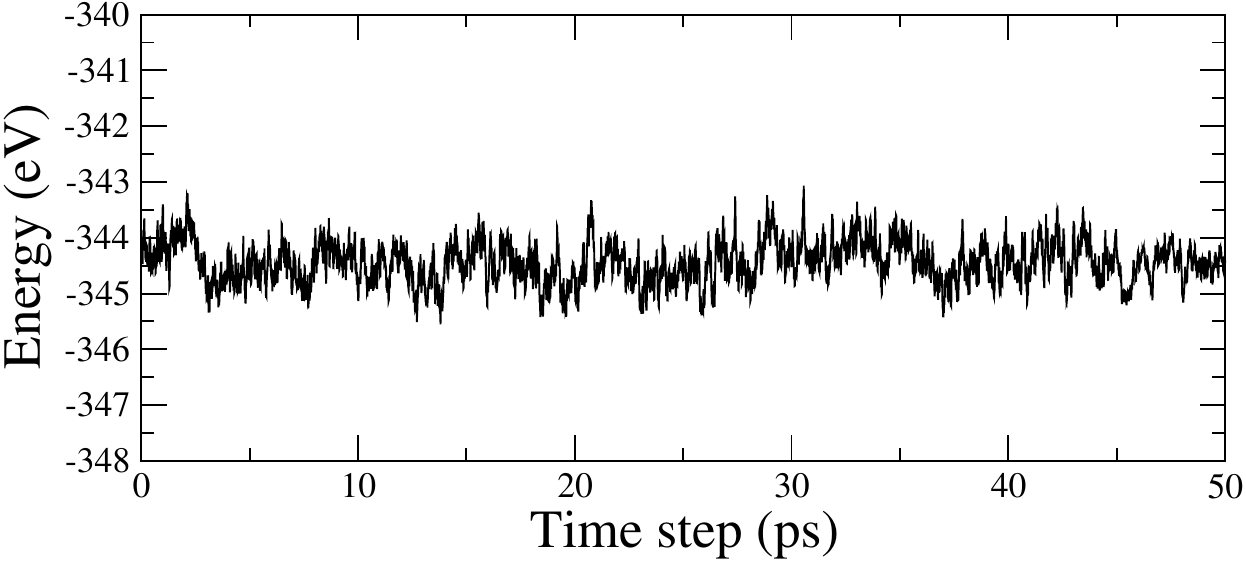}}  \\
\subfloat[]{\includegraphics[width = 6cm, scale=1, clip = true]{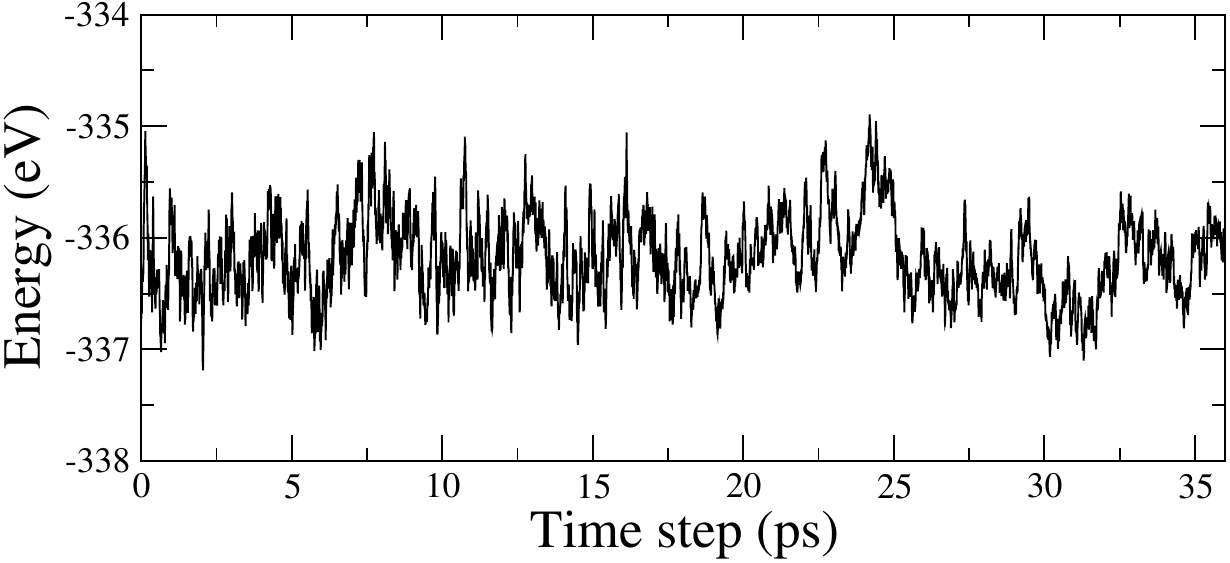}} 
\subfloat[]{\includegraphics[width = 6cm, scale=1, clip = true]{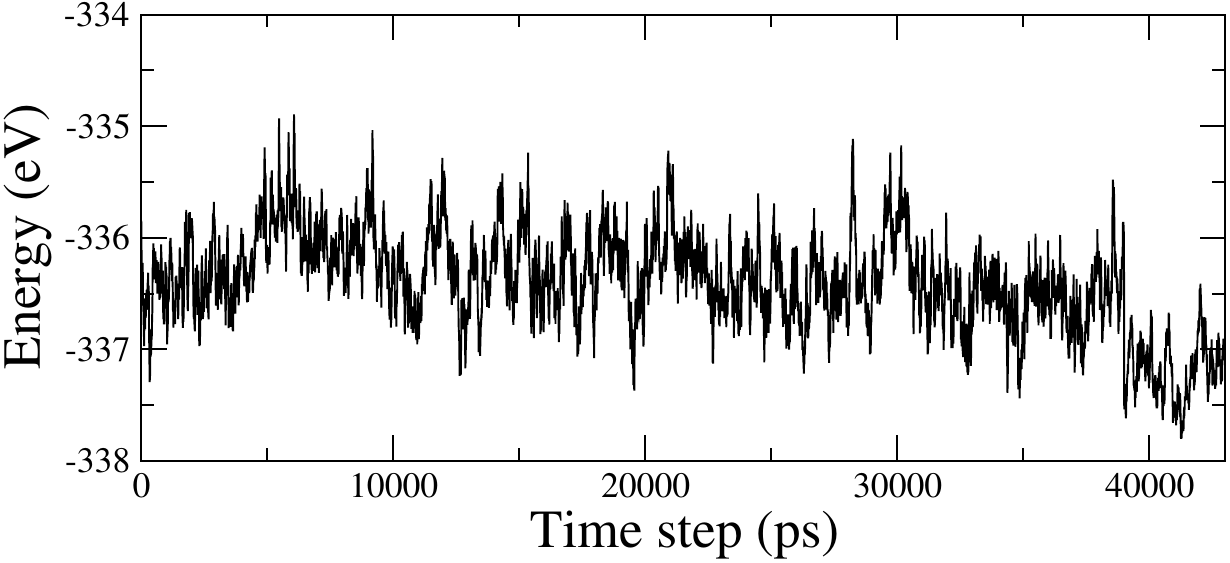}} \\
\subfloat[]{\includegraphics[width = 6cm, scale=1, clip = true]{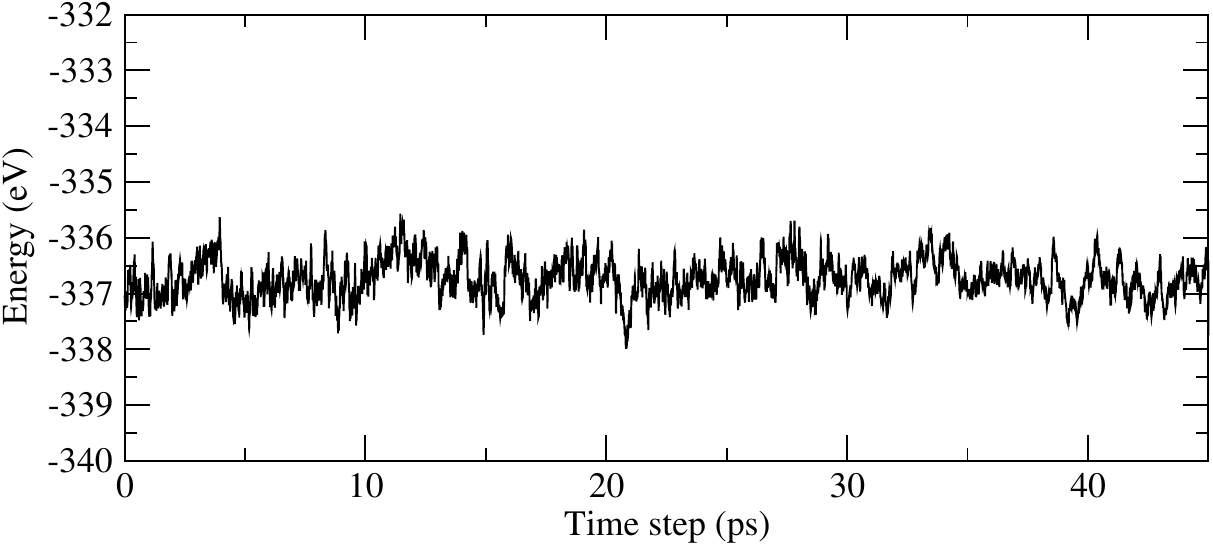}} 
\subfloat[]{\includegraphics[width = 6cm, scale=1, clip = true]{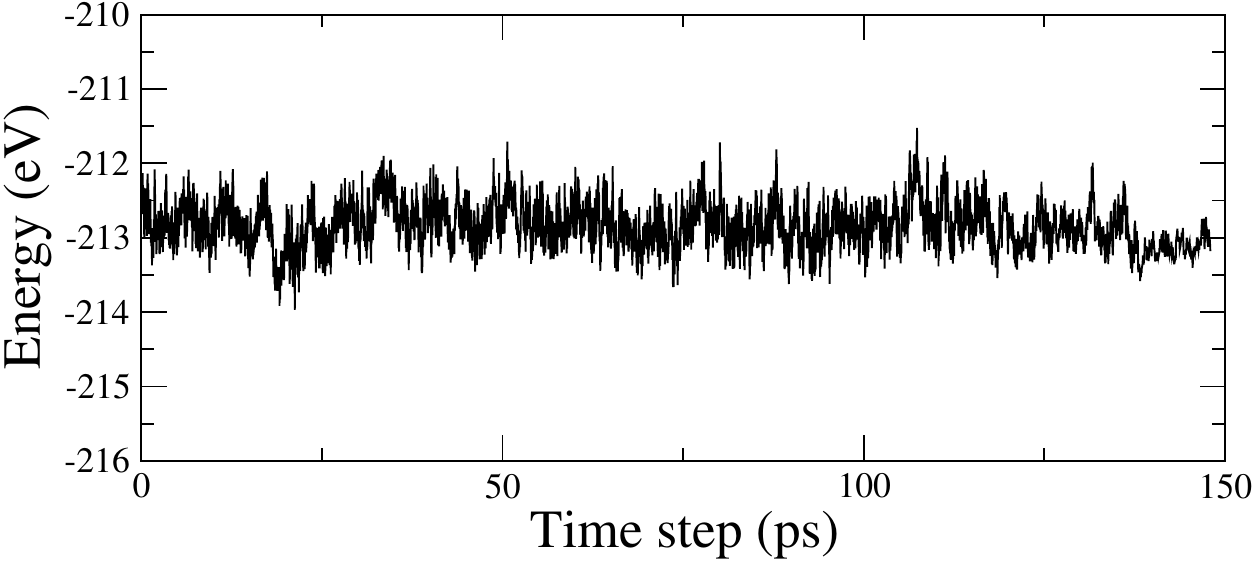}} 
\caption{Ab initio molecular dynamics simulations profiles of energy as a function of time of bismuthene nanopores. a) divacancy, b) tetravacancy, c) hexagonal vacancy bare P6a, d) hexagonal vacancy bare P6b, e) hexagonal vacancy bare P6c, f) P555-777, g)  bare hexagonal pore type-2, and  h)  hydrogenated hexagonal pore type-2.}
\label{fig:S1}
\end{figure*}

\begin{figure*}[ht!]
  \centering
  \subfloat[]{\includegraphics[width = 3cm, scale=1, clip = true]{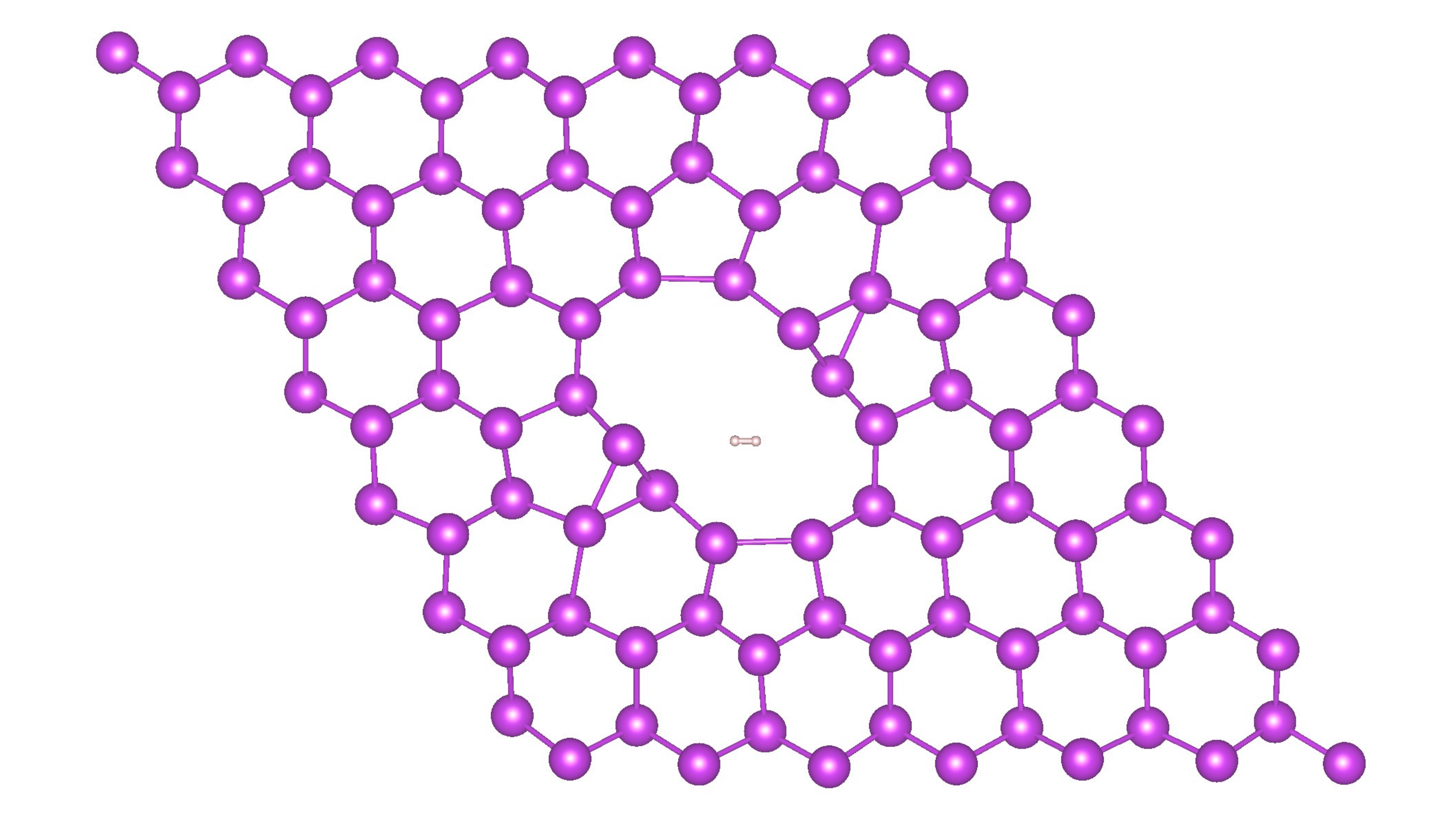}}
  \subfloat[]{\includegraphics[width = 3cm, scale=1, clip = true]{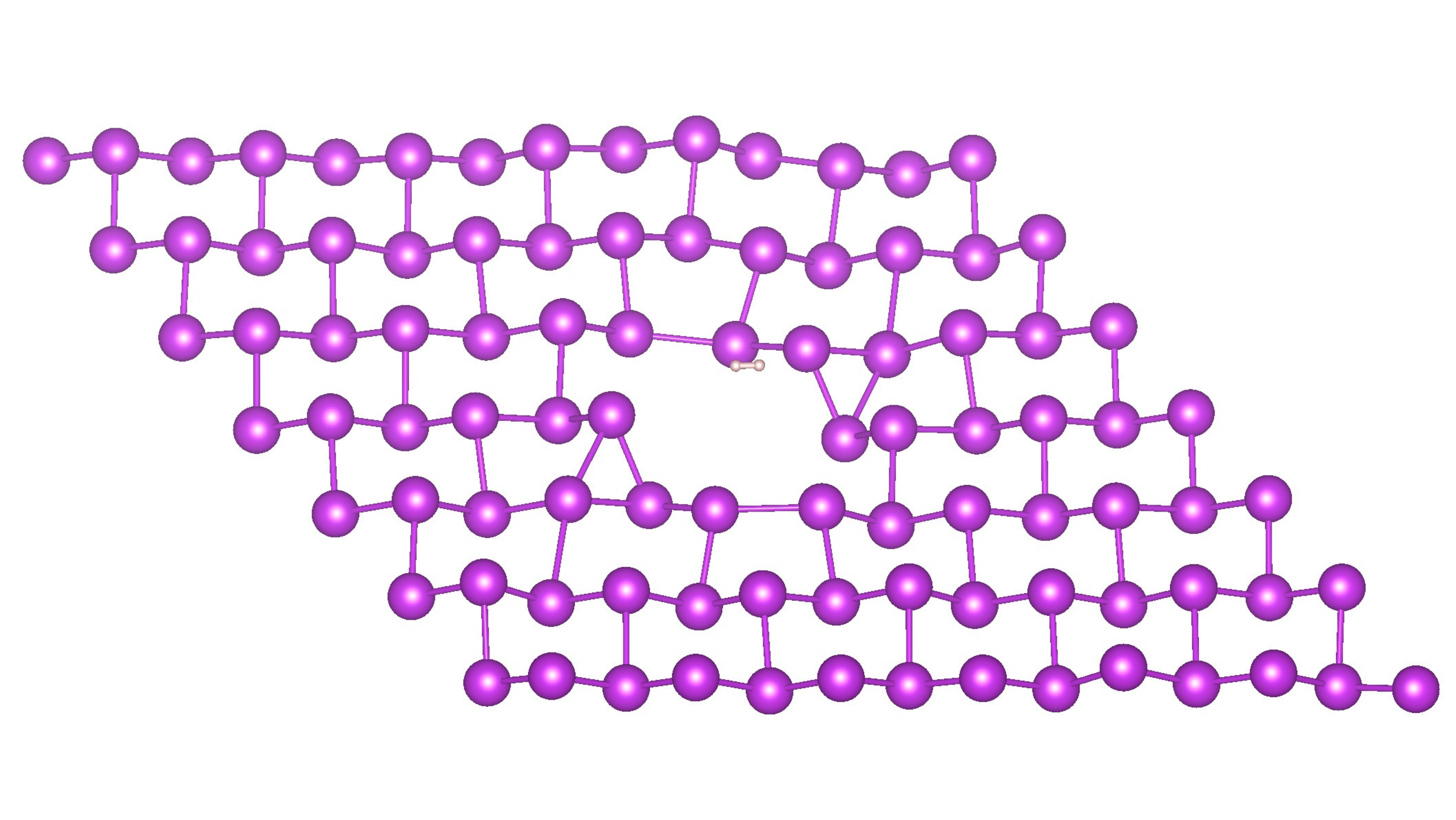}}
  \subfloat[]{\includegraphics[width = 3cm, scale=1, clip = true]{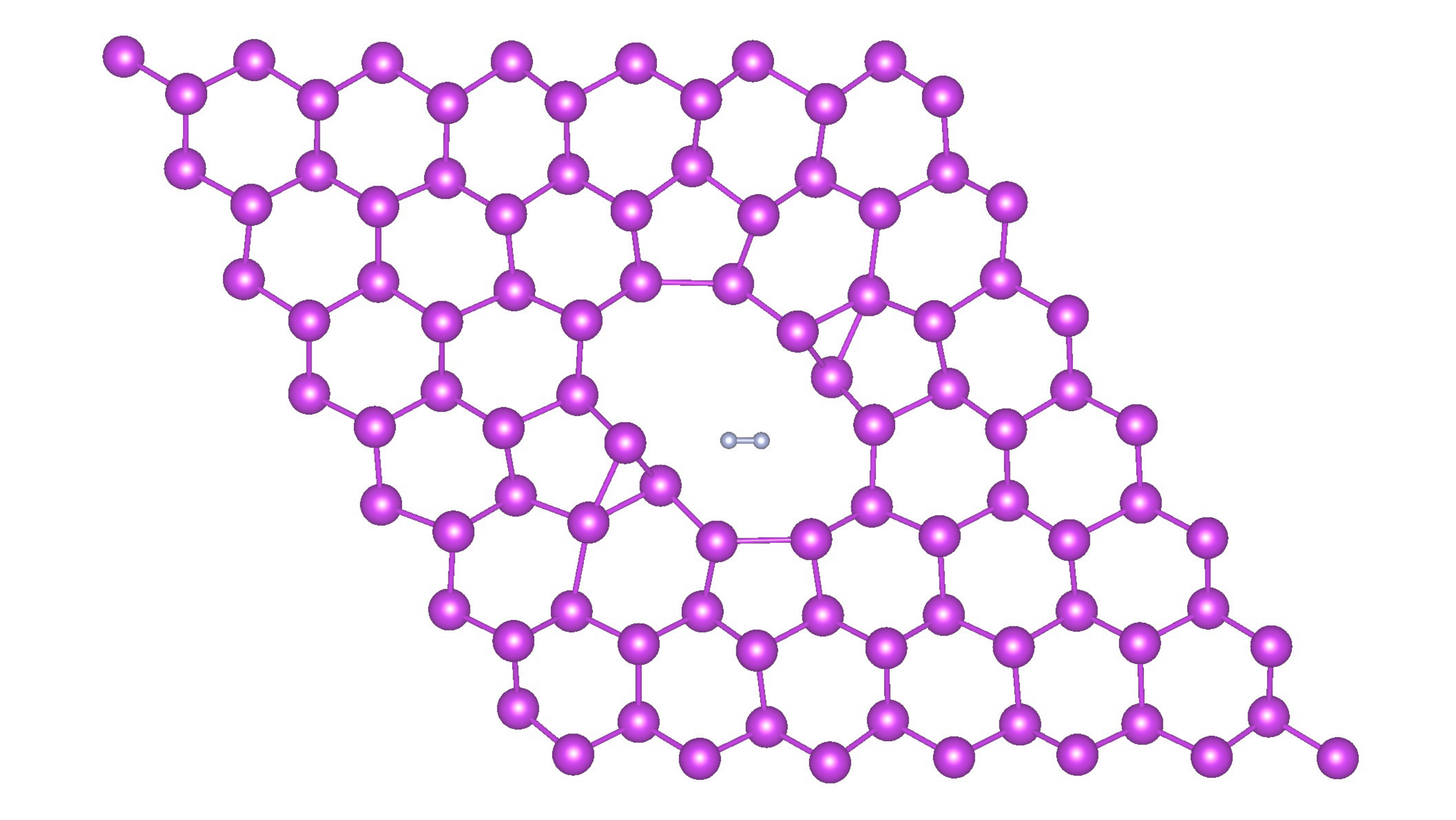}}
  \subfloat[]{\includegraphics[width = 3cm, scale=1, clip = true]{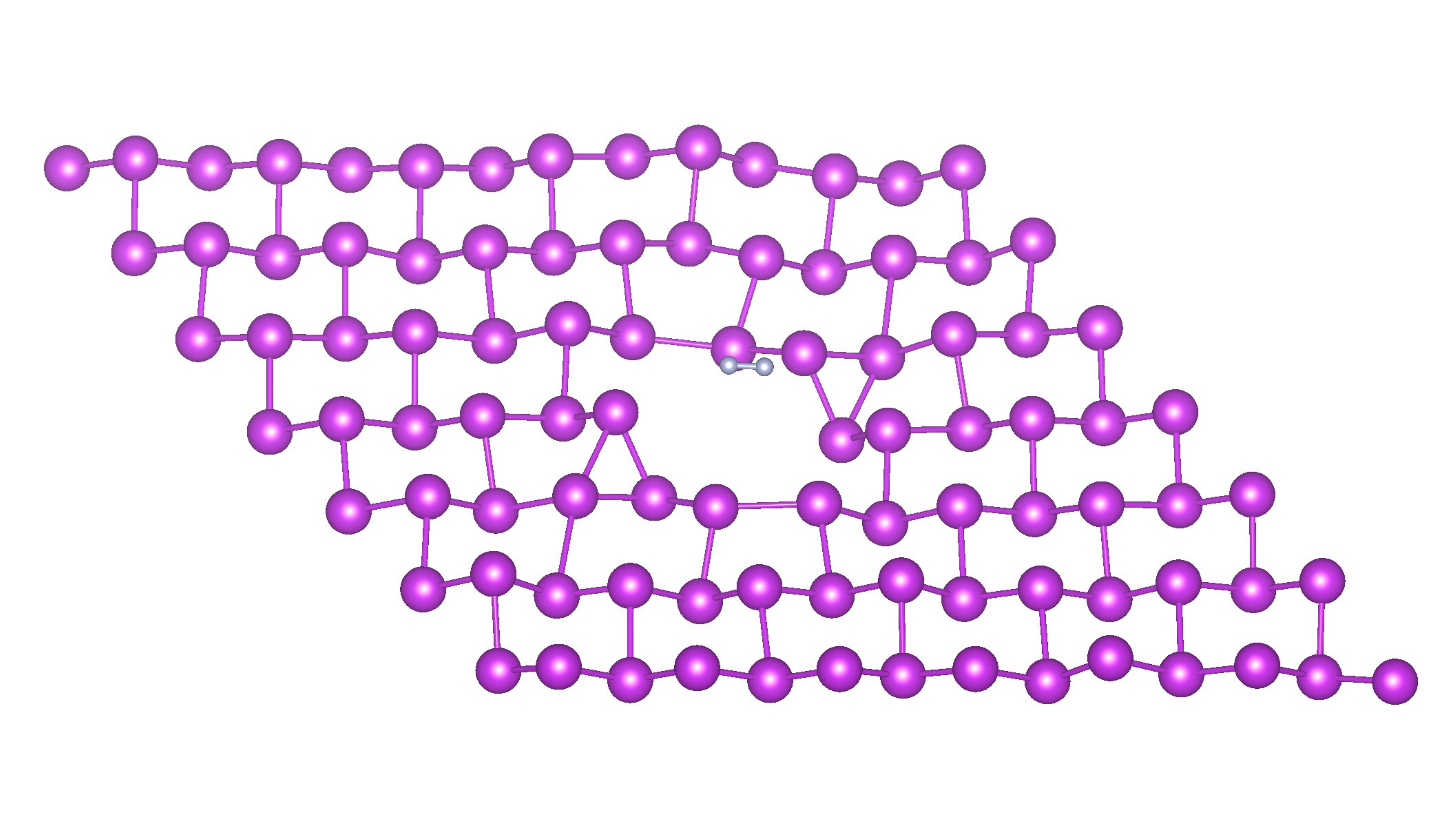}} \\
  \subfloat[]{\includegraphics[width = 3cm, scale=1, clip = true]{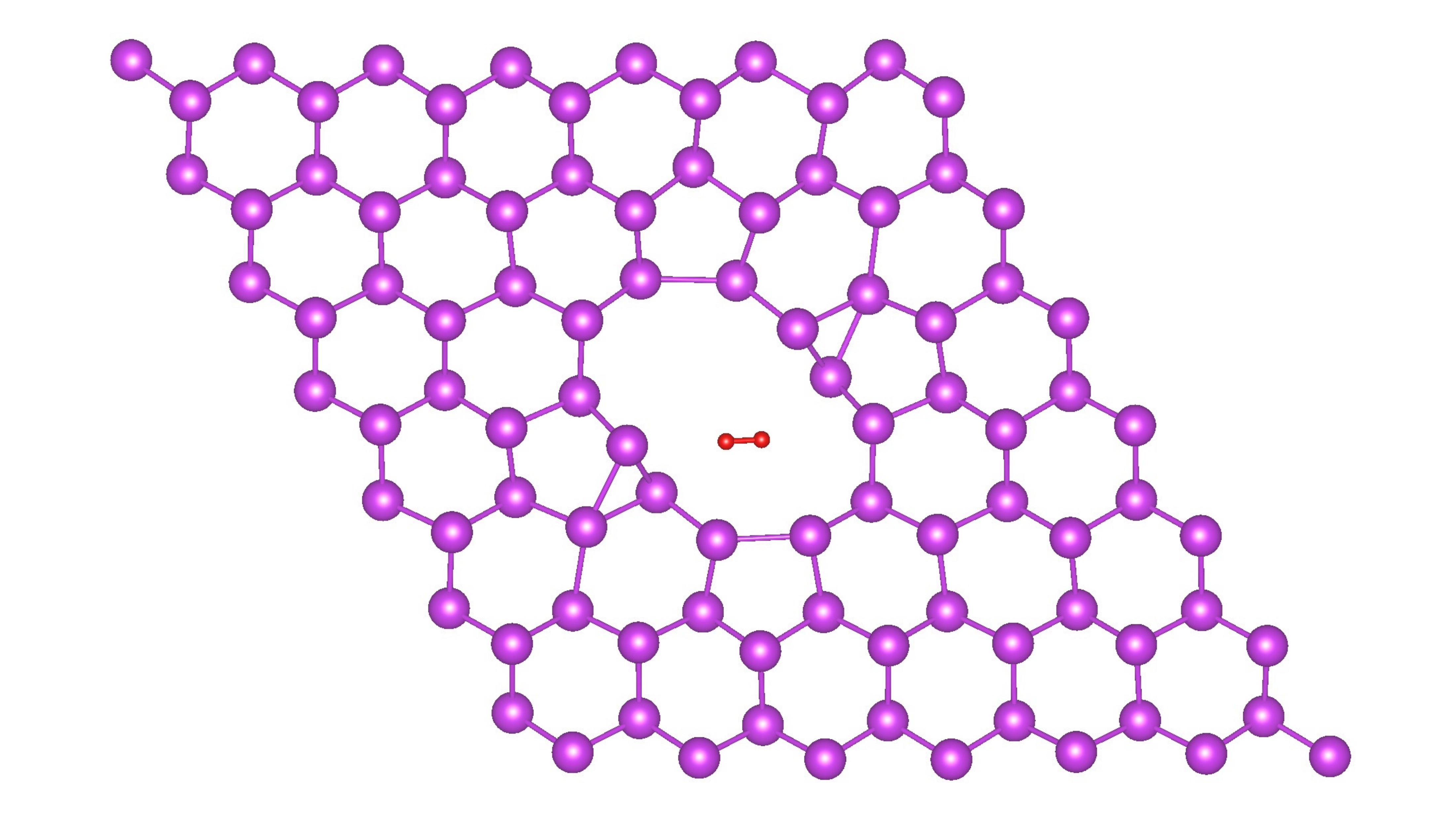}}
  \subfloat[]{\includegraphics[width = 3cm, scale=1, clip = true]{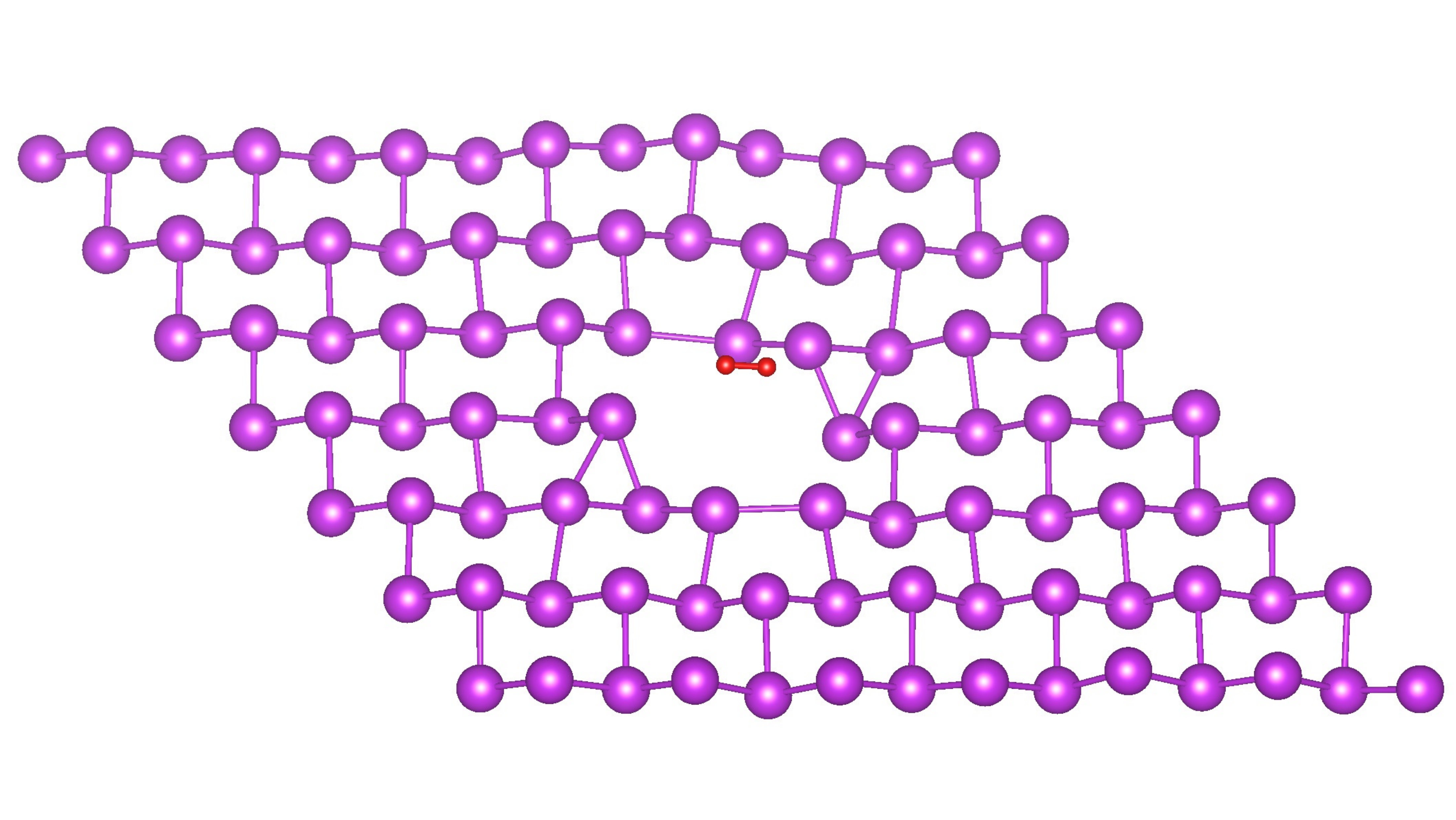}}
  \subfloat[]{\includegraphics[width = 3cm, scale=1, clip = true]{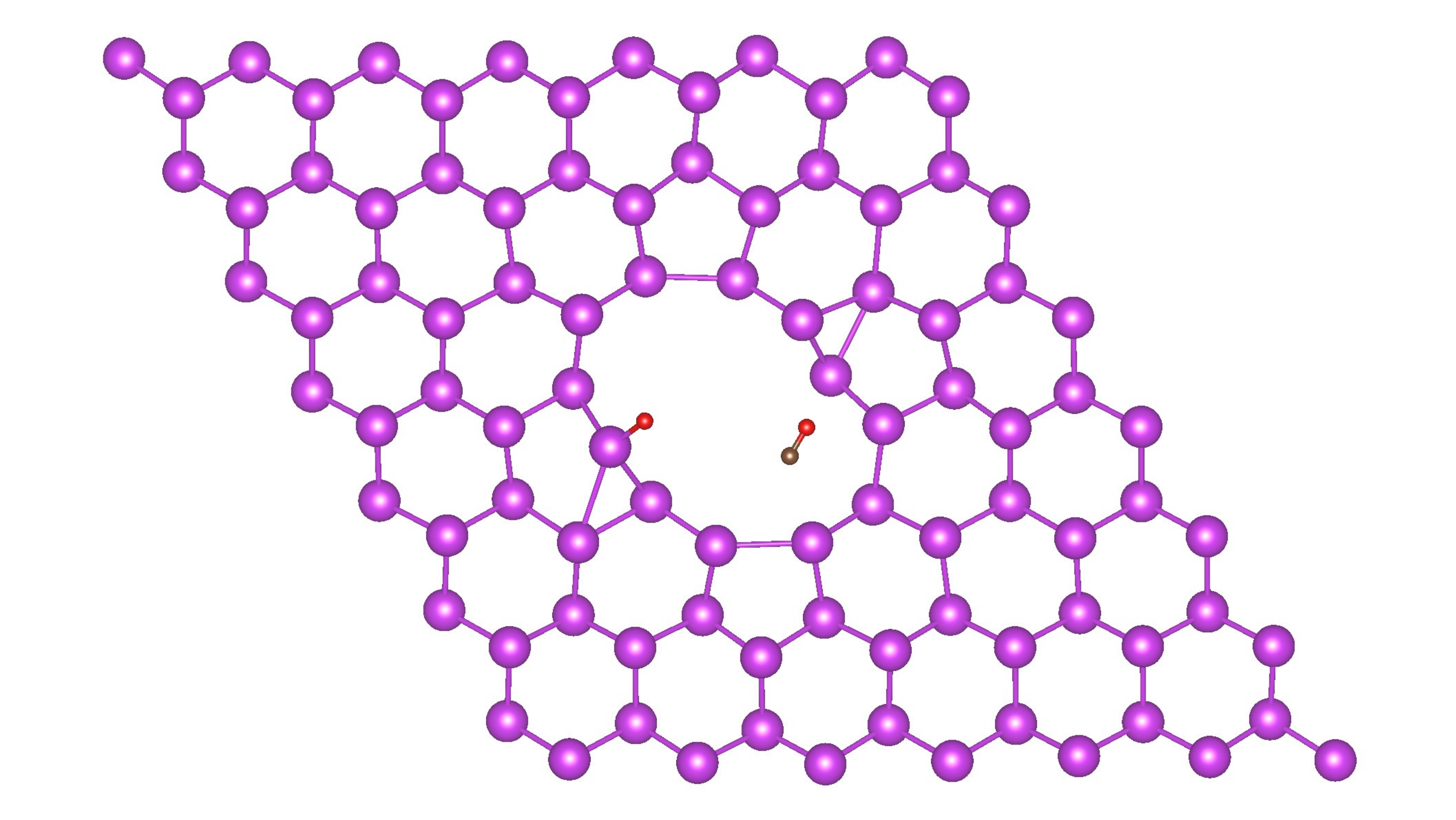}}
  \subfloat[]{\includegraphics[width = 3cm, scale=1, clip = true]{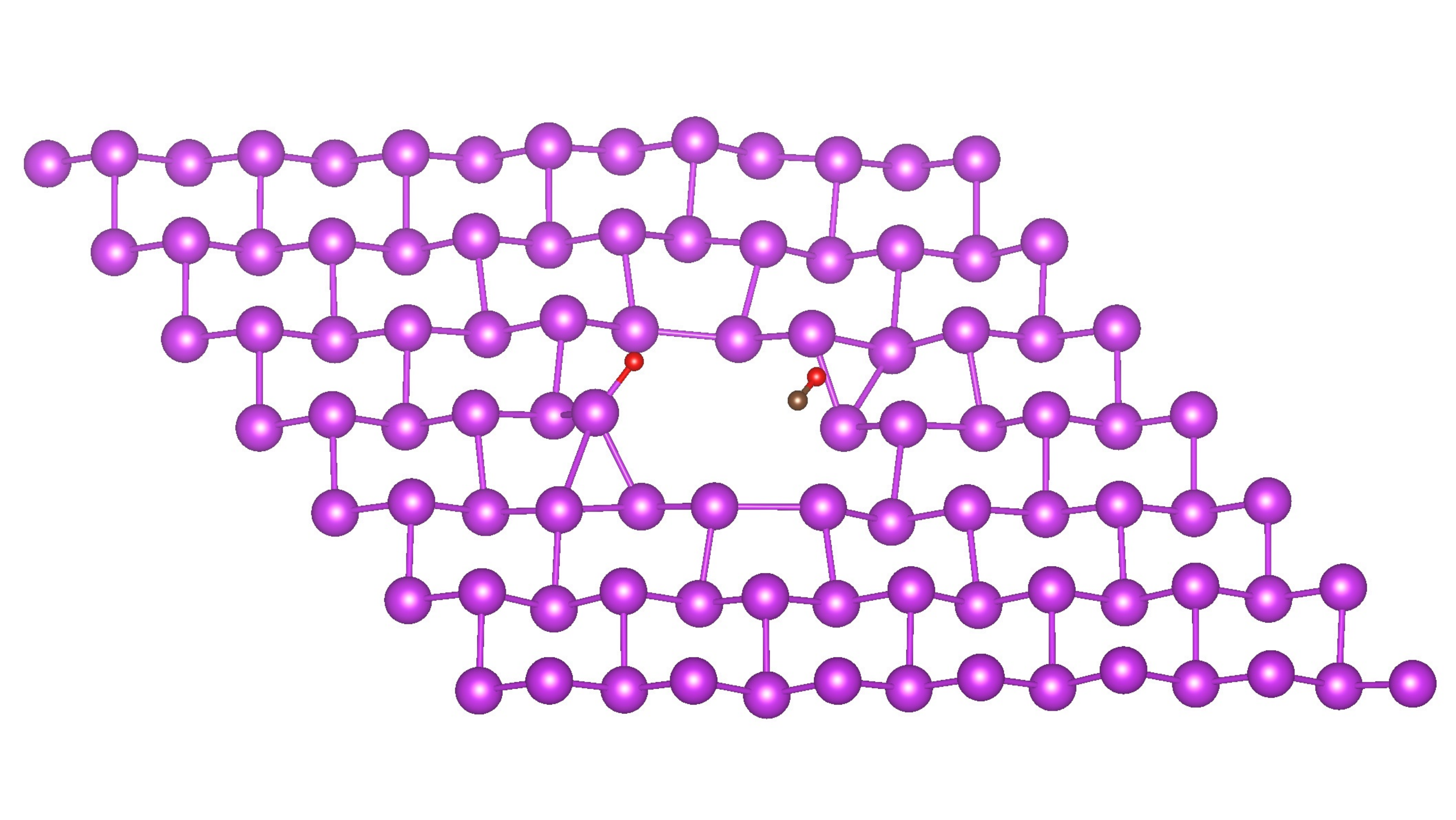}}\\
  \subfloat[]{\includegraphics[width = 3cm, scale=1, clip = true]{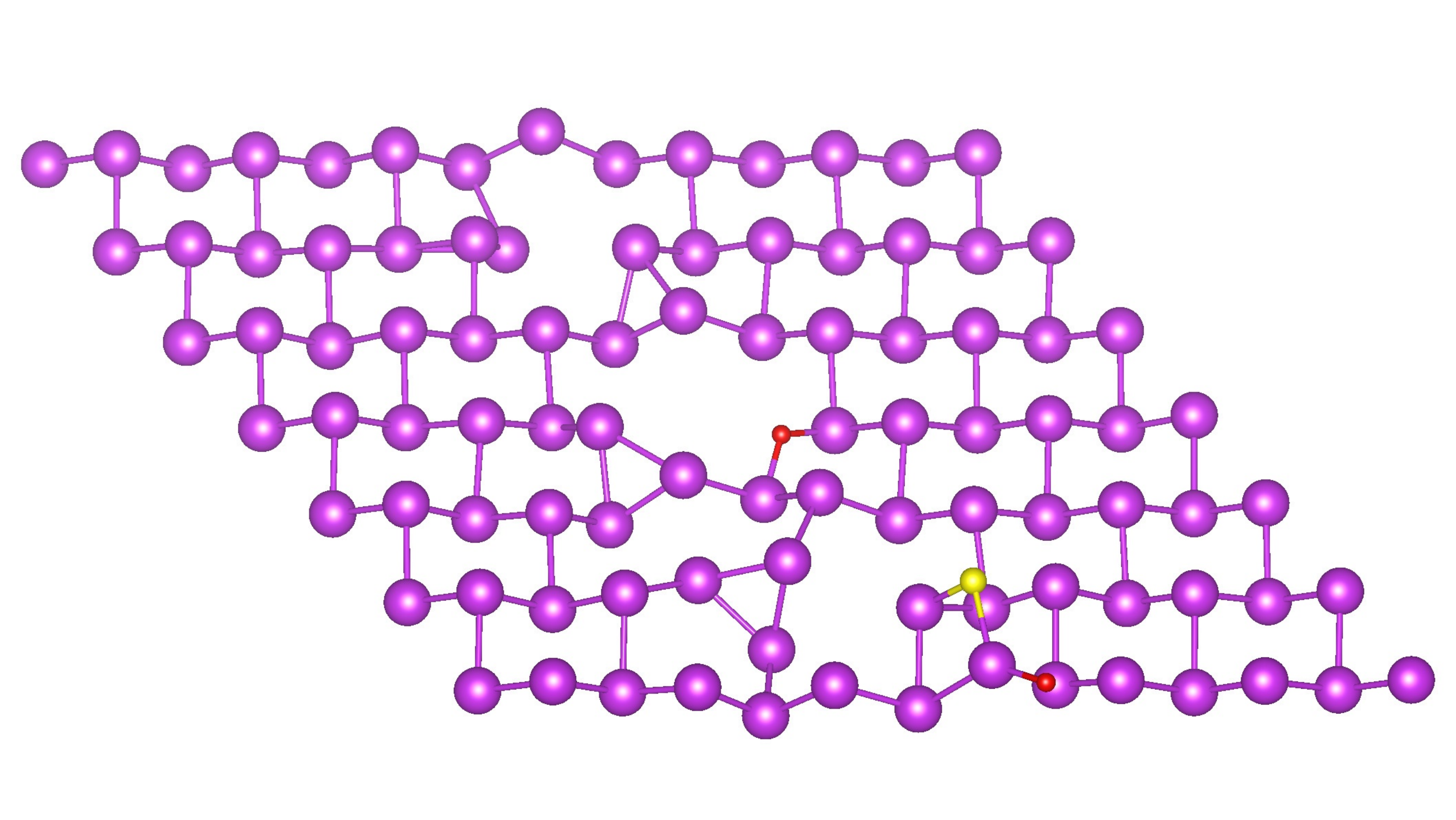}}
  \subfloat[]{\includegraphics[width = 3cm, scale=1, clip = true]{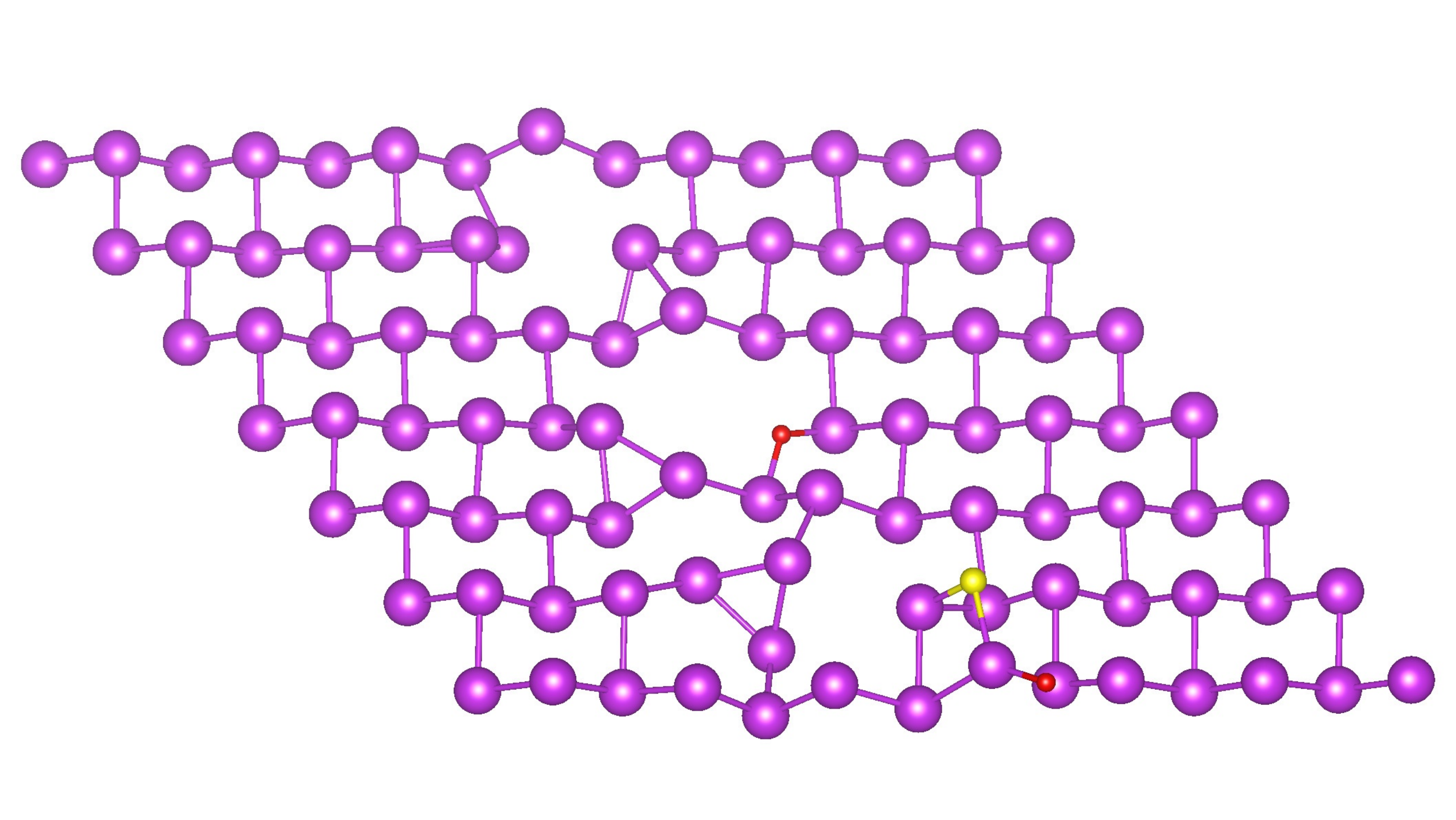}}
  \subfloat[]{\includegraphics[width = 2.8cm, scale=1, clip = true]{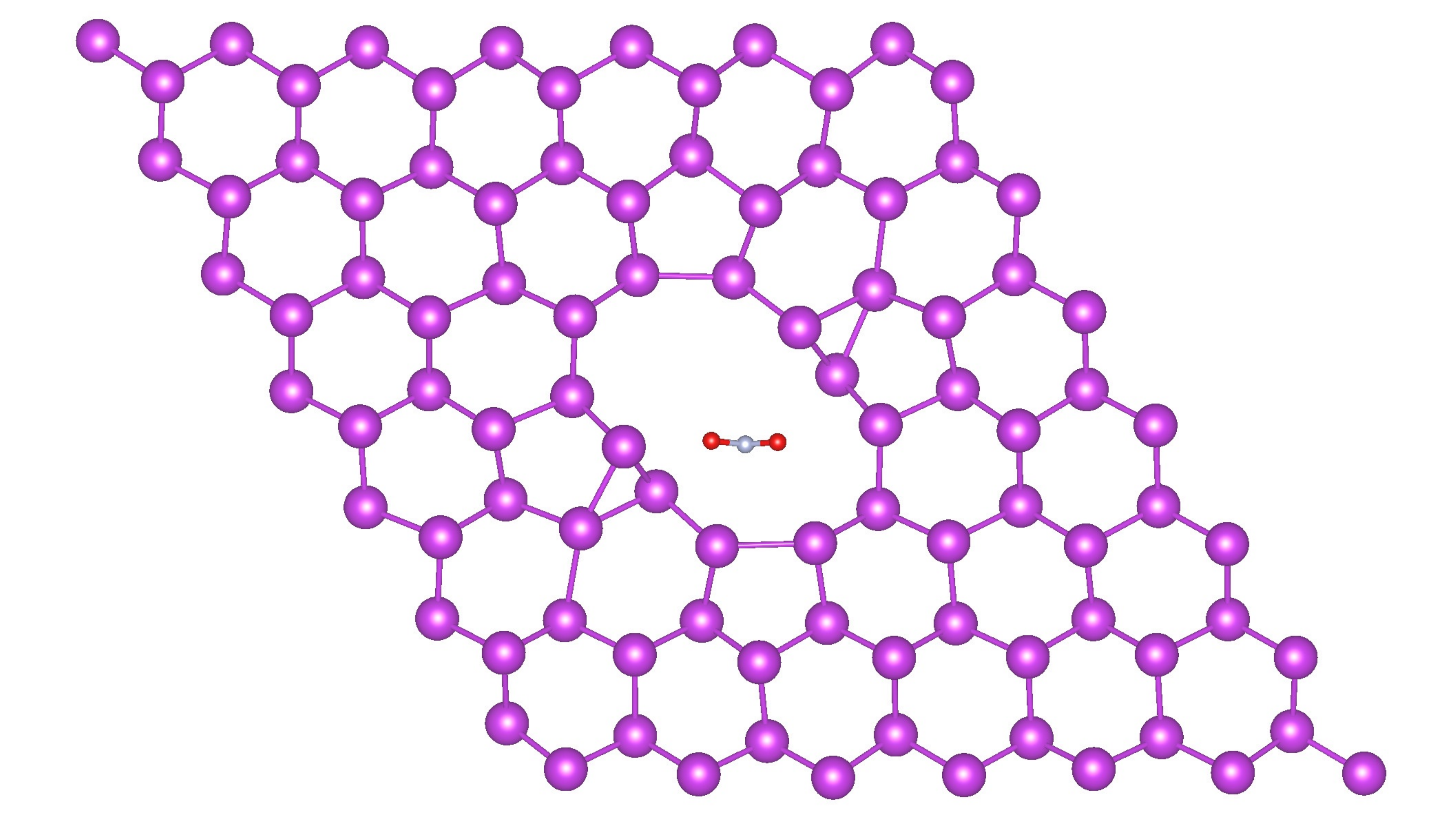}}
  \subfloat[]{\includegraphics[width = 2.8cm, scale=1, clip = true]{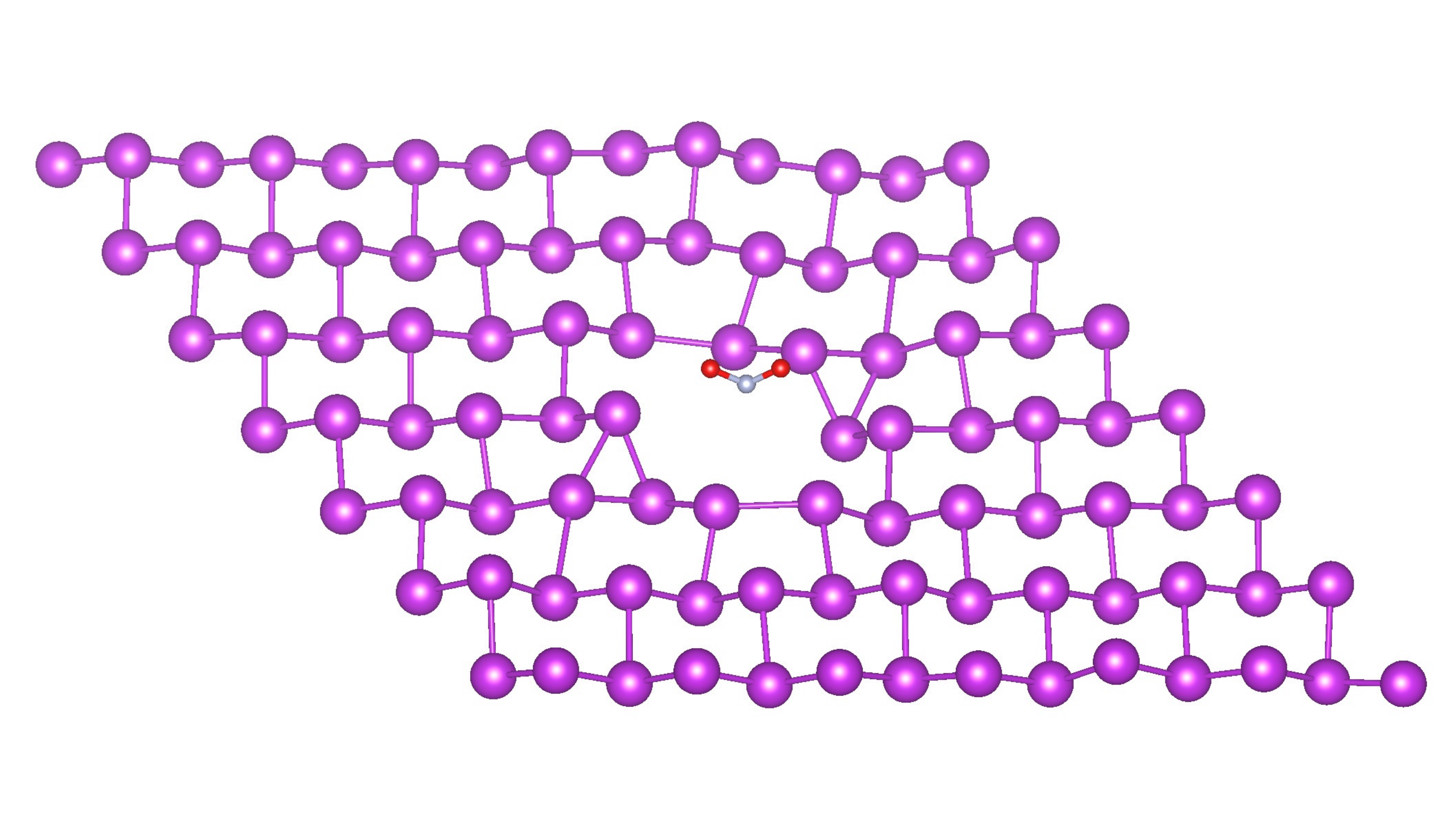}} \\
  \subfloat[]{\includegraphics[width = 3cm, scale=1, clip = true]{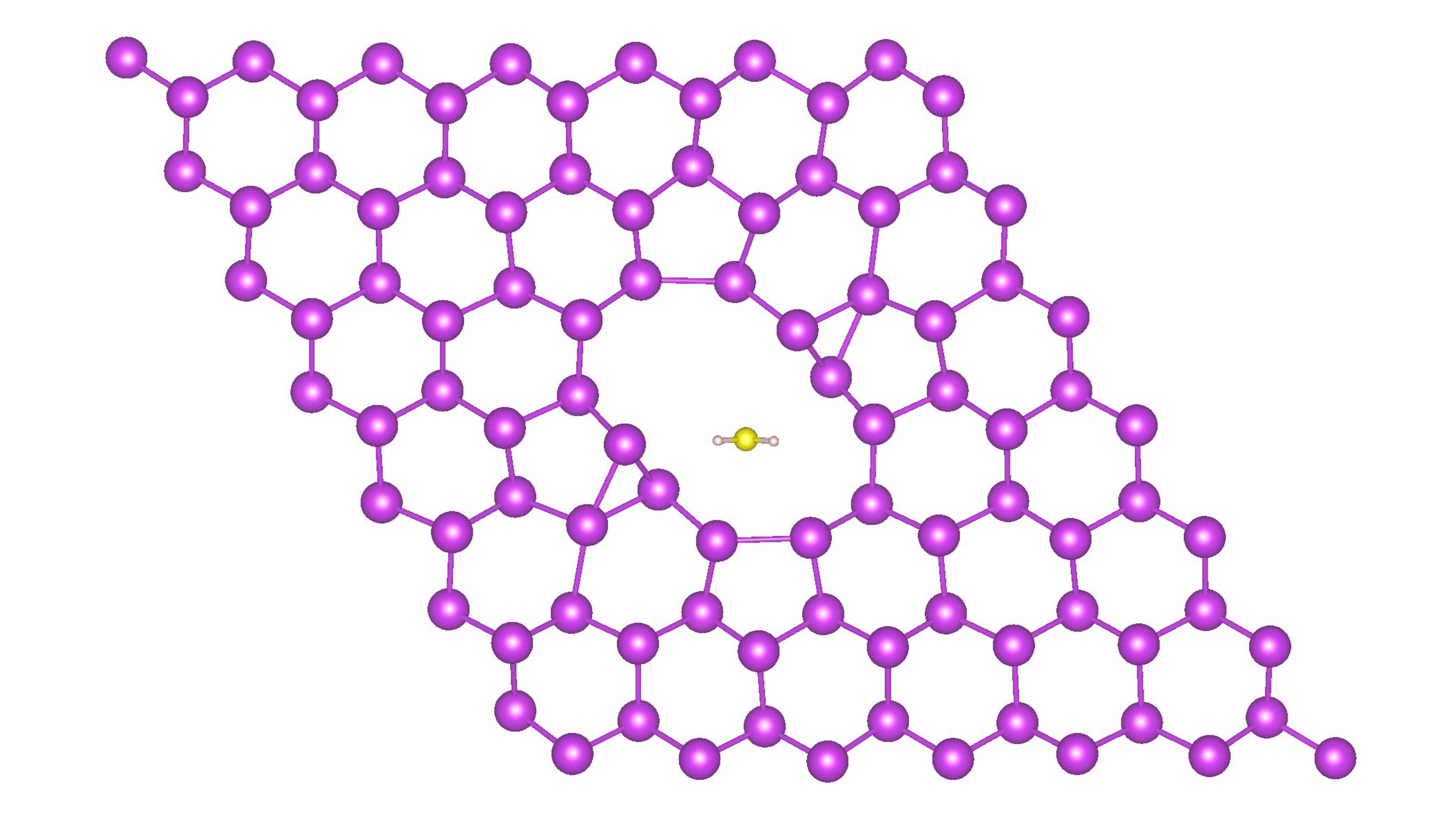}}
  \subfloat[]{\includegraphics[width = 3cm, scale=1, clip = true]{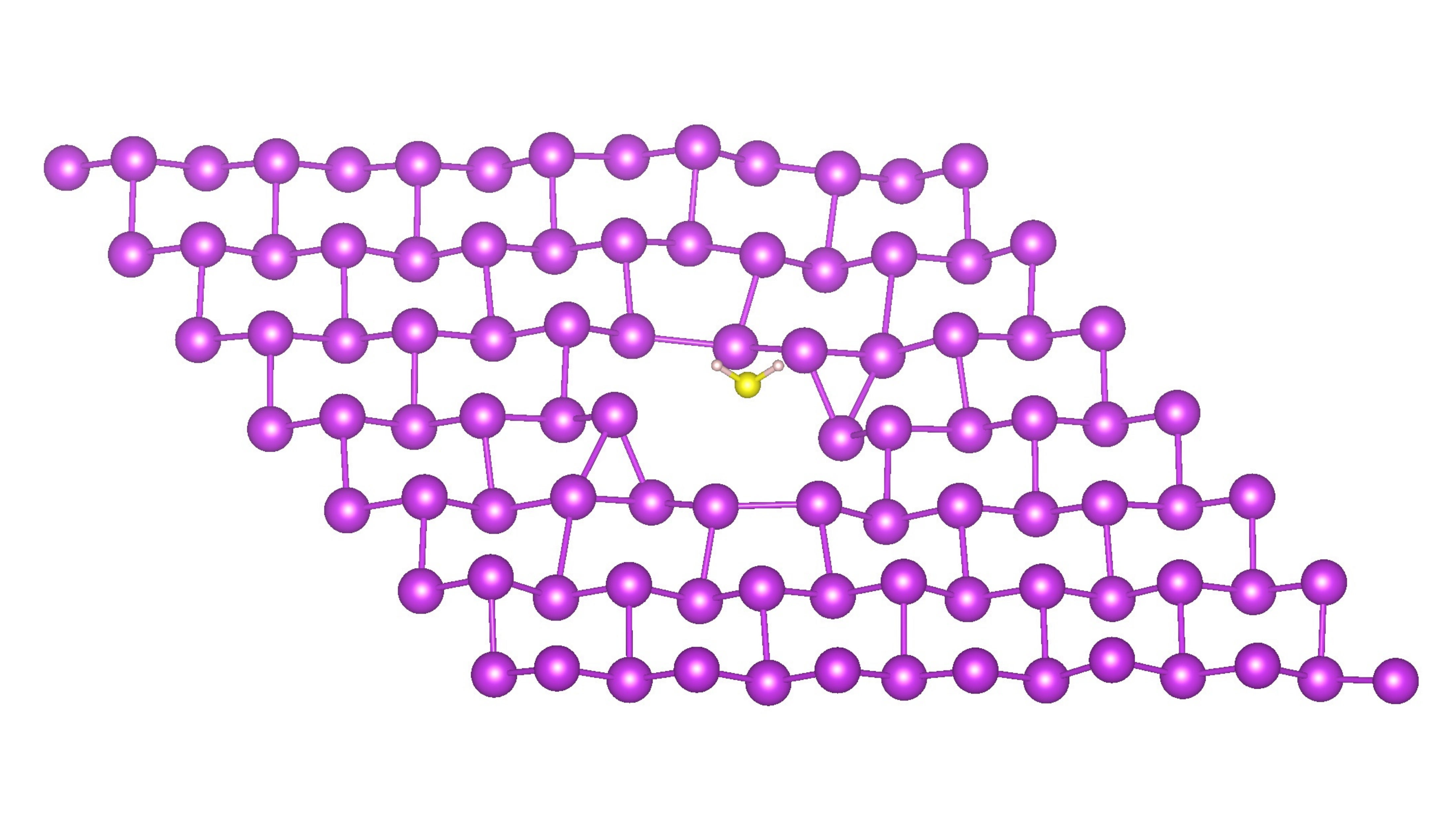}}
  \subfloat[]{\includegraphics[width = 3cm, scale=1, clip = true]{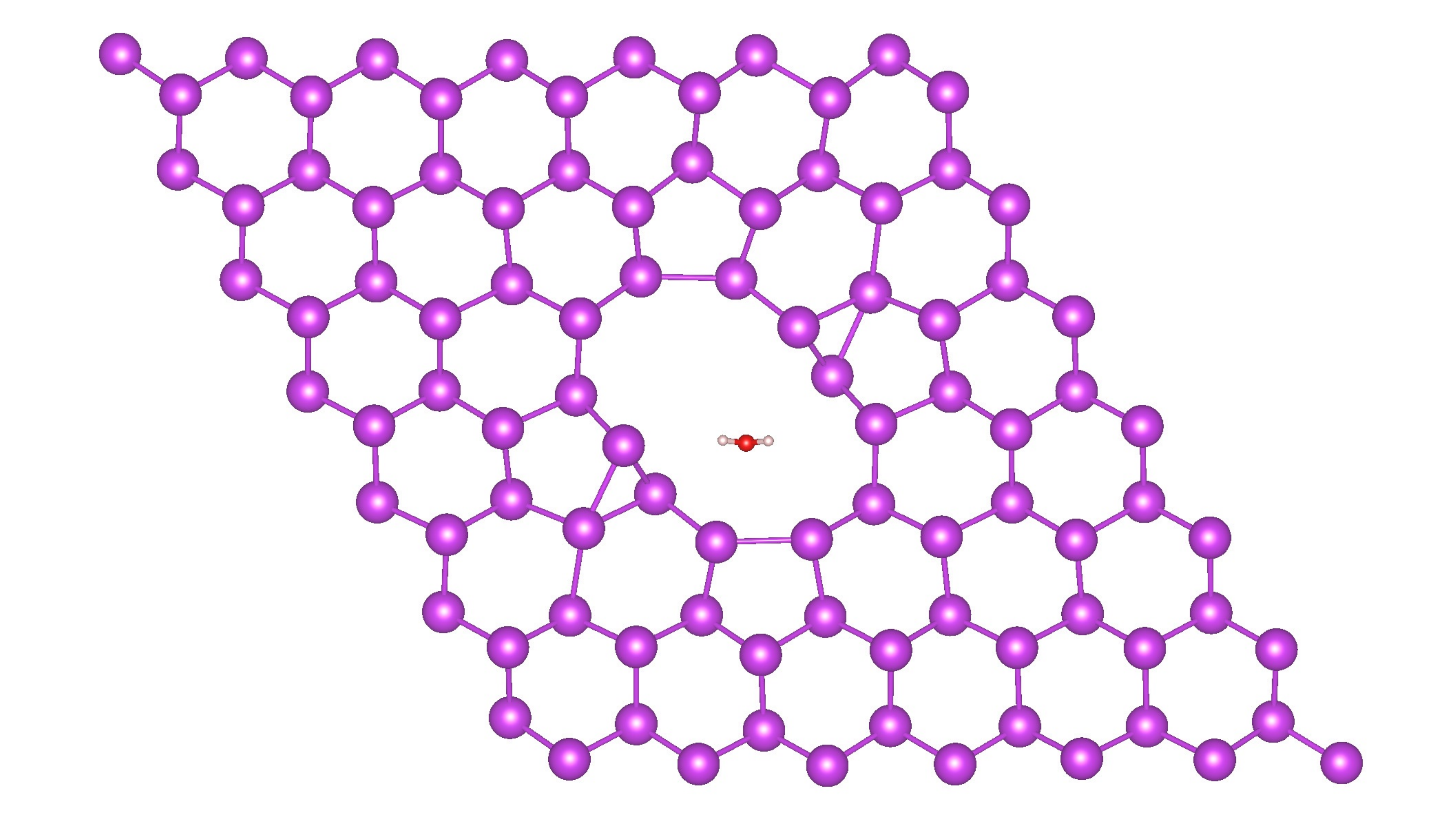}}
  \subfloat[]{\includegraphics[width = 3cm, scale=1, clip = true]{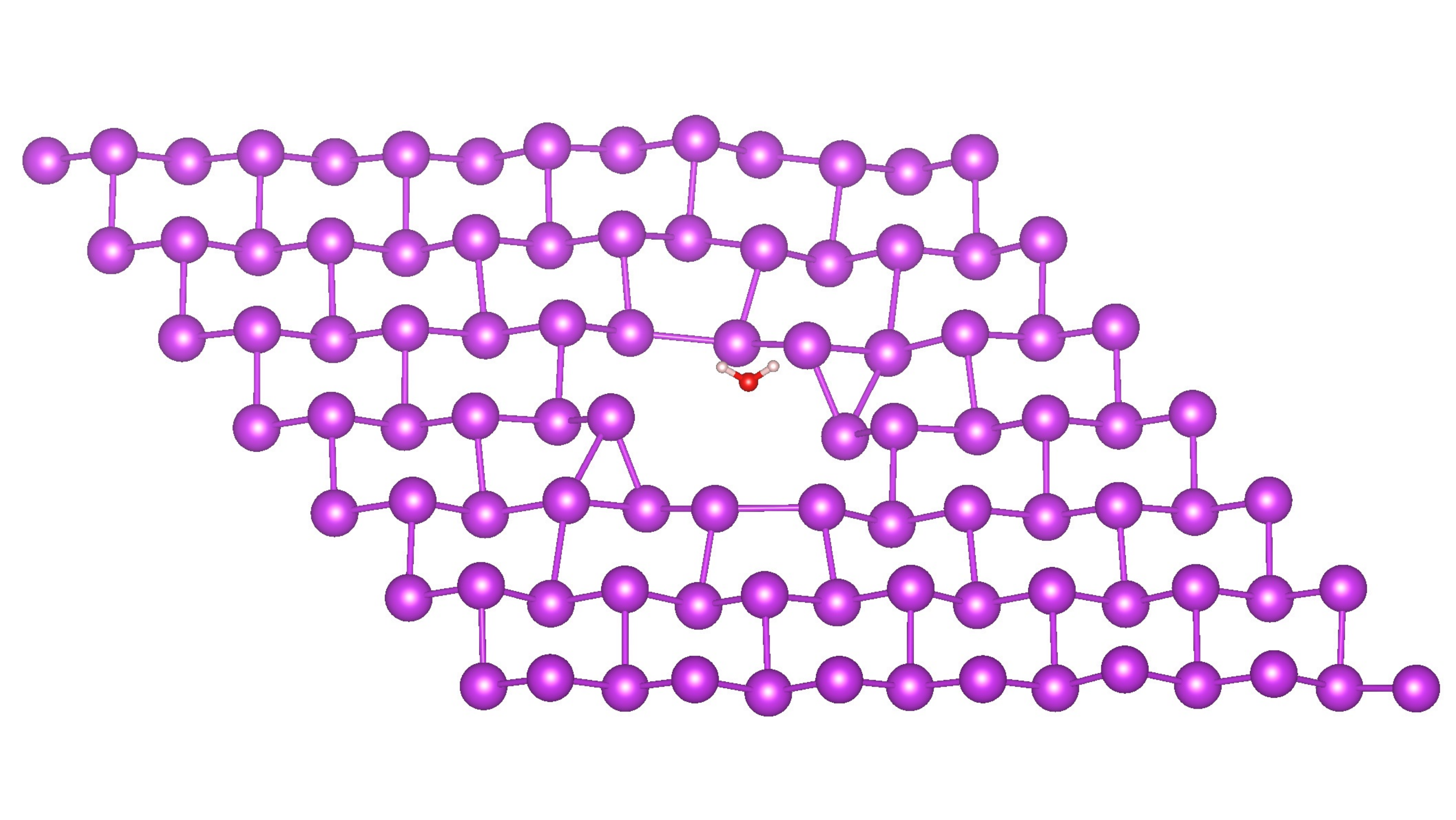}} \\
  \subfloat[]{\includegraphics[width = 3cm, scale=1, clip = true]{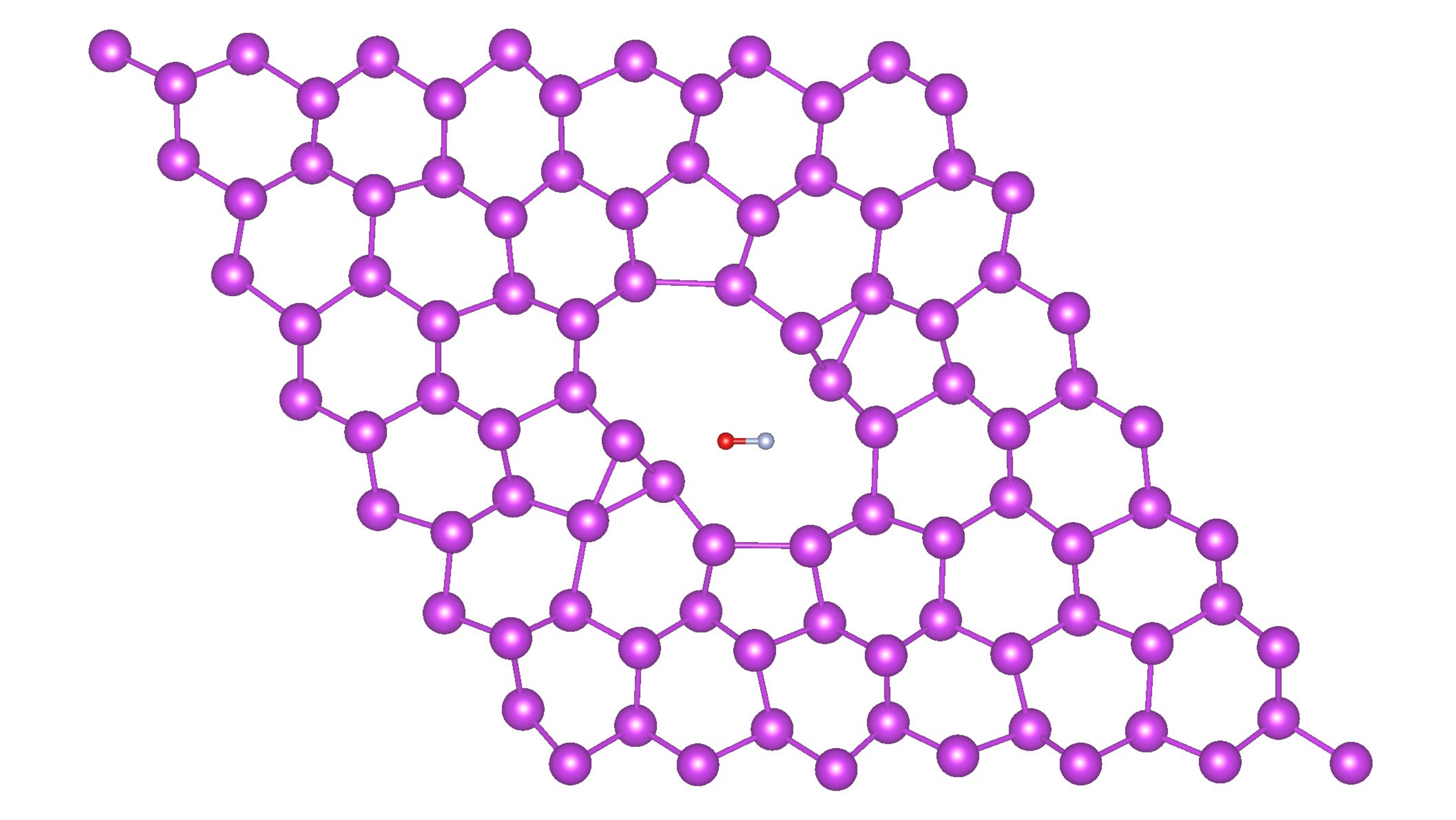}}
  \subfloat[]{\includegraphics[width = 3cm, scale=1, clip = true]{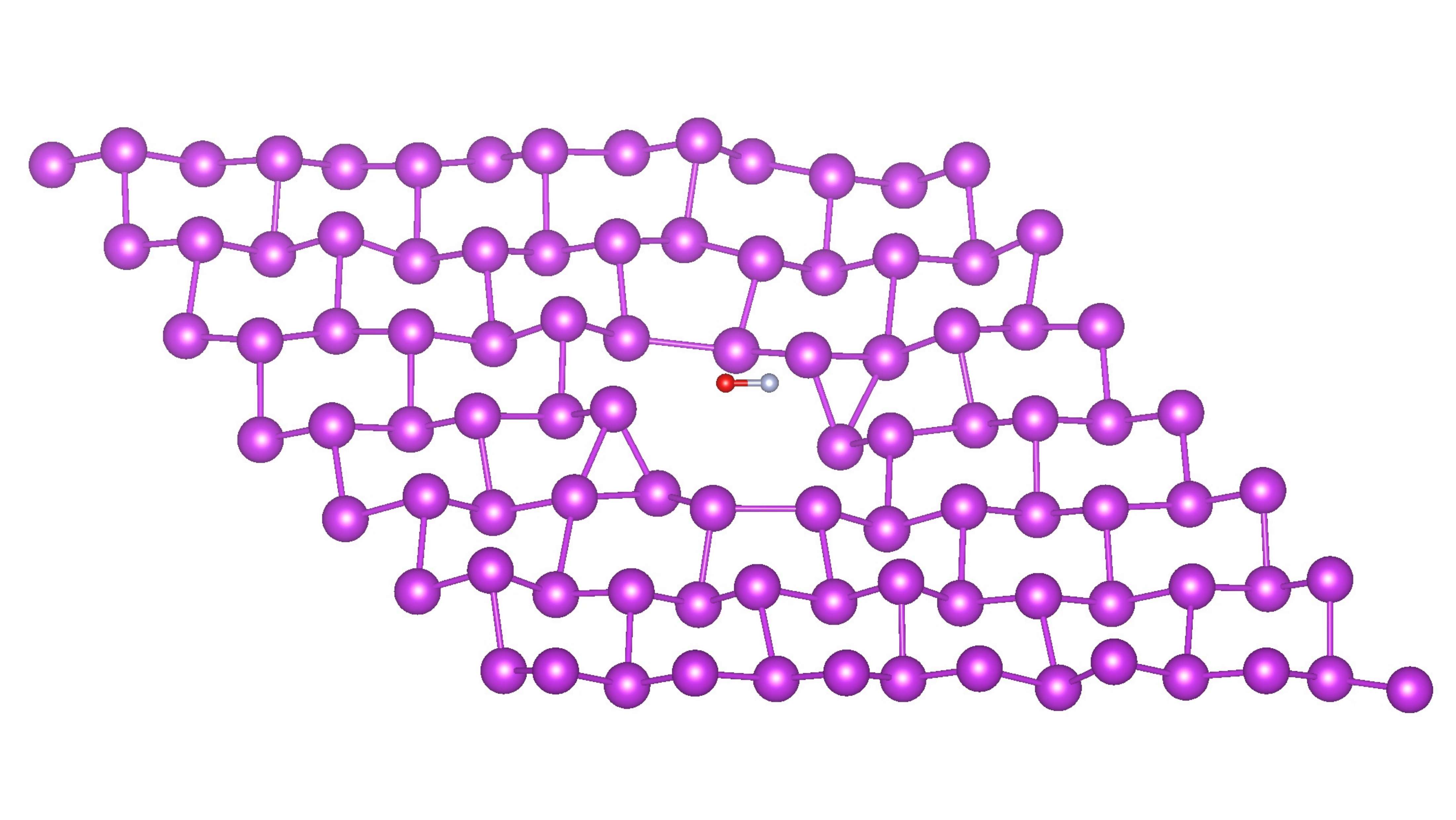}} \
  \subfloat[]{\includegraphics[width = 3cm, scale=1, clip = true]{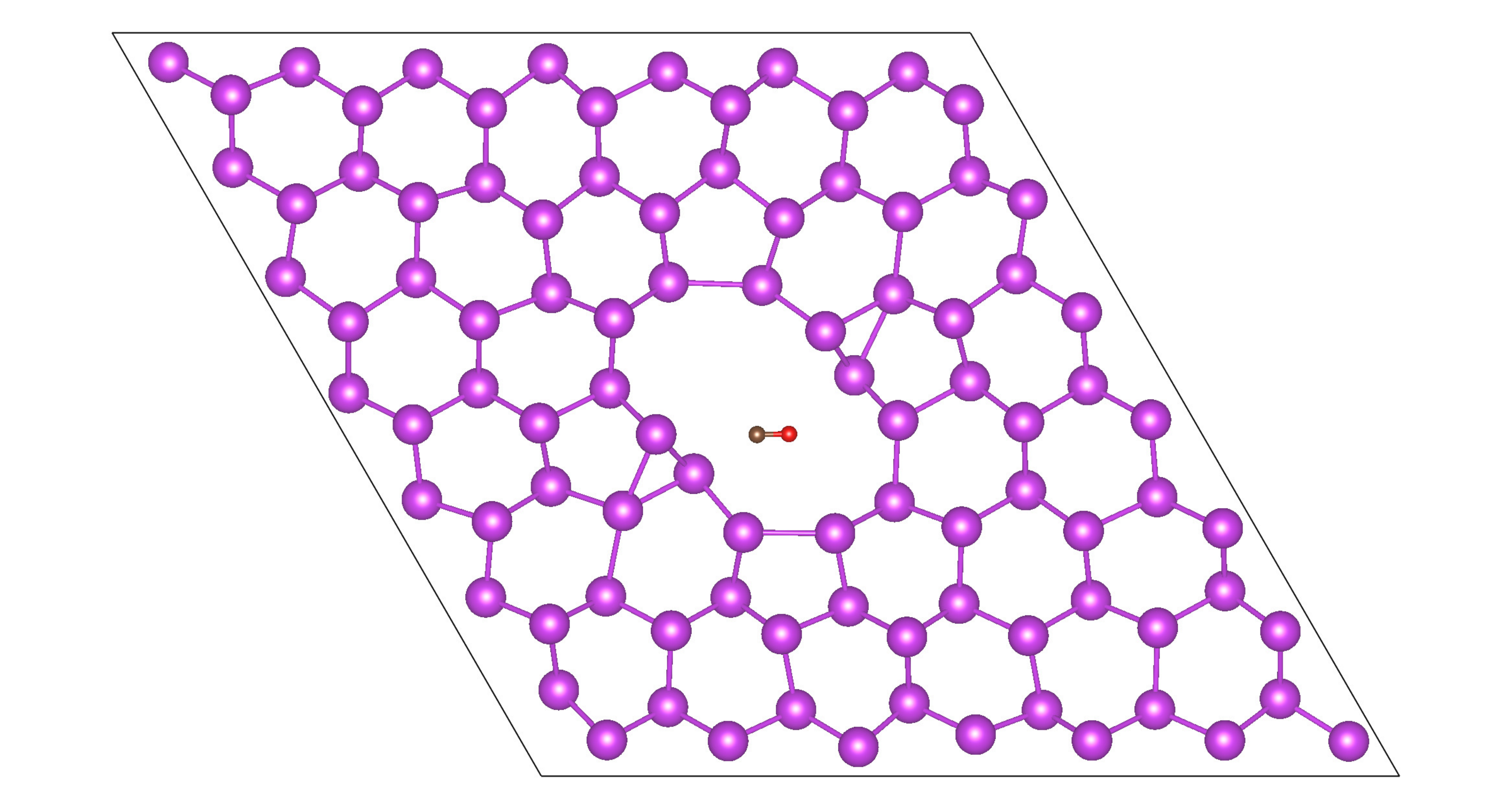}}
  \subfloat[]{\includegraphics[width = 3cm, scale=1, clip = true]{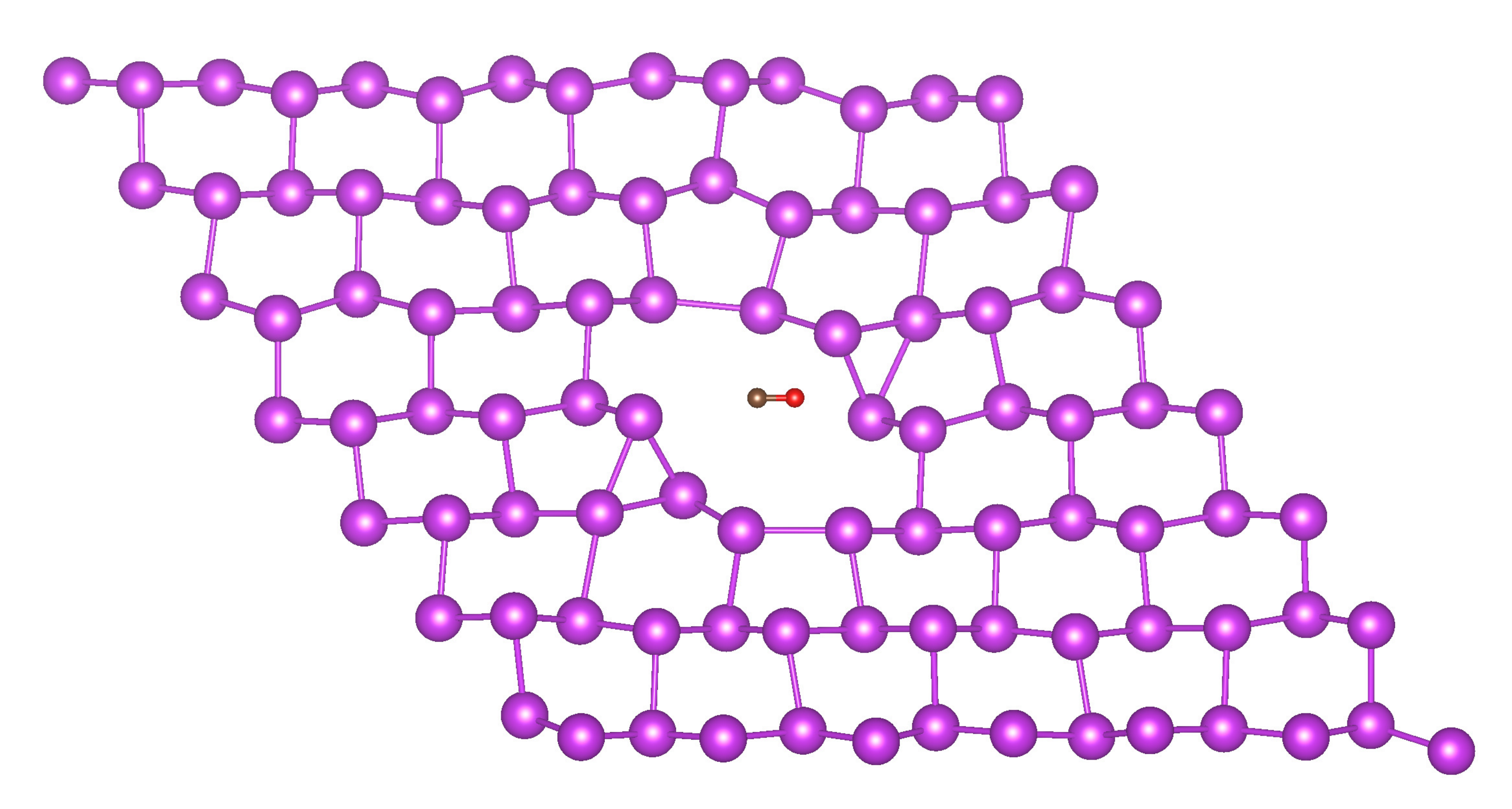}}
  \caption{Adsorption of small molecules in bismuthene nanopores on a tetravacancy (P4) calculated within GGA-PBE. a) H$_2$, b) N$_2$, c) O$_2$, d) CO$_2$, e) SO$_2$, f) NO$_2$, g) H$_2$S, h) H$_2$O and i) NO and j) CO.}
  \label{fig:S2}
\end{figure*}

\begin{figure*}[ht!]
  \centering
  \subfloat[]{\includegraphics[width = 3cm, scale=1, clip = true]{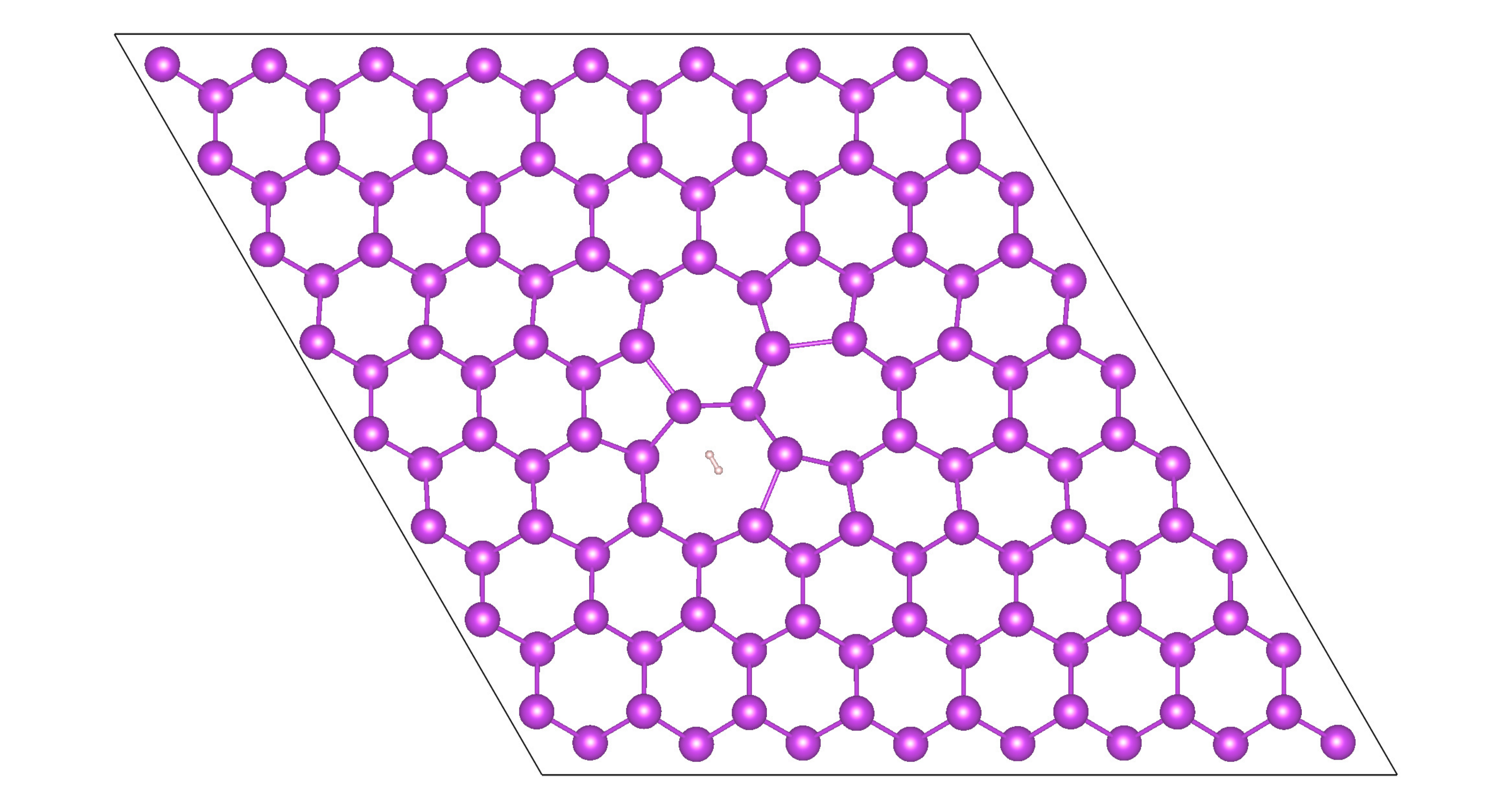}}
  \subfloat[]{\includegraphics[width = 3cm, scale=1, clip = true]{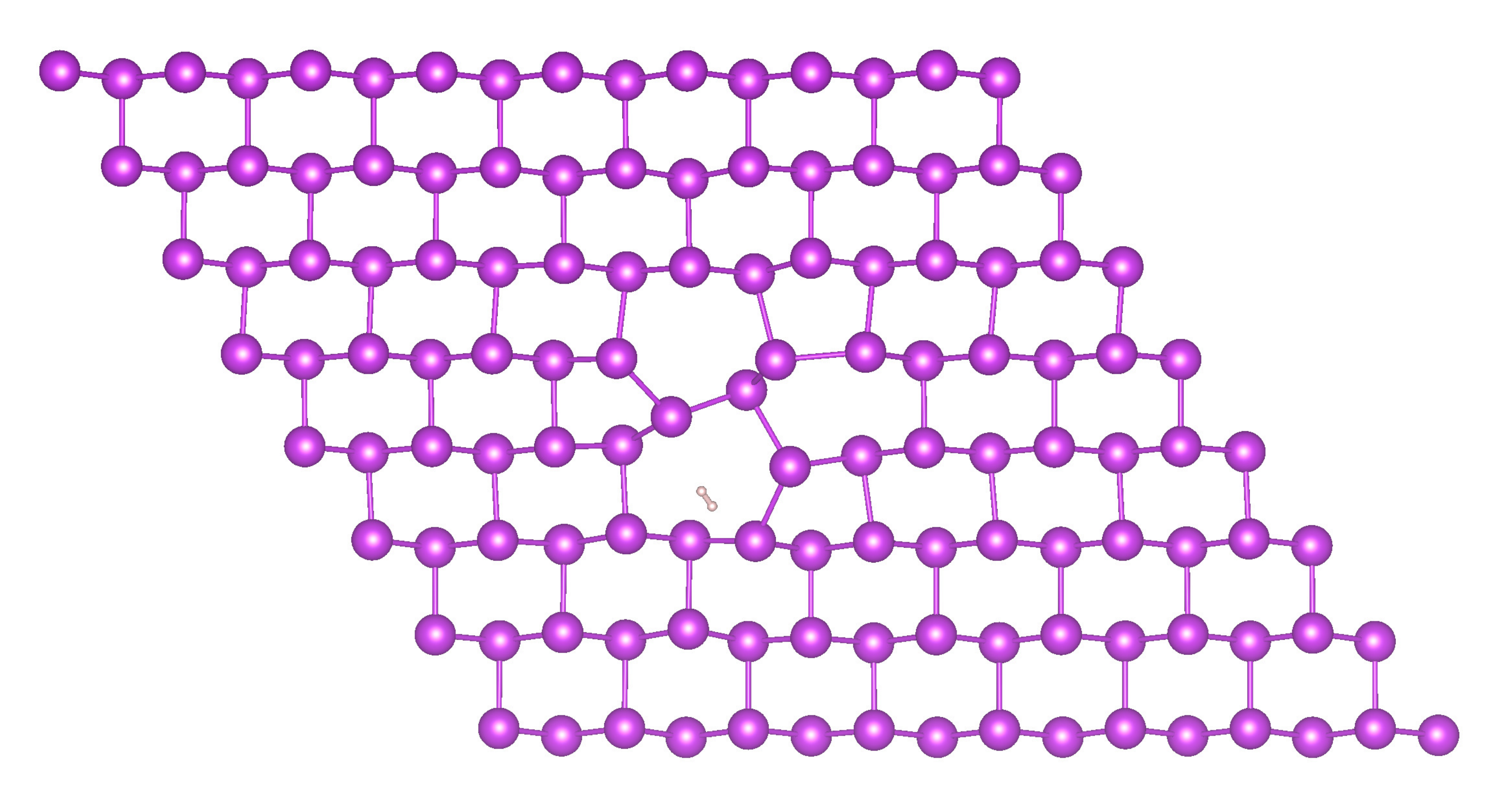}}
  \subfloat[]{\includegraphics[width = 3cm, scale=1, clip = true]{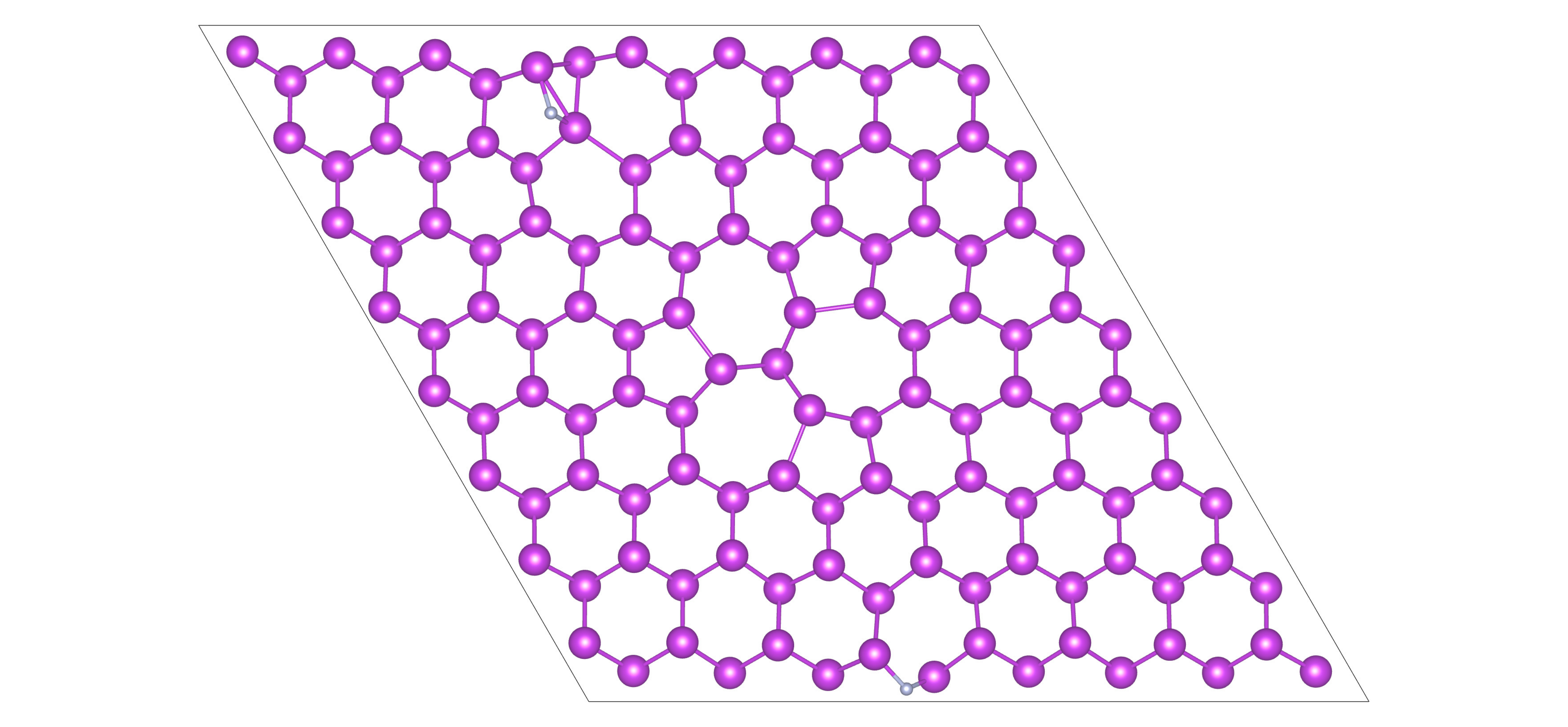}}
  \subfloat[]{\includegraphics[width = 3cm, scale=1, clip = true]{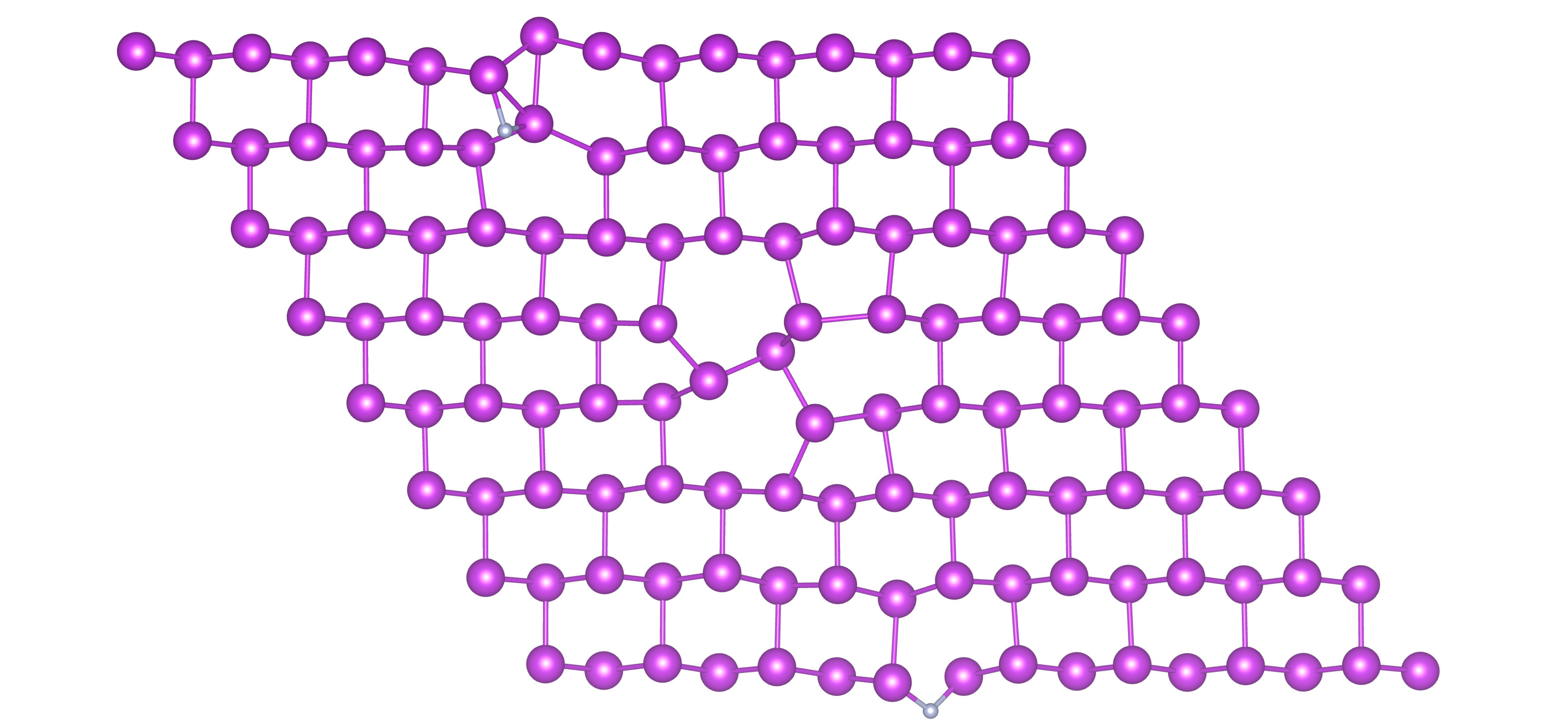}} \\
  \subfloat[]{\includegraphics[width = 3cm, scale=1, clip = true]{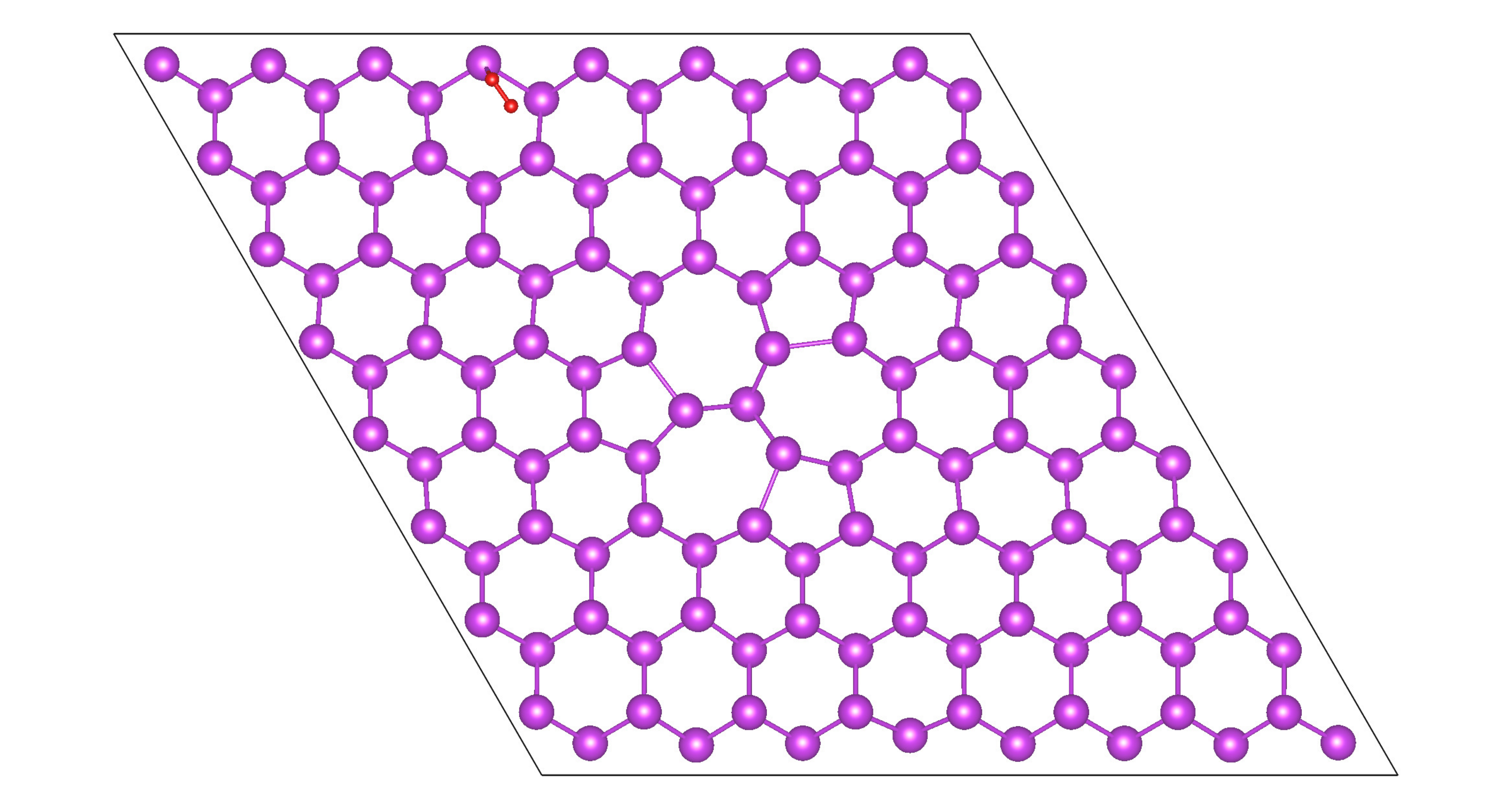}}
  \subfloat[]{\includegraphics[width = 3cm, scale=1, clip = true]{CONTCAR_555_777_O2_topview.pdf}}
  \subfloat[]{\includegraphics[width = 3cm, scale=1, clip = true]{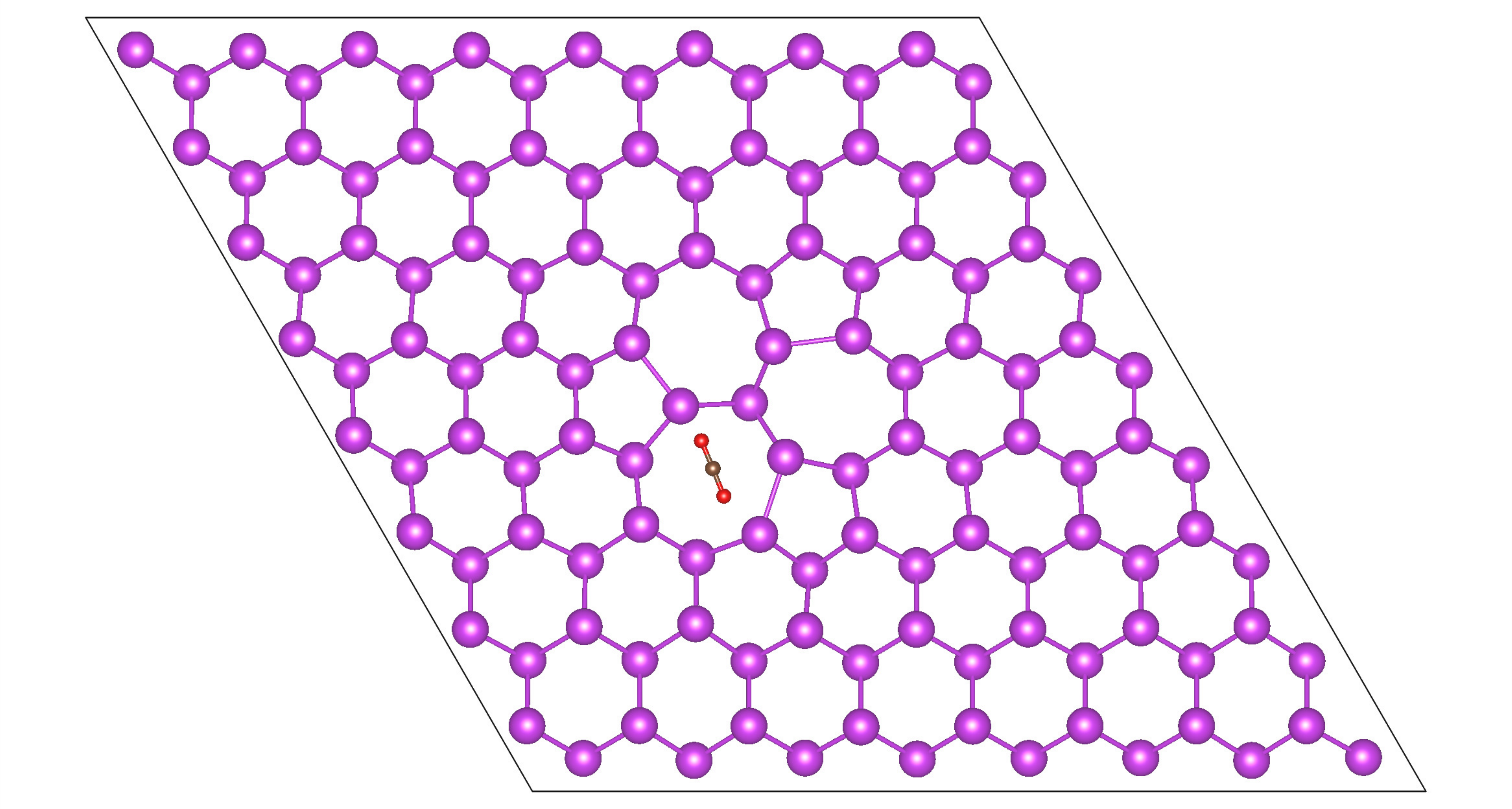}}
  \subfloat[]{\includegraphics[width = 3cm, scale=1, clip = true]{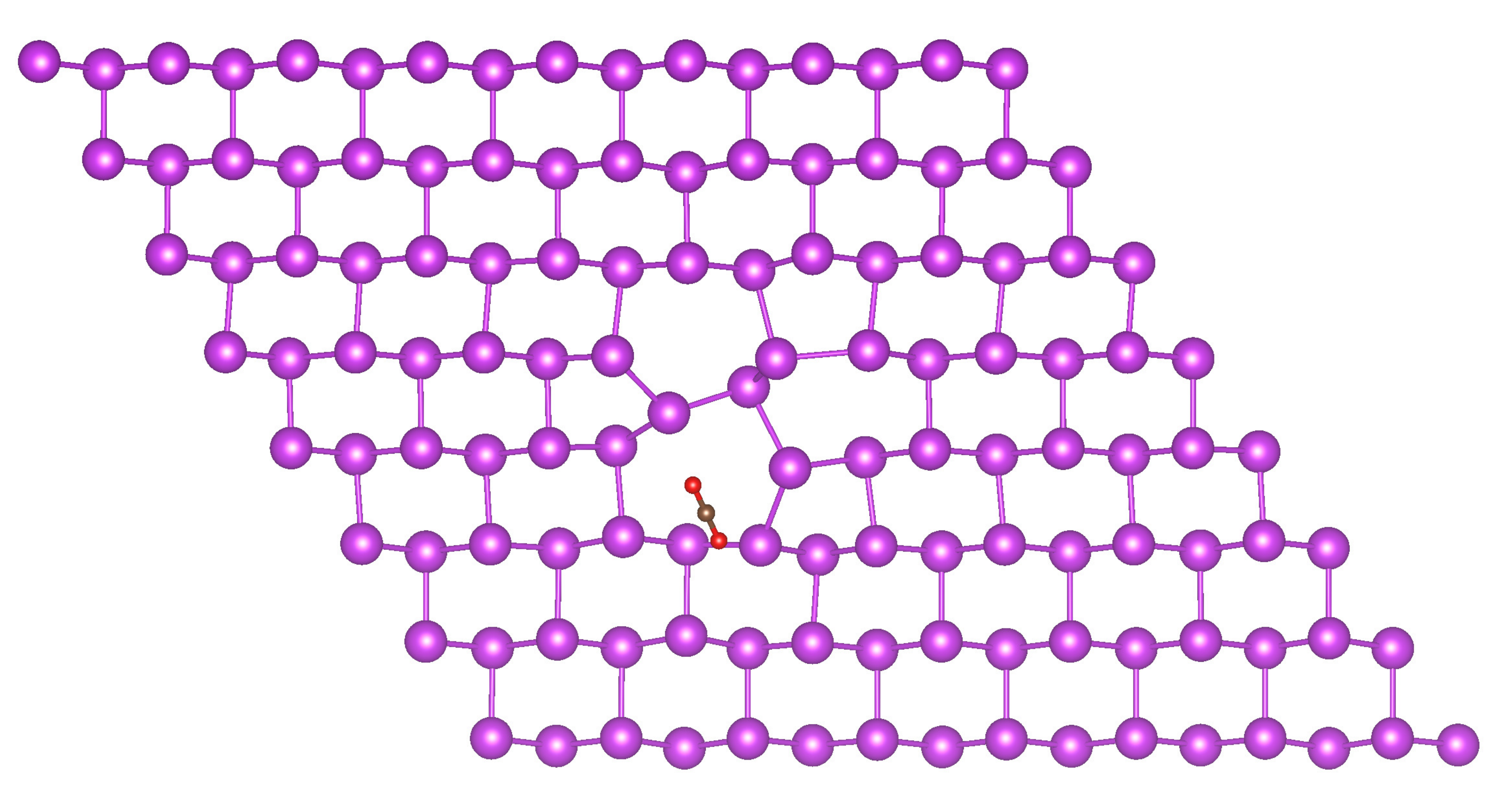}}\\
  \subfloat[]{\includegraphics[width = 3cm, scale=1, clip = true]{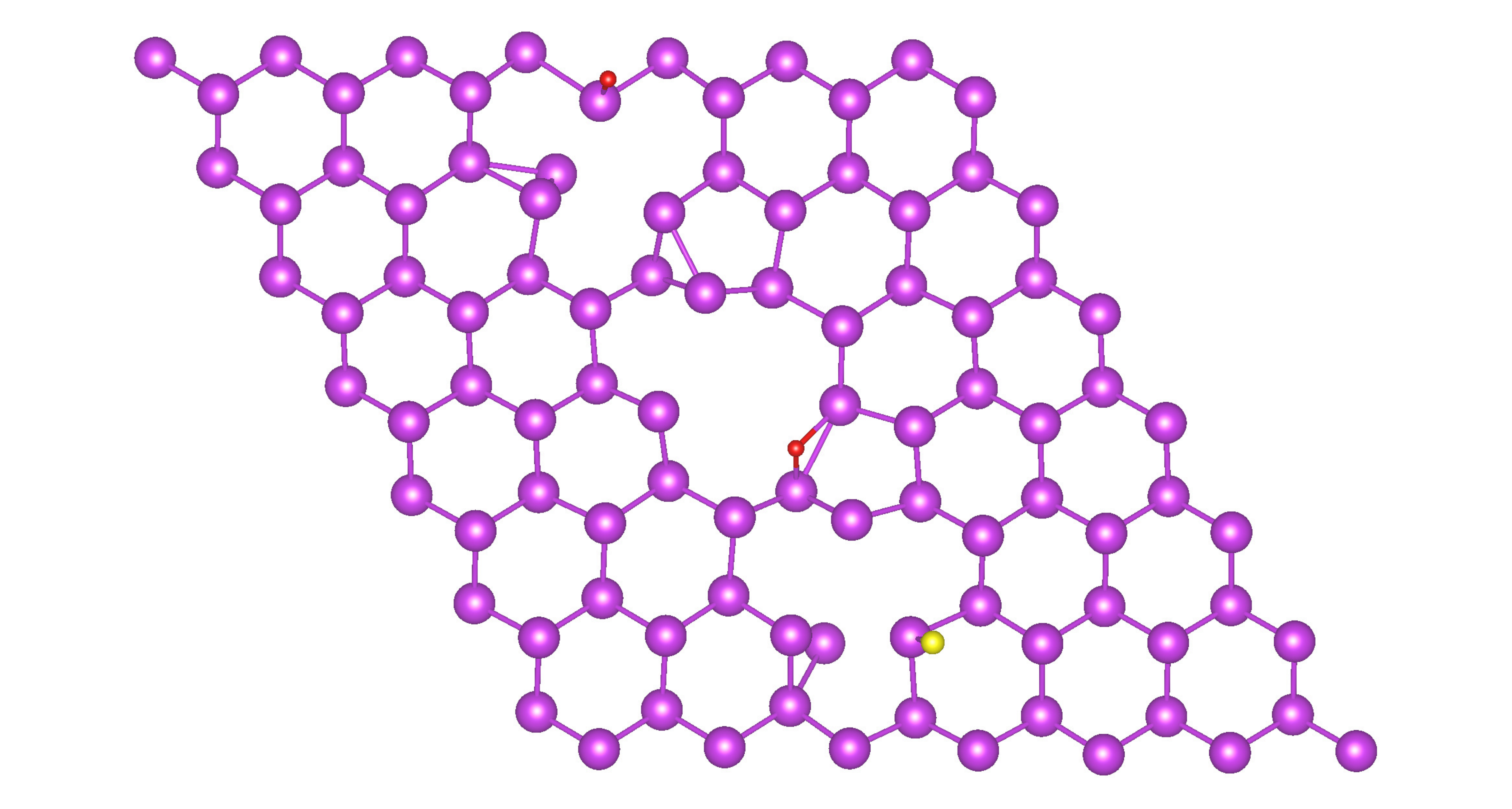}}
  \subfloat[]{\includegraphics[width = 3cm, scale=1, clip = true]{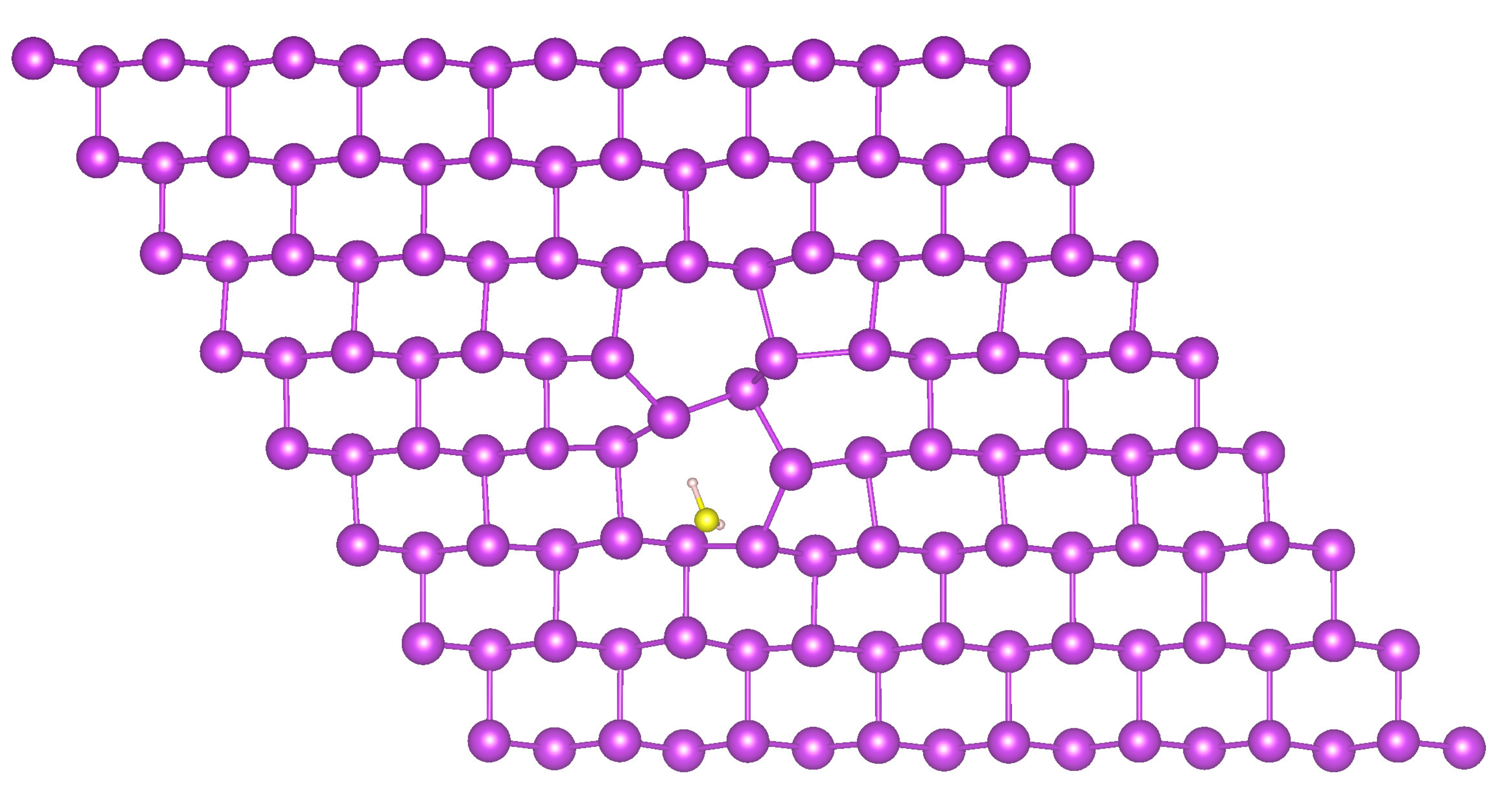}}
  \subfloat[]{\includegraphics[width = 3cm, scale=1, clip = true]{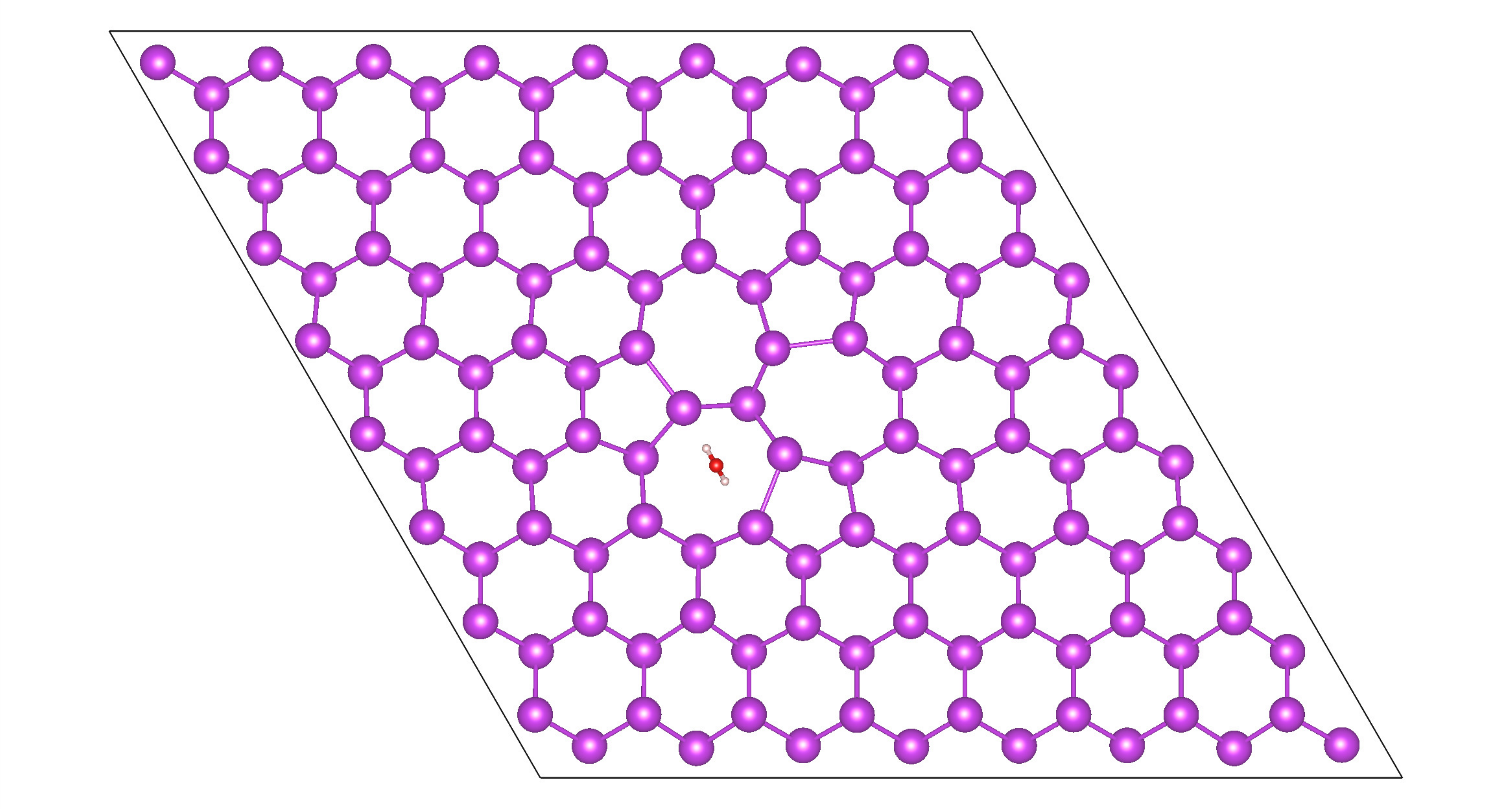}}
  \subfloat[]{\includegraphics[width = 3cm, scale=1, clip = true]{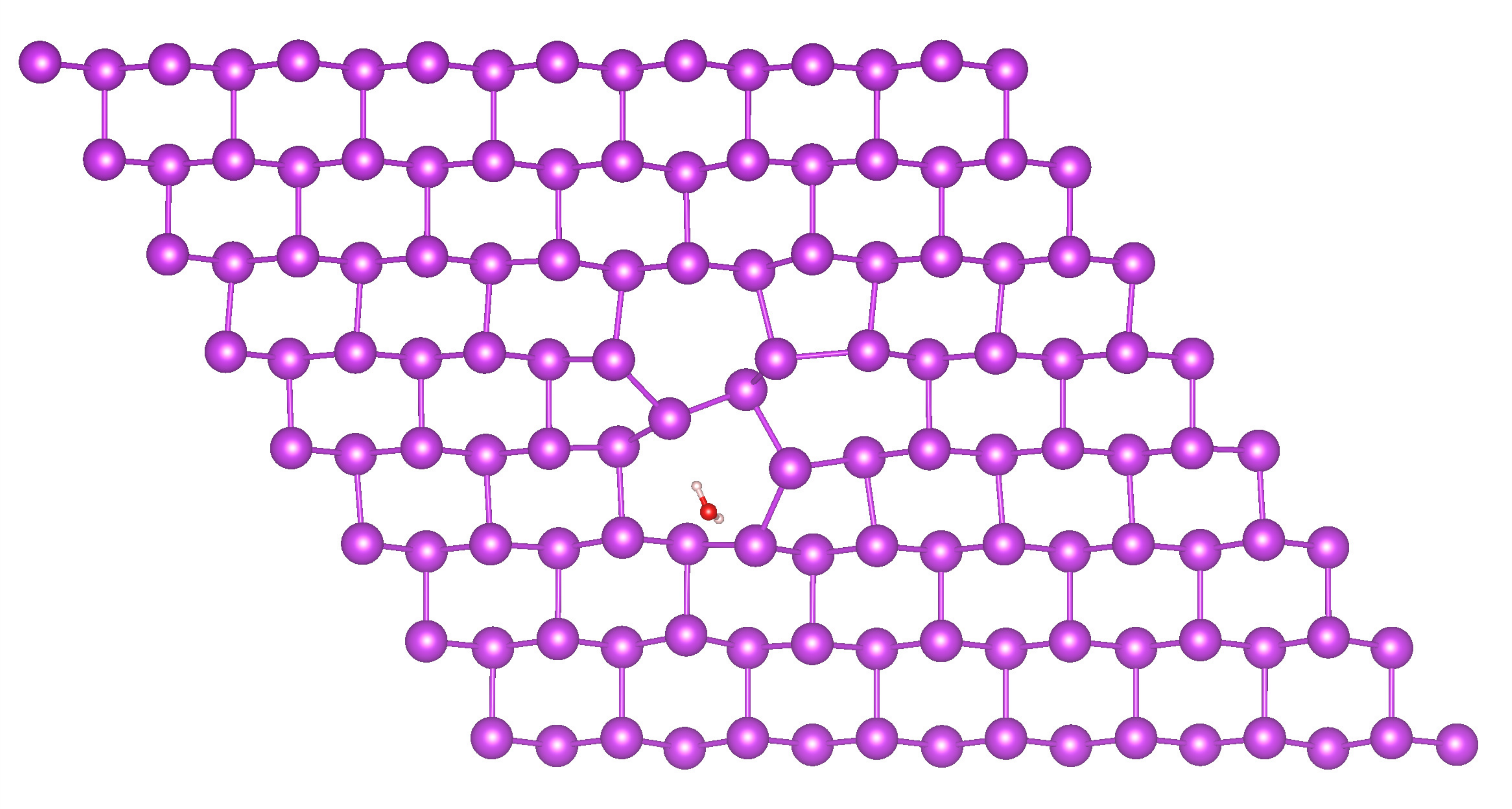}} \\
  \subfloat[]{\includegraphics[width = 3cm, scale=1, clip = true]{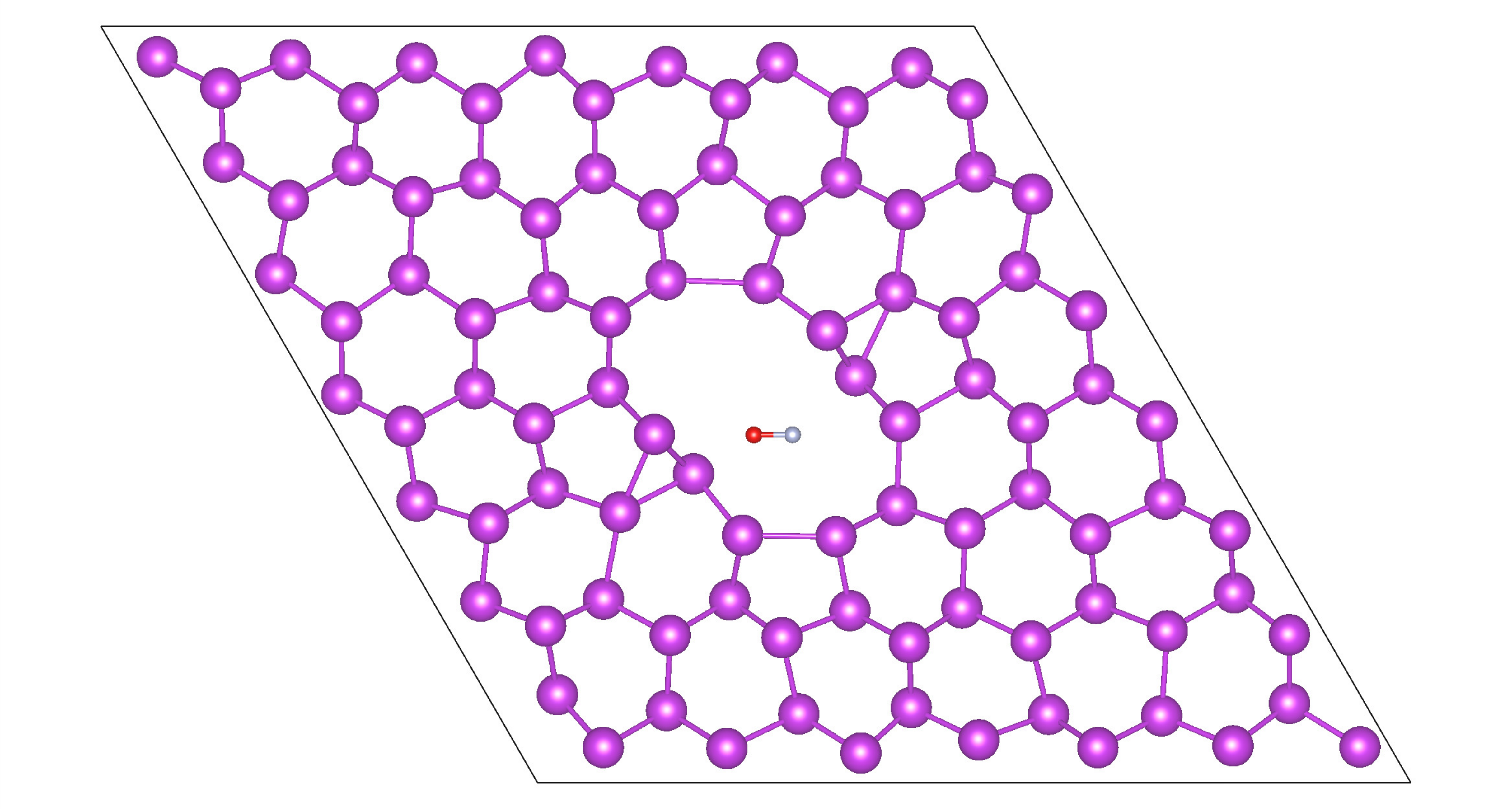}}
  \subfloat[]{\includegraphics[width = 3cm, scale=1, clip = true]{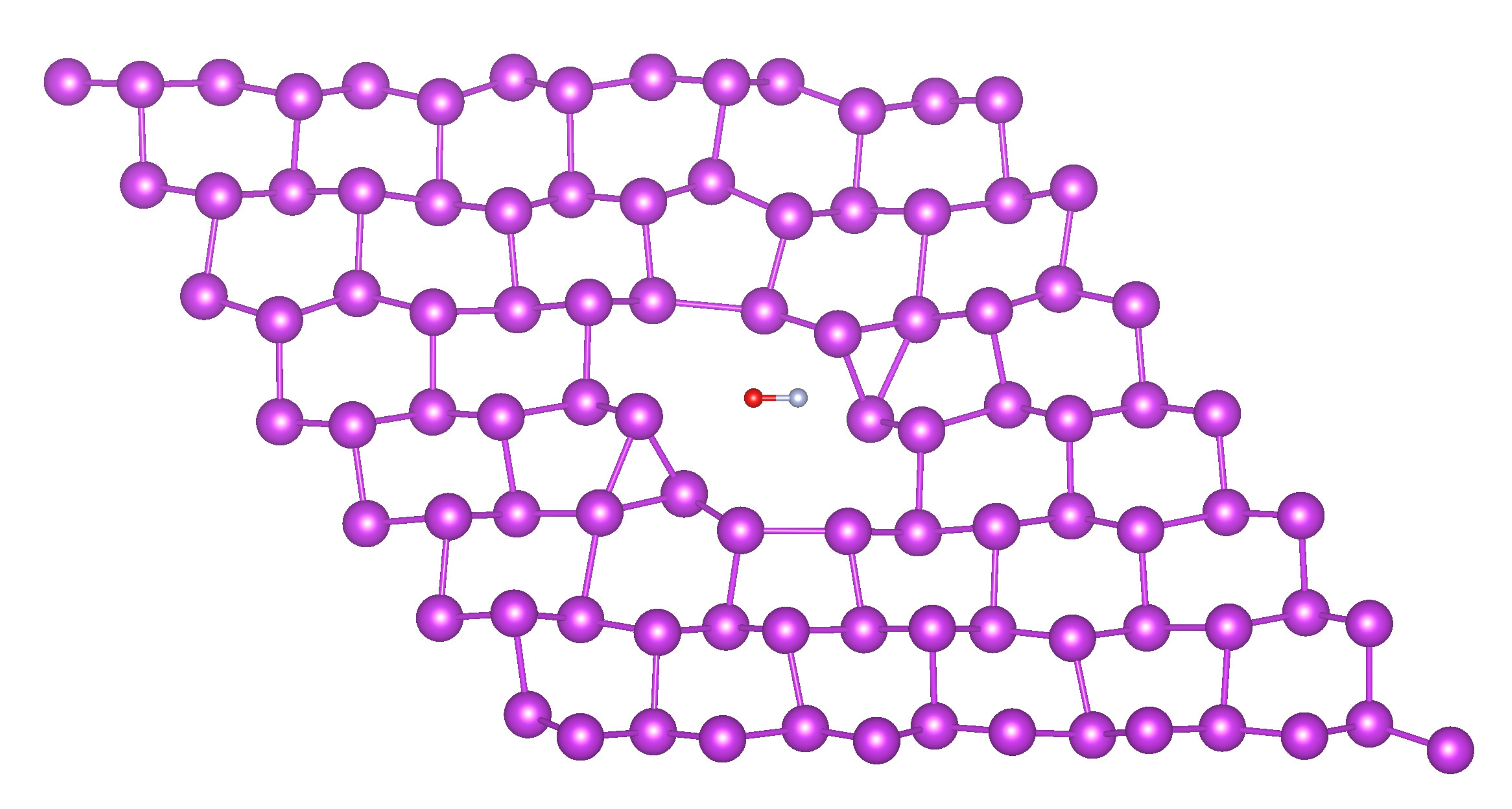}}
  \subfloat[]{\includegraphics[width = 3cm, scale=1, clip = true]{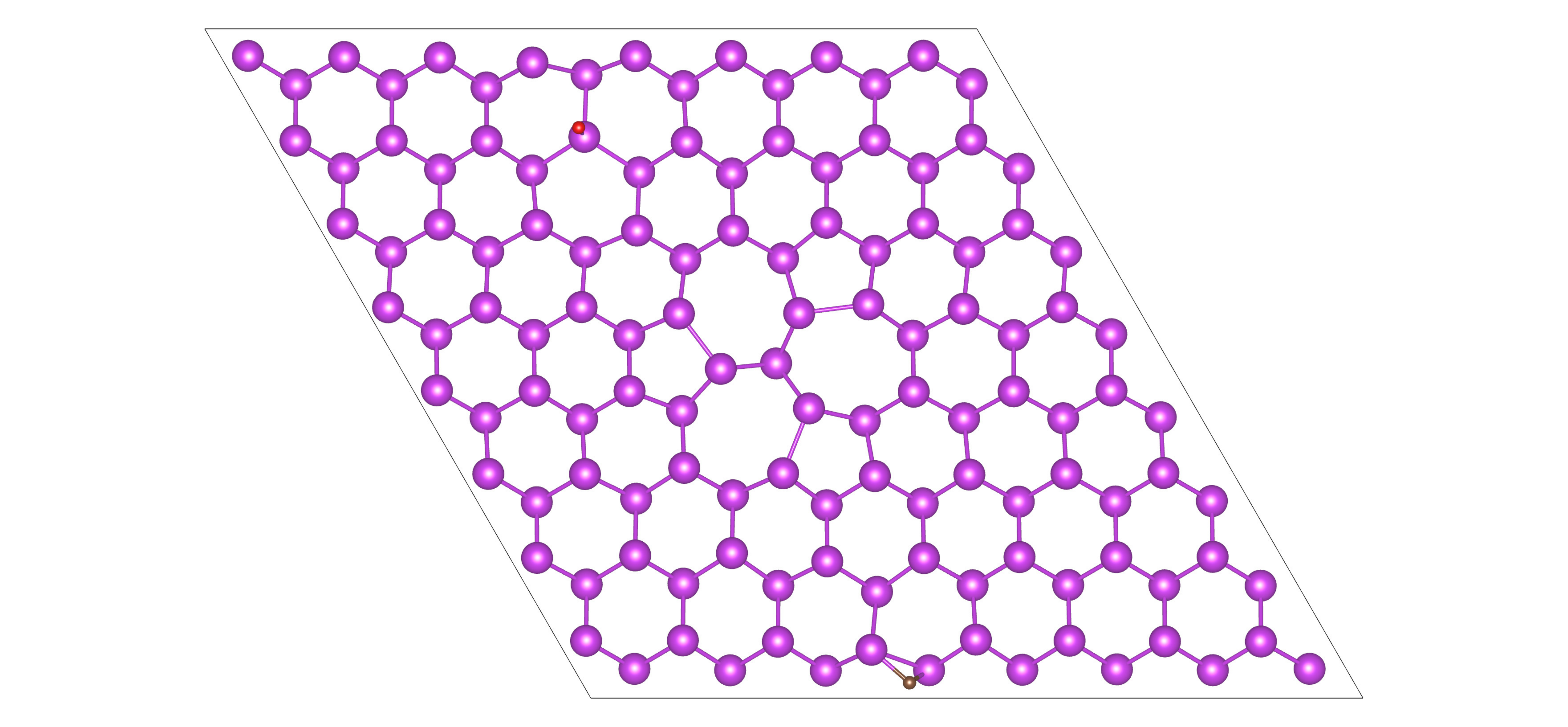}}
  \subfloat[]{\includegraphics[width = 3cm, scale=1, clip = true]{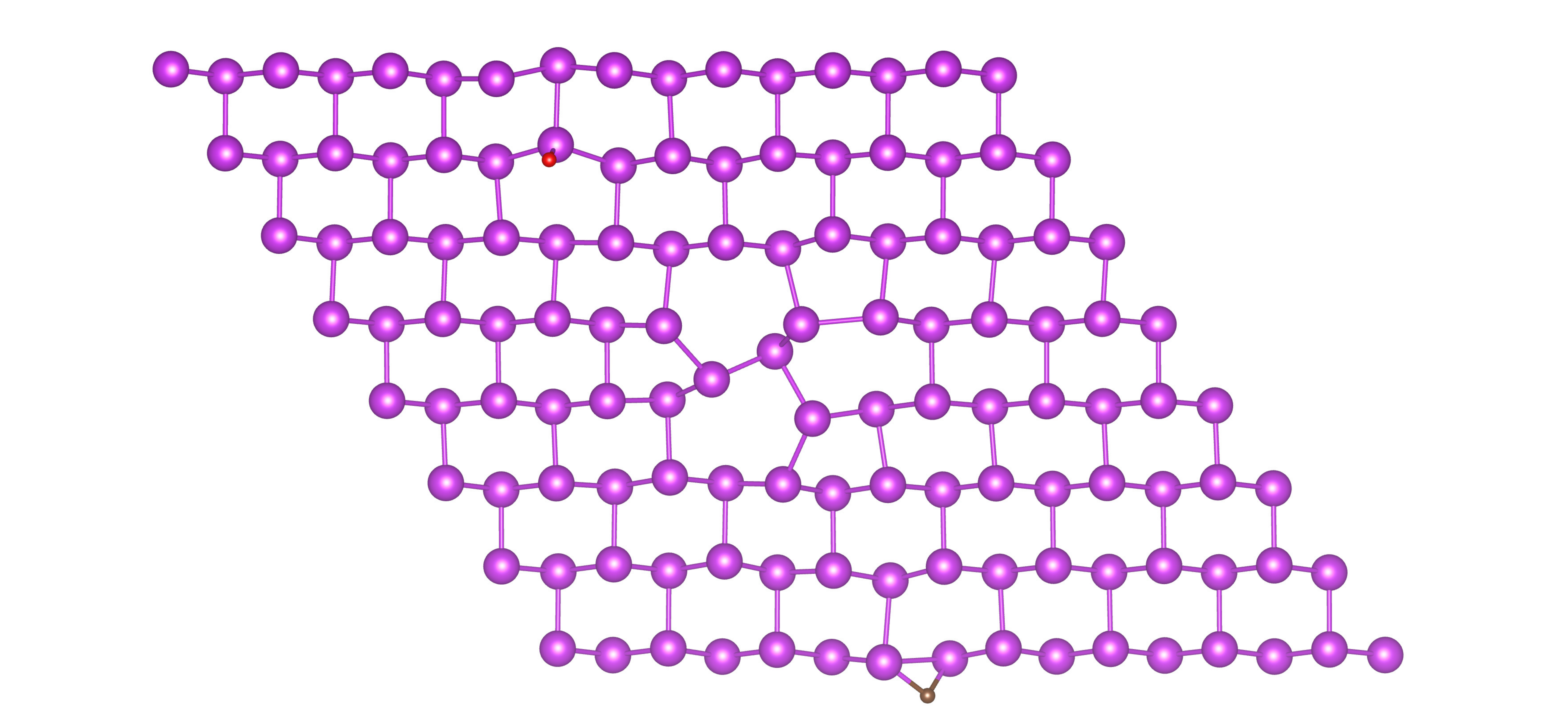}}
  \caption{Adsorption of small molecules in bismuthene nanopores on a tetravacancy (P4) calculated within GGA-PBE. a) H$_2$, b) N$_2$, c) O$_2$, d) CO$_2$, e) SO$_2$, f) NO$_2$, g) H$_2$S, h) H$_2$O and i) NO and j) CO.}
  \label{fig:S2}
\end{figure*}

\end{document}